\documentclass[aps, 10pt, prd,
               notitlepage, twocolumn, superscriptaddress,
               longbibliography, tightenlines,
               nofootinbib, floatfix]{revtex4-2}
\usepackage{tabularx}
\usepackage{graphicx}
\usepackage{epsfig}
\usepackage{amssymb}
\usepackage{amsmath}
\usepackage{tensor}
\usepackage{newtxtext}
\usepackage[varg]{newtxmath}
\usepackage{enumerate}
\usepackage{mathtools}
\usepackage[normalem]{ulem}
\usepackage{bm}
\usepackage[OT1]{fontenc}
\usepackage{multirow}
\usepackage{makecell}
\usepackage{soul}
\usepackage{anyfontsize}
\usepackage{microtype}
\usepackage[dvipsnames]{xcolor}
\usepackage[colorlinks,linktocpage,breaklinks]{hyperref}
\usepackage[capitalize]{cleveref}
\usepackage{booktabs}

\hypersetup{colorlinks=true,
            citecolor=NavyBlue,
            linkcolor=NavyBlue,
            urlcolor=NavyBlue}

\newcommand{\dd}{{\rm d}}

\newcommand{\HEOB}{\hat{H}_{\mathrm{EOB}}}
\newcommand{\Heff}{\hat{H}_{\mathrm{eff}}}

\newcommand{\pr}{p_{r_*}}

\newcommand{\nnl}{\nonumber\\}
\newcommand{\nnlq}{\nonumber\\&\quad}
\def\vct#1{{\bm{#1}}}

\begin{document}

\title{Relativistic tidal response and mode excitation in neutron-star binaries}

\date{November 10, 2026}

\author{Yong Gao}
\email{yong.gao@aei.mpg.de}
\affiliation{Max Planck Institute for Gravitational Physics (Albert Einstein Institute), 14476 Potsdam, Germany}

\author{Jan Steinhoff}
\email{jan.steinhoff@aei.mpg.de}
\affiliation{Max Planck Institute for Gravitational Physics (Albert Einstein Institute), 14476 Potsdam, Germany}

\begin{abstract}
Tidal excitation links gravitational-wave emission to the internal
dynamics of neutron stars. We study the dynamical tidal response of nonrotating stars with composition
stratification, elastic crusts and possible first-order phase transitions.
Using analytical Mano--Suzuki--Takasugi exterior solutions, we match
stellar perturbations to an effective field theory and extract the poles
and residues of the relativistic tidal response. Over the inspiral band,
this response is accurately approximated by a sum of harmonic
oscillators, as in Newtonian theory, each specified by a frequency
$\omega_\alpha$ and a mode-specific tidal deformability $\lambda_\alpha$,
or equivalently a tidal overlap $I_\alpha$. We couple
these oscillators to relativistic effective-one-body dynamics,
including gravitational redshift and frame dragging, to follow mode
excitation and gravitational-wave emission. A decomposition into
driven and free oscillations gives analytical mode amplitudes accurate to
a few percent during inspiral. We also reconstruct the time-dependent
crustal strain. In a representative $1.20+1.50\,M_\odot$
hadronic--hybrid binary, gravity, interface and shear modes produce a
gravitational-wave phase shift of $\lesssim10^{-2}\,{\rm rad}$. Despite this small phase shift, gravity-mode excitation
can overstrain the crust near its base, while the lowest shear mode
produces widespread overstrain.
\end{abstract}
\maketitle

\section{Introduction}

Tidal interactions excite stellar oscillations in binaries
\citep{Cowling:1941,Zahn:1970AA,Alexander:1987MNRAS} and close encounters
\citep{Fabian:1975MNRAS,Press:1977ApJ} and imprint the stars'
internal structure on the gravitational-wave (GW) signal during the late
inspiral of a neutron-star binary
\citep{Shibata:1993qc,Reisenegger:1994a,Lai:1993di,Kokkotas:1995xe}.
To leading order in the adiabatic approximation, this response is
characterized by quadrupolar tidal deformabilities, which relate the
induced mass quadrupoles to the applied tidal fields and depend on the
dense-matter equation of state
(EOS)~\citep{Hinderer:2007mb,Damour:2009vw,Binnington:2009bb}.
The resulting changes in the orbital dynamics and GW emission accumulate
in the waveform phase, allowing GW observations to constrain the
EOS~\citep{Flanagan:2007ix,Hinderer:2009ca,VinesWaveform2011,Wade:2014vqa}.
Analyses of GW170817 have constrained the tidal deformabilities, stellar
radii, and pressure of matter at supranuclear
densities~\citep{LIGOScientific:2017vwq,LIGOScientific:2018cki,LIGOScientific:2018hze,Annala:2017llu,De:2018uhw,Most:2018hfd,Essick:2019ldf}.

As the binary inspirals, the increasing tidal frequency probes the
finite-frequency response of the star, the so-called dynamical tides. In Newtonian theory, dynamical tides
are described by the excitation of stellar normal modes
\citep{Cowling:1941,Zahn:1970AA,Press:1977ApJ,Alexander:1987MNRAS,Shibata:1993qc,Reisenegger:1994a,Lai:1993di,Kokkotas:1995xe}.
Let $Q_{\ell m}$ and $E_{\ell m}$ denote the spherical-harmonic components
of the induced mass multipole and the applied tidal moment, with
$\ell\ge2$ and $-\ell\le m\le\ell$. Their Fourier amplitudes, denoted by
tildes, are related by the linear tidal response function
$\tilde F_\ell(\omega)$. For a nonrotating, spherical star in the absence of dissipation,
the response takes the form
\citep{Chakrabarti:2013xza,Andersson:2019ahb,Passamonti:2020fur,Passamonti:2022yqp}
\begin{equation}
 \tilde Q_{\ell m}=-\tilde F_\ell\,\tilde E_{\ell m}\,,\quad
 \tilde F_\ell(\omega)=\sum_n
 \frac{\lambda_{n\ell}\omega_{n\ell}^2}
 {\omega_{n\ell}^2-\omega^2}\,.
 \label{eq:intro-response}
\end{equation}
Here $\omega$ is the tidal driving angular frequency, $n$ labels the
normal modes at fixed $\ell$, and $\lambda_{n\ell}$ is the contribution
of the mode with frequency $\omega_{n\ell}$ to the static multipolar
deformability. Spherical symmetry makes $\tilde F_\ell$, $\omega_{n\ell}$,
and $\lambda_{n\ell}$ independent of $m$. Summing over all modes at fixed
$\ell$ gives $\tilde F_\ell(0)=\sum_n\lambda_{n\ell}$ in the static limit.
The poles at $\omega=\omega_{n\ell}$ connect the adiabatic tide to
resonant mode excitation. The Newtonian derivation is given in
\cref{sec:newton_amplitude,sec:tidal-response}.

For the leading quadrupolar tide in typical neutron-star models,
the fundamental ($f$) mode contributes most of the static deformability
and dominates the nonresonant response
\citep{Kokkotas:1995xe,Andersson:2019ahb,Passamonti:2020fur,Passamonti:2022yqp}.
Its frequency dependence enhances tidal effects and changes the GW phase
even when resonance is not reached before merger
\citep{Hinderer:2016eia,Steinhoff:2016rfi,Schmidt:2019wrl,Andersson:2019dwg,Kuan:2022etu,Kuan:2023qxo,Gao:2025esj}.
This imprint can constrain the $f$-mode frequency
\citep{Pratten:2019sed,Williams:2022vct}, while neglecting dynamical tides can bias
EOS inference \citep{Pratten:2021pro}.
Rapid spin antialigned with the orbital angular momentum can lower the
relevant inertial-frame mode frequency enough to allow resonance before
merger \citep{Ho:1998hq,Ma:2020rak,Steinhoff:2021dsn,Yu:2024uxt}, with large GW phase
shifts demonstrated in fully relativistic simulations
\citep{Kuan:2024jnw}.

Low-frequency modes below the $f$ mode offer complementary
information about the stellar interior. Each mode that couples to the
tide contributes an additional pole to \cref{eq:intro-response}, with a
mode-specific tidal deformability $\lambda_{n\ell}$. Composition gradients provide
buoyancy and support gravity ($g$) modes
\citep{Reisenegger:1992ApJ,Lai:1993di}, provided the relevant reactions
are slow on the oscillation timescale
\citep{Lai:1993di,Andersson:2019mxp,Counsell:2023pqp}.
Examples include lepton gradients
\citep{Kantor:2014lja,Passamonti:2015oia,Yu:2016ltf,Counsell:2024pua},
hyperon gradients
\citep{Dommes:2015wul,Yu:2017cxe,Tran:2022dva,Thakur:2026cac}, and stratification
in quark--hadron mixed phases
\citep{Wei:2018tts,Jaikumar:2021jbw,Zhao:2022toc}.
Sharp density discontinuities can support interface ($i$) modes, a class
of $g$ modes whose buoyancy is concentrated at the interface
\citep{Finn:1987ads}. Such modes have been studied for first-order
hadron--quark transitions
\citep{Lau:2020bfq,Miao:2023jqe,Counsell:2025hcv,Pereira:2025xsi}, and can
also arise from intrahadronic transitions involving $\Delta$ resonances
\citep{Sun:2026agt,Canullan-Pascual:2026fyb}.
The inclusion of a solid crust adds shear ($s$) modes
\citep{McDermott:1988ApJ,Passamonti:2020fur} and can support elastic
interface modes at fluid--solid boundaries
\citep{McDermott:1988ApJ,Piro:2004bh,Tsang:2011ad,Zhu:2022pja}.
In stratified stars, interface oscillations can mix with composition
$g$ modes \citep{Kruger:2014pva,Gao:2025aqo}.

The tidal driving frequency can sweep through these low-frequency modes
during the inspiral. Resonant excitation exchanges energy and angular
momentum between the orbit and the star, leaving a free oscillation
after the crossing
\citep{Lai:1993di,Reisenegger:1994a,Shibata:1993qc,Kokkotas:1995xe,Ho:1998hq,Flanagan:2006sb,Pnigouras:2025muo}.
For linear tides in nonrotating stars on quasicircular orbits, many
low-frequency modes couple weakly to the tide and produce small GW phase
shifts \citep{Lai:1993di,Yu:2016ltf,Yu:2017cxe,Gao:2025aqo}.
Rotation modifies mode frequencies and tidal couplings and can enhance
the resonant phase shift \citep{Ho:1998hq,Xu:2017hqo,Pnigouras:2022zpx}.
Eccentric orbits introduce additional tidal harmonics and allow repeated
mode excitation
\citep{Alexander:1987MNRAS,Gold:2011df,Tsang:2013mca,Chirenti:2016xys,Yang:2019kmf,Vick:2019cun}, which can
enhance the detectability of $g$-mode resonances \citep{Takatsy:2026doc}.
Nonlinear dynamics can also change the excitation and GW phase shift
\citep{Yu:2022fzw,Kwon:2024zyg,Kwon:2025zbc,Reboul-Salze:2025gyi,Pitre:2025qdf,Guo:2026jix}.
Sensitivity estimates
\citep{Andersson:2017iav,Ho:2023shr}, waveform-mismatch calculations
\citep{Revilla-Pena:2025ztm}, and Bayesian forecasts
\citep{Gittins:2026ntx} have assessed their detectability with current and
future GW detectors.

The local deformation of the crust provides another measure of
tidal excitation \citep{Penner:2011br,Pan:2020tht}. A mode with a small effect on the orbit can generate
large strain where its displacement varies rapidly, potentially driving
the crust beyond its elastic limit
\citep{Tsang:2011ad,Passamonti:2020fur,Kuan:2021jmk,Gao:2025aqo}.
Precursor emission has been observed before short gamma-ray bursts
\citep{Troja:2010zm}. Crust failure and the subsequent transfer of energy
to the magnetosphere have been proposed as a source of this emission
\citep{Tsang:2011ad,Suvorov:2020tmk,Kuan:2021sin,Suvorov:2022ldw,Kuan:2023kif,Neill:2021lat,Neill:2022psd,Suvorov:2024cff,Most:2024eig}.

Interpreting the GW signatures of dynamical tides and the associated
crust dynamics requires a description in general relativity (GR) at three levels:
(i) stellar oscillations with detailed microphysics, (ii) the tidal
response that determines their excitation, and (iii) their coupling to
relativistic orbital dynamics. The stellar oscillation problem has a
well-established foundation
\citep{Thorne:1967ApJ,Lindblom:1983ps,Detweiler:1985zz}, with subsequent
calculations incorporating viscous damping
\citep{Cutler:1987ApJ,Cutler:1990ApJ}, stratification, crust elasticity and sharp
interfaces
\citep{Finn:1990ads,Kruger:2014pva,Kuan:2021jmk,Kuan:2022bhu,Tonetto:2020bie,Miao:2023jqe,Kruger:2024fxn,Gao:2025aqo,Dong:2026qrn}.

The tidal-response problem is more subtle.
Relativistic calculations of static tides include composition
stratification, elastic crusts and sharp density interfaces
\citep{Penner:2011pd,Gittins:2020mll,Pereira:2020jgv,Shen:2026ehk}. In Newtonian gravity,
linearity separates the applied tidal potential from the stellar
self-field, and a complete set of orthogonal modes yields the exact
oscillator sum in \cref{eq:intro-response}, with weights fixed by overlap integrals
\citep{Reisenegger:1994b,Chakrabarti:2013xza,Andersson:2019ahb,Passamonti:2020fur,Passamonti:2022yqp}. For a radiating
relativistic star, the quasinormal-mode problem is not self-adjoint, and
the modes alone do not span the full causal response
\citep{Kokkotas:1999bd,Pitre:2023xsr,Miao:2026rie}. Moreover, extracting
an applied tide and an induced multipolar field from a dynamical metric
requires a specified convention
\citep{Poisson:2020vap,HegadeKR:2024agt,Katagiri:2024wbg}.

Studies of crustal and phase-transition resonances have often combined
relativistic stellar models with Newtonian perturbation equations and
tidal forcing, or prescribed relativistic extensions of Newtonian
overlap integrals
\citep{Tsang:2011ad,Lau:2020bfq,Kuan:2021jmk,Miao:2023jqe,Counsell:2025hcv,Pereira:2025xsi,Gao:2025aqo}.
Such hybrid treatments violate the mode orthogonality underlying the
Newtonian overlap construction, even when the perturbation equations
are Newtonian. Consequently, weak couplings can be contaminated by the
strongly tide-coupled $f$-mode
\citep{Reisenegger:1994b,Miao:2023jqe}.
Even when the mode problem is solved fully relativistically,
determining the forced response requires more than the free
eigenfunctions. It also requires the interior tidal field and its
projection onto the modes. In general relativity, however, the
separation between the applied tidal field and the star's self-field
is not unique, so a relativistic overlap cannot be specified by the
eigenfunction alone \citep{HegadeKR:2026kku}.
A consistent treatment of tidal resonances must therefore formulate
the modes, the interior tidal field, and their coupling within a
common relativistic framework.

The effective field theory (EFT) of extended objects
\citep{Goldberger:2004jt,Goldberger:2005cd} offers a systematic route to
the response. The star is represented by a worldline carrying dynamical
multipole moments, whose response is determined by matching to
relativistic stellar perturbations.
Chakrabarti, Delsate, and
Steinhoff (hereafter CDS)~\citep{Chakrabarti:2013xza,Chakrabarti:2013lua}
derived this effective description in
Newtonian gravity, recovering the exact mode sum in
\cref{eq:intro-response}. Their relativistic matching calculation showed
that the response of a polytropic star can be accurately approximated
by the same oscillator form \citep{Chakrabarti:2013lua}. (See also Ref.~\cite{Saketh:2026inprep} for a very recent and more accurate matching following Ref.~\cite{Saketh:2026trm}.)
Research on relativistic dynamical tidal response has advanced
rapidly since GW170817.
Time-derivative expansions provide a description without explicit modes
\citep{Poisson:2020vap,Pitre:2023xsr}, while direct frequency-domain
calculations resolve fluid-mode resonances
\citep{HegadeKR:2024agt,Andersson:2025iyd}.
Further studies examine dissipation and dense-matter dependence
\citep{HegadeKR:2026iou}, as well as quasi-universal relations among
low-frequency response coefficients \citep{Saes:2025jvr}.
Near-zone matching has also
yielded relativistic mode sums and overlap constructions
\citep{HegadeKR:2025qwj,HegadeKR:2026kku}, and scattering approaches
relate the response and resonant excitation to asymptotic wave
amplitudes \citep{Miao:2025utd,Saketh:2026trm,Saketh:2026inprep}. Complementary analyses
identify regular contributions beyond the pole sum in \cref{eq:intro-response}
\citep{Martinez-Rodriguez:2026omk,Apostolidis:2026qsg} and the direct
and late-time tail contributions to the causal response \citep{Miao:2026rie}.

The third ingredient is to couple the stellar response to
relativistic orbital dynamics. Post-Newtonian corrections enhance the
driving tidal field \citep{Vines:2010ca,Bini:2012gu}, while gravitational
redshift and frame dragging modify the mode evolution and resonance
conditions \citep{Steinhoff:2016rfi}. The effective-one-body (EOB)
description of relativistic binary motion \citep{Buonanno:1998gg} was
extended to adiabatic tidal interactions by \citet{Damour:2009wj}. A relativistic formulation of
dynamical tides, based on a covariant oscillator action and an EOB
Hamiltonian including tidal spin, was developed in 2016
\citep{Hinderer:2016eia,Steinhoff:2016rfi}. Subsequent work incorporated stellar-spin
effects \citep{Steinhoff:2021dsn} and further developed and tested tidal
EOB waveform models
\citep{Nagar:2018zoe,Gamba:2022mgx,YuLau2025,Haberland:2025luz}.

An advantage of EFT is that the same covariant action determines tidal
forcing and orbital backreaction, allowing the matched stellar response
to enter relativistic binary dynamics with consistent definitions
\citep{Chakrabarti:2013lua,Steinhoff:2016rfi,Apostolidis:2026qsg}.
In this work, we extend EFT response matching to nonrotating neutron
stars with elastic crusts, composition stratification and possible
first-order phase transitions. We determine the frequencies and
mode-specific tidal deformabilities of weakly coupled $g$, $s$ and $i$
modes directly from the poles and residues of the relativistic response,
without a Newtonian overlap prescription. We verify the oscillator
approximation in \cref{eq:intro-response} over the inspiral band and use
the same couplings and exterior normalization to connect relativistic
EOB mode excitation and GW emission to the internal strain.

We work in
geometric units with $G=c=1$. Lower-case Latin indices from the start of the
alphabet, $a,b,\dots$, denote four-dimensional spacetime indices running over
$0$--$3$; we use these rather than Greek letters because $\nu$, $\lambda$, and
$\mu$ already stand for the metric potentials and the reduced mass. Lower-case
Latin indices $i,j,\dots$ run over $1,2,3$ and denote either spatial coordinate
indices, when carried by position variables, or components in a local Euclidean
frame comoving with the center of the star, when carried by the tidal,
quadrupole, and spin tensors. Capitalized Latin indices $A,B,\dots$ likewise run
over $1,2,3$ and denote components in the spatial frame co-rotating with the
orbit. Boldface denotes three-vectors carrying such indices. Round, square, and
angle brackets on indices denote symmetrization, antisymmetrization, and
symmetric-tracefree projection, respectively.

\section{Relativistic tidal response}\label{sec:response-action}

The orbital dynamics require the mode frequencies and tidal
couplings. Reconstructing the deformation inside the star also requires
the displacement carried by a given mode quadrupole. We first establish
this normalization in Newtonian theory, then determine its relativistic
counterpart by matching the exterior stellar perturbation to the EFT.
The same matching supplies the mode parameters used in
\cref{sec:eob-tides} and the normalization of the internal strain in
\cref{sec:strain}.

\subsection{Normal modes and tidal response in Newtonian gravity}
\label{sec:newton_amplitude}

Consider a spherical, nonrotating star with density $\rho(r)$ in
an external tidal potential $\Phi(\mathbf x,t)$. To quadratic order in
the Lagrangian displacement $\bm\xi(\mathbf x,t)$, the dynamics follow
from the Lagrangian \citep{Friedman:1978Lagrangian,Shapiro:1983CompactObjects}
\begin{equation}
    \label{eq:la1}
    L_{\rm DT} = \tfrac{1}{2}\langle\dot{\bm\xi}, \dot{\bm\xi}\rangle
              - \tfrac{1}{2}\langle\bm\xi, \mathcal D\bm\xi\rangle
              + \langle\bm f^{\rm ext}, \bm\xi\rangle\,,
\end{equation}
Here
$\langle\bm\xi,\bm\eta\rangle\equiv\int d^3x\,\rho\,\bm\xi^*\!\cdot\!\bm\eta$
is the mass-weighted inner product and $\bm f^{\rm ext}=-\nabla\Phi$ is
the tidal acceleration. The operator $\mathcal D$ includes pressure,
self-gravity and elastic restoring forces. With the appropriate surface
and interface conditions it is self-adjoint under this inner product
\citep{Friedman:2013xza,Passamonti:2020fur}.

The eigenfunctions $\bm\xi_{n\ell m}(\mathbf x)$ of $\mathcal D$ are
labelled by a radial order $n$ and by the angular indices $(\ell, m)$ of
their spherical-harmonic dependence, and satisfy
\begin{equation}
    \mathcal D\,\bm\xi_{n\ell m} = \omega_{n\ell}^2\,\bm\xi_{n\ell m}\,,
    \, \,
    \langle\bm\xi_{n\ell m}, \bm\xi_{n'\ell'm'}\rangle
    = \delta_{nn'}\,\delta_{\ell\ell'}\,\delta_{mm'}\,.
    \label{eq:eigen}
\end{equation}
For a stable background the frequencies $\omega_{n\ell}$ are
real and, by spherical symmetry, independent of $m$. With the
orthonormalization in \cref{eq:eigen}, the tidally induced displacement
has the normal-mode expansion
\begin{equation}
    \bm\xi(\mathbf x, t) = \sum_{n\ell m} q_{n\ell m}(t)\,\bm\xi_{n\ell m}(\mathbf x)\,,
    \,\,
    q_{n\ell m}(t) = \langle\bm\xi_{n\ell m}, \bm\xi\rangle\,.
    \label{eq:modeexp}
\end{equation}
Reality requires $q_{n\ell,-m}=(-1)^m q_{n\ell m}^*$. All mode
sums below include both signs of $m$. Substituting \cref{eq:modeexp}
into \cref{eq:la1} gives
\begin{equation}
    \begin{aligned}
    L_{\rm DT} = \sum_{n\ell m}\!\bigl[&
        \tfrac{1}{2}|\dot q_{n\ell m}|^2
      - \tfrac{1}{2}\,\omega_{n\ell}^2\,|q_{n\ell m}|^2\\
      &+\operatorname{Re}\!\left(
      \langle\bm f^{\rm ext},\bm\xi_{n\ell m}\rangle q_{n\ell m}
      \right)\bigr]\,.
    \end{aligned}
    \label{eq:Lmode}
\end{equation}

The tidal field is raised by a companion of mass $m_2$ at position
$\mathbf r(t)$, with the orbit in the plane $\theta=\pi/2$ of spherical
coordinates $(|\mathbf x|,\theta,\varphi)$ centered on the star. Its
potential at a point $\mathbf x$ inside the star,
$\Phi(\mathbf x, t) = -m_2/|\mathbf x - \mathbf r(t)|$ with
$|\mathbf x| < r(t)$, has the multipole expansion
\begin{equation}
    \Phi(\mathbf x, t) = -m_2\sum_{\ell\ge 2,\, m}\frac{4\pi}{2\ell+1}\,
    \frac{|\mathbf x|^\ell}{r(t)^{\ell+1}}\,
    Y_{\ell m}\!\left(\tfrac{\pi}{2}, \phi(t)\right)\,
    Y^*_{\ell m}(\theta,\varphi)\,,
    \label{eq:Phi-harmonic}
\end{equation}
where $\phi(t)$ is the orbital phase,
and the $\ell=0$ and $\ell=1$ terms, which do not deform the star, have
been dropped.
The coupling term $\langle\bm f^{\rm ext},\,\bm\xi_{n\ell m}\rangle$ in \cref{eq:Lmode}
can then be written as
\begin{equation}
    \langle\bm f^{\rm ext},\,\bm\xi_{n\ell m}\rangle
    = -\frac{N_\ell}{\ell!}\,I_{n\ell}\,E_{\ell m}^{*}(t)\,,
    \label{eq:coupling}
\end{equation}
where $I_{n\ell}$ is the tidal overlap integral~\citep{Press:1977ApJ,Reisenegger:1994b},
\begin{equation}
    I_{n\ell} \equiv
    \langle\,\bm\xi_{n\ell m},\, \nabla(|\mathbf x|^\ell\,Y_{\ell m})\,\rangle\,,
    \label{eq:overlap}
\end{equation}
which is independent of $m$ for a spherical star. We choose the
mode phases so that $I_{n\ell}\geq0$. The tidal moment is
\begin{equation}
    E_{\ell m}(t) \equiv -(2\ell-1)!!\,N_\ell\,
    \frac{m_2}{r(t)^{\ell+1}}\,
    Y_{\ell m}^{*}\!\left(\tfrac{\pi}{2}, \phi(t)\right)\,,
    \label{eq:Elm-orbital}
\end{equation}
and the normalization factor $N_\ell$ is given by
\begin{equation}
    N_\ell = \sqrt{4\pi\,\ell!/(2\ell+1)!!}\,.
\end{equation}
Varying \cref{eq:Lmode} with respect to $q_{n\ell m}^*$ gives
\begin{equation}
    \ddot q_{n\ell m} + \omega_{n\ell}^2\,q_{n\ell m}
    = -\frac{N_\ell}{\ell!}\,I_{n\ell}\,E_{\ell m}(t)\,.
    \label{eq:oscEOM}
\end{equation}
Each mode is a harmonic oscillator driven by $E_{\ell m}$, with
coupling set by $I_{n\ell}$. With the mass-multipole normalization of
\cref{eq:intro-response}, its contribution to the static deformability is
\begin{equation}
    \lambda_{n\ell}
      =\frac{N_\ell^2 I_{n\ell}^2}{\ell!\,\omega_{n\ell}^2}\,,\qquad
    \lambda_\alpha
      =\frac{4\pi I_\alpha^2}{15\omega_\alpha^2}
      \quad (\ell=2)\,.
    \label{eq:lambda-alpha}
\end{equation}
Here $\alpha$ labels quadrupolar modes, with
$\omega_\alpha=\omega_{n2}$ and $I_\alpha=I_{n2}$. Their deformabilities
sum to the static quadrupolar deformability
\citep{Andersson:2019ahb}.

For the quadrupolar tide, it is convenient to normalize each mode
by the mass quadrupole it carries. We therefore replace the displacement
amplitude $q_{\alpha m}$ by
\begin{equation}
    Q_{\alpha m}(t)
      \equiv N_2 I_\alpha q_{\alpha m}(t)
      =\sqrt{2\lambda_\alpha}\,\omega_\alpha q_{\alpha m}(t)\,,
    \label{eq:Q-rescaling}
\end{equation}
retaining modes with nonzero tidal overlap and keeping the azimuthal
index $m$ explicit.
In terms of $Q_{\alpha m}$, the Lagrangian \cref{eq:Lmode} becomes
\begin{equation}
    L_{\rm DT} = \sum_{\alpha,m}\!\left[
        \frac{|\dot Q_{\alpha m}|^2}{4\,\lambda_\alpha\,\omega_\alpha^2}
        - \frac{|Q_{\alpha m}|^2}{4\,\lambda_\alpha}
      - \tfrac{1}{2}\,\operatorname{Re}
          \!\left(E_{2m}^{*}Q_{\alpha m}\right)
    \right]\,.
    \label{eq:Lmode-Q}
\end{equation}
The coupling is now that of a quadrupole in a tidal field, and each
mode enters through $\omega_\alpha$ and $\lambda_\alpha$ only.

\subsubsection{Transformation to the basis of STF tensors}

The effective action of \cref{sec:gr_eft} and the orbital dynamics
of \cref{sec:eob-tides} are written with STF tensors~\citep{Thorne:1980ru} rather than spherical
harmonics, so we now rewrite the Lagrangian in that basis. Expanding
$\Phi$ about the center of the star, again without the $\ell=0$ and
$\ell=1$ terms,
\begin{equation}
    \Phi(t, \mathbf{x}) = \sum_{\ell\geq 2}\frac{1}{\ell!}\,
    x^L\,E_L(t)\,,
    \qquad
    E_L \equiv \partial_L\Phi\big|_{\mathbf{x}=0}\,,
    \label{eq:Phi-expansion}
\end{equation}
where $L = i_1 i_2 \cdots i_\ell$ is a multi-index of length~$\ell$
and $E_L$ is the rank-$\ell$ STF tidal moment.%
\footnote{
    The STF moment $E_L$ is related to its spherical-harmonic
    counterpart $E_{\ell m}$ in \cref{eq:Elm-orbital} through
    $E_{\ell m} = N_\ell\,\mathcal{Y}_L^{*\,\ell m}\,E_L$,
    where the coefficients $\mathcal{Y}_L^{\ell m}$ are defined by
    $Y_{\ell m} = \mathcal{Y}_L^{\ell m}\,n^{\langle L\rangle}$, and
    $N_\ell$ enters through the inverse
    conversion $n^{\langle L\rangle} = N_\ell^{2}\sum_m
    \mathcal{Y}_L^{*\,\ell m}\,Y_{\ell m}$.
}

For $\ell=2$ the tidal moment is the Newtonian tidal tensor,
\begin{equation}
    \label{eq:Eij-newton}
    E_{ij} = \partial_i\partial_j\Phi(\boldsymbol{x})
    \big|_{\boldsymbol{x}=0}\,.
\end{equation}
For the $\ell=2$ component we likewise introduce the STF
mode quadrupole
\begin{equation}
    Q_\alpha^{ij} \equiv N_2\sum_{m=-2}^{2}\mathcal{Y}_{ij}^{2m}\,Q_{\alpha m}\,,
    \label{eq:QSTF-mode}
\end{equation}
which collects the five $m$ components of a given mode $\alpha$
into a single rank-two STF tensor. The total tidally induced
quadrupole is then the sum over all stellar modes,
\begin{equation}
    Q^{ij} = \sum_\alpha Q_\alpha^{ij}\,.
\end{equation}
In this basis the Lagrangian \cref{eq:Lmode-Q} takes the form
\begin{equation}
    L_{\rm DT} = \sum_\alpha\!\left[
        \frac{\dot Q_\alpha^{ij}\,\dot Q_\alpha^{ij}}
             {4\,\lambda_\alpha\,\omega_\alpha^{2}}
      - \frac{Q_\alpha^{ij}\,Q_\alpha^{ij}}{4\,\lambda_\alpha}
      - \tfrac{1}{2}\,E^{ij}\,Q_\alpha^{ij}
    \right]\,.
    \label{eq:Lmode-STF}
\end{equation}
With the canonical momentum
\begin{equation}
P_{\alpha}^{ij} = \frac{\partial L}{\partial \dot{Q}_{\alpha}^{ij}} = \frac{\dot{Q}_{\alpha}^{ij}}{2\lambda_{\alpha}\omega_{\alpha}^2}\,,
\end{equation}
 the Hamiltonian of the dynamical tides reads
\begin{equation}
    H_{\mathrm{DT}} = \sum_\alpha\left[\lambda_\alpha\omega_\alpha^2\, P_\alpha^{ij}P_\alpha^{ij} + \frac{1}{4\lambda_\alpha}\, Q_\alpha^{ij}Q_\alpha^{ij} + \frac{1}{2}\,E^{ij} Q_\alpha^{ij}\right]\,.
\label{eq:H-STF}
\end{equation}
This is the tidal sector that \cref{sec:eob-tides} adds to the
orbital Hamiltonian, with the Newtonian $\omega_\alpha$ and
$\lambda_\alpha$ replaced by their relativistic counterparts.

\subsubsection{Tidal response function}
\label{sec:tidal-response}

The Euler--Lagrange equation of
\cref{eq:Lmode-STF} is
\begin{equation}
    \ddot Q_\alpha^{ij} + \omega_\alpha^{2}\,Q_\alpha^{ij}
    = -\lambda_\alpha\,\omega_\alpha^{2}\,E^{ij}(t)\,,
    \label{eq:mode-EOM}
\end{equation}
which is \cref{eq:oscEOM} in the new variables, with the overlap
integral absorbed into $\lambda_\alpha$.

The frequency dependence of the driven response follows by Fourier
transformation. With the convention
\begin{equation}
    f(t) = \int\frac{d\omega}{2\pi}\,\tilde f(\omega)\,
    e^{-i\omega t}\,,
\end{equation}
the time derivative becomes
$-i\omega$ and \cref{eq:mode-EOM} becomes algebraic,
\begin{equation}
    \bigl(\omega_\alpha^{2} - \omega^{2}\bigr)\,
    \tilde Q_\alpha^{ij}(\omega)
    = -\lambda_\alpha\,\omega_\alpha^{2}\,\tilde E^{ij}(\omega)\,.
\end{equation}
Summed over modes, the quadrupole responds to the tidal field through
a single function of frequency~\citep{Chakrabarti:2013xza},
\begin{equation}
    \tilde Q^{ij}(\omega) = -\tilde F(\omega)\,
    \tilde E^{ij}(\omega)\,,
    \label{eq:Q-response}
\end{equation}
where, using \cref{eq:lambda-alpha}, the tidal response function reads
\begin{equation}
    \tilde F(\omega) = \sum_\alpha
    \frac{\lambda_\alpha\,\omega_\alpha^{2}}
         {\omega_\alpha^{2} - \omega^{2}}
    = \frac{4\pi}{15}\sum_\alpha
    \frac{I_\alpha^{2}}{\omega_\alpha^{2} - \omega^{2}}\,.
    \label{eq:F-total}
\end{equation}
Each term has poles at $\omega=\pm\omega_\alpha$, with residues
fixed by $\lambda_\alpha$ or, equivalently, $I_\alpha$. Thus the pair
$\{\omega_\alpha,\lambda_\alpha\}$ specifies the mode's contribution
to the response.

In the static limit $\omega\to 0$ the response reduces to
$\tilde F(0)=\sum_\alpha\lambda_\alpha$, the adiabatic tidal deformability
of the star~\citep{Flanagan:2007ix,Hinderer:2007mb,Damour:2009vw,Binnington:2009bb}, which is the sense in which
$\lambda_\alpha$ is the tidal deformability of mode $\alpha$.

\subsection{Formulation in general relativity}\label{sec:gr_eft}

The Newtonian derivation uses both a normal-mode basis and a
separate external tidal potential. In GR, outgoing-wave boundary
conditions give a non-self-adjoint quasinormal-mode problem, and its
modes alone do not span the causal response \citep{Kokkotas:1999bd}.
The metric perturbation also contains both the applied tide and the
stellar response, whose separation must be specified
\citep{Pitre:2023xsr,HegadeKR:2024agt,Katagiri:2024wbg,Andersson:2025iyd}.
We therefore determine the response by matching the exterior stellar
perturbation to the EFT of a worldline with dynamical multipoles.
CDS~\citep{Chakrabarti:2013lua}
developed this construction and found
that a relativistic polytrope admits an accurate oscillator
representation, as in \cref{eq:F-total}. Later work has refined the
matching and its interpretation for perfect-fluid stars
\citep{HegadeKR:2024agt,Katagiri:2024wbg,HegadeKR:2025qwj,Saketh:2026trm,Apostolidis:2026qsg,Saketh:2026inprep}.
Below we specify the stellar perturbation problem before constructing
and matching its exterior field.

\subsubsection{Oscillations of relativistic stars}\label{sec:relativistic-osc}

The equilibrium star is described by the static, spherically symmetric line element
\begin{equation}
    \label{eq:metric}
    \dd s^2 = -e^\nu\,\dd t^2 + e^\lambda\,\dd r^2
    + r^2\,\dd\theta^2 + r^2 \sin^2\!\theta\,\dd\phi^2\,,
\end{equation}
where the metric functions $\nu(r)$ and $\lambda(r)$ follow from the Tolman--Oppenheimer--Volkoff (TOV)
equations for a given equation of
state (EOS). The background matter is a perfect
fluid, the crust being unstrained in equilibrium, with energy-momentum
tensor
\begin{equation}
    T^{\rm fluid}_{a b} = \left(\epsilon+p\right) u_a u_b + p g_{ab}\,,
\end{equation}
where $\epsilon$ and $p$ are the energy density and pressure, $u^a$ the four-velocity, and
$g_{ab}$ the metric.

Linear perturbations are treated in the relativistic Lagrangian
formalism~\cite{Friedman:1975ApJ,Andersson:2004nv,Andersson:2020phh,
Friedman:2013xza}, in which the Lagrangian variation $\Delta$ and
the Eulerian variation $\delta$ of any field
are related by
\begin{equation}
    \label{eq:lagrangian_perturbation}
    \Delta = \delta + \mathcal{L}_{\boldsymbol\xi} \,,
\end{equation}
with $\mathcal{L}_{\boldsymbol\xi}$ the Lie derivative along the
displacement $\xi^a$. We decompose all perturbations into spherical
harmonics with harmonic time dependence $e^{i\omega t}$, where
$\omega$ is the (complex) mode frequency. This is the convention of
the stellar-perturbation literature and is opposite to the Fourier
convention of \cref{sec:tidal-response}. It matters only for complex
frequencies, and \cref{eq:mst-qnm-condition} below states how the two
are related. In Regge--Wheeler
gauge~\cite{Regge:1957td} the polar-parity metric perturbation is
parametrized by four radial functions $\{H_0, H_1, H_2, K\}$,
\begin{equation}
    \label{eq:metric_perturbation}
    \delta g_{ab} = -r^\ell\!\left(\!\begin{array}{cccc}
    H_0 e^\nu & i\omega r H_1 & 0 & 0 \\
    i\omega r H_1 & H_2 e^\lambda & 0 & 0 \\
    0 & 0 & r^2 K & 0 \\
    0 & 0 & 0 & r^2\sin^2\!\theta\, K
    \end{array}\!\right)\!Y_{\ell m}\, e^{i\omega t}\,,
\end{equation}
and the displacement vector by two further functions $\{W, V\}$,
\begin{equation}
    \label{eq:displacement_vector}
    \xi^a = r^\ell\!\left(0,\;
    W r^{-1} e^{-\lambda/2},\;
    -V r^{-2}\partial_\theta,\;
    -\frac{V}{r^2\sin^2\!\theta}\partial_\phi\right)
    Y_{\ell m}\, e^{i\omega t}\,.
\end{equation}
For a perfect-fluid star, the linearized Einstein equations $\delta
G_{ab} = 8\pi\,\delta T_{ab}^{\rm fluid}$ determine these six radial functions~\citep{Thorne:1967ApJ,Lindblom:1983ps,Detweiler:1985zz}.

The perturbed energy-momentum tensor
$\delta T_{ab}^{\rm fluid}$ is constructed from the perturbations of
the four-velocity, energy density, and pressure. The four-velocity
perturbation follows from the normalization $u^a u_a = -1$ together
with~\cref{eq:lagrangian_perturbation},
\begin{equation}
    \delta u^a = q_b{}^a\,\mathcal{L}_{\boldsymbol u}\xi^b
    + \tfrac{1}{2}\, u^a u^b u^c\,\delta g_{bc}\,,
\end{equation}
where $q^{ab} = g^{ab} + u^a u^b$ projects orthogonal to $u^a$.

Baryon conservation fixes the Lagrangian variation of the rest-mass
density $\rho$ by the projection of $\Delta g_{ab}$ orthogonal to the
four-velocity,
\begin{equation}
    \label{eq:delta_rho}
    \frac{\Delta\rho}{\rho} = -\tfrac{1}{2}\, q^{ab}\,\Delta g_{ab}\,,
\end{equation}
and the first law of thermodynamics for cold matter
then gives the energy-density perturbations,
\begin{equation}
    \label{eq:delta_epsilon}
    \Delta\epsilon = (\epsilon + p)\,\frac{\Delta\rho}{\rho}\,,
    \qquad
    \delta\epsilon = \Delta\epsilon - \epsilon'\,\xi^r\,.
\end{equation}
The pressure perturbation is fixed by the adiabatic relation
between $\Delta p$ and $\Delta\epsilon$,
\begin{equation}
    \label{eq:adiabatic}
    \frac{\Delta p}{p} = \Gamma\,\frac{\Delta\epsilon}{\epsilon + p}\,,
    \qquad
    \delta p = \Delta p - p'\,\xi^r\,,
\end{equation}
where the adiabatic index depends
on the timescale of the perturbation relative to that of the weak interactions. Matter that stays in
$\beta$-equilibrium has $\Gamma=\Gamma_0$, the index of the background
EOS, while matter of frozen composition has $\Gamma=\Gamma_1$. The
difference between the two measures the composition
stratification and is what supports
the $g$-modes~\citep{Finn:1987ads,Reisenegger:1992ApJ}. The oscillation periods relevant here are much shorter than
the weak-reaction timescales in an
inspiralling neutron star~\citep{Haensel:2002qw,Kantor:2014lja,Zhao:2025pgx}, so we use $\Gamma_1$ throughout.

To account for the elastic crust, we follow the relativistic theory
of elasticity of Carter and Quintana~\citep{Carter:1972,Carter:1973zz,Karlovini:2002fc,Beig:2002pk}
and add to the perturbed energy-momentum tensor a Hookean shear
stress
\begin{equation}
    \label{eq:stress}
    \delta\pi_a{}^b = -2\mu_{\rm s}\,\Delta s_a{}^b\,,
\end{equation}
where $\mu_{\rm s}$ is the isotropic shear modulus (the subscript
distinguishes it from the reduced mass $\mu$ of
\cref{sec:eob-tides}) and $\Delta s_a{}^b$ is the Lagrangian
strain. Because the crust is
unstrained in equilibrium, the Lagrangian and Eulerian variations of the
strain and of the stress coincide, and the strain is the symmetric,
traceless, transverse part of $\Delta g_{ab}$,
\begin{equation}
    \label{eq:strain}
    \Delta s_a{}^b = \tfrac{1}{2}\!\left(q_a{}^c\, q^{db}
    - \tfrac{1}{3}\, q_a{}^b\, q^{cd}\right)\!\Delta g_{cd}\,.
\end{equation}
 Following Refs.~\citep{Finn:1990ads,McDermott:1988ApJ,Kruger:2014pva},
we define the traction functions by
\begin{equation}
    \label{eq:tractions}
    r^{\ell-2}T_1 Y_{\ell m} e^{i\omega t} \equiv \delta\pi_r{}^r,\qquad
    r^{\ell-3}T_2\,\partial_\theta Y_{\ell m} e^{i\omega t}
    \equiv \delta\pi_r{}^\theta\,,
\end{equation}
These functions
vanish in the fluid region and supply the junction conditions
at the solid--fluid interface. The remaining components of
$\delta\pi_a{}^b$ are determined algebraically by $\{T_1, T_2, V\}$,
$\mu_{\rm s}$, and the background metric. The crust thus raises the number of radial
functions to eight, $\{H_0, H_1, H_2, K, W, V, T_1,
T_2\}$. The full equations and
junction conditions are given in
Appendix~\ref{app:interior}.

Outside the star, the polar metric perturbations obey the
Zerilli equation~\citep{Zerilli:1970se,Fackerell:1971ApJ,Lindblom:1983ps}.
We apply the Chandrasekhar transformation~\citep{Chandrasekhar:1975nkd}
to work with the Regge--Wheeler (RW) function $X$, whose analytic MST
solutions will be used below. It satisfies
\begin{equation}\label{eq:RW}
    \frac{\dd^2 X}{\dd r_*^2}
    +\!\left[\omega^2 - \!\left(1 - \tfrac{2 M}{r}\right)\!
              \frac{\ell(\ell+1) - 6 M/r}{r^2}\right]\! X
    = 0\,,
\end{equation}
where $r_* = r + 2M\log(r/2M-1)$ is the tortoise coordinate and $M$
is the mass of the star. At the surface $r=R$, the same
transformation determines $X(R)$ and $X'(R)$ from the interior metric
perturbations $H_1(R)$ and $K(R)$.

Far from the star the two independent solutions of~\cref{eq:RW}
asymptote to outgoing and ingoing gravitational waves. For a generic
real frequency $\omega$ the interior solution matches onto
a mixture of the two, so that the star is driven by a
wave incident from infinity. The quasinormal modes are the discrete frequencies
at which the ingoing component vanishes, so that the star oscillates
freely and radiates only outgoing waves, the imaginary part of $\omega$ being the gravitational-wave
damping rate.

\subsubsection{Effective action and the exterior solution}
\label{sec:eft-action}

In the EFT the star is replaced by a point
particle carrying a dynamical quadrupole, with worldline action~\citep{Goldberger:2004jt,Chakrabarti:2013lua}
\begin{equation}\label{eq:Stidal}
    S_{\rm eff} = \int \dd\tau\left[-M - \tfrac{1}{2}\,E_{ij}\,Q^{ij}
    + \cdots\right]\,,
\end{equation}
where $\tau$ is the proper time, $E_{ij}$ is the electric part of the
Weyl tensor in the rest frame of the particle, the GR analogue of the
Newtonian tidal tensor, and $Q^{ij}$ is the dynamical quadrupole. Its
linear response has the same form as \cref{eq:Q-response},
$Q=-\tilde F(\omega)E$ for each harmonic component, with the components
normalized as in \cref{sec:newton_amplitude}. This action specifies which tidal field and induced quadrupole
define $\tilde F$. The numerical stellar solution supplies their combined
exterior field $X$, so extracting the response requires matching that
field to the point-particle description. The Newtonian limit illustrates
the separation, since the applied tide and induced multipole have
distinct radial dependence.

We first illustrate the matching in Newtonian gravity, retaining
finite-frequency stellar dynamics~\citep{Chakrabarti:2013xza}. In the
exterior, the weak-field, near-zone limit $M/r\ll1$ and $|\omega|r\ll1$
neglects wave propagation. The Newtonian potential $\Phi$ is then
reconstructed from $X$ as
\begin{equation}
    \Phi = -\frac{1}{2\pi}\int \dd\omega \sum_{\ell m} e^{i\omega t}\, Y_{\ell m}\, \frac{1}{2}\!\left[\frac{\dd}{\dd r} + \frac{\ell(\ell+1)}{2r}\right]\! X_{\ell m\omega}\,,
\end{equation}
where $X_{\ell m\omega}$ are solutions to the Newtonian limit of \cref{eq:RW}. The generic solution can be written as
\begin{equation}
    \label{eq:X_newton}
    X_{\ell m\omega} = C_1(\omega)\,r^{\ell+1} + C_2(\omega)\,r^{-\ell}\,,
\end{equation}
The coefficients $C_1(\omega)$ and $C_2(\omega)$ are fixed by
matching to the finite-frequency interior solution.
The first term is the growing
solution, which corresponds to the external tidal field,
and the second is the $2^\ell$-pole
field of the star. The response
function is the ratio of the two,
\begin{equation}
    \label{eq:newton_c21}
    \tilde F(\omega) = \frac{\ell(\ell-1)}{(\ell+1)(\ell+2)(2\ell-1)!!}\,
    \frac{C_2}{C_1}
    \;\xrightarrow{\ell=2}\;\frac{1}{18}\,\frac{C_2}{C_1}\,,
\end{equation}
In this Newtonian problem, $C_1(\omega)=0$ selects a free
oscillation with no applied tide. These are the normal modes of
\cref{sec:newton_amplitude}, and the ratio reproduces the mode sum
\cref{eq:F-total}.

By contrast, setting $\omega=0$ in the relativistic exterior
problem gives the static solution, expressed in terms of the Gauss
hypergeometric function $_2F_1$,
\begin{align}
X = & C_1\, r^{\ell+1}\, {}_2F_1(-\ell-2,\, 2-\ell,\, -2\ell;\, 2M/r) \nonumber\\
     &+ C_2\, r^{-\ell}\, {}_2F_1(\ell-1,\, \ell+3,\, 2(\ell+1);\, 2M/r)\,.
\end{align}
For $M/r \to 0$ both hypergeometric
functions tend to unity and the solution reduces to the Newtonian form
\cref{eq:X_newton}, so the
integration constants keep their meaning as the amplitudes of the external
field and of the response. The static response function is the
adiabatic tidal deformability $\lambda_{\ell}$,
\begin{equation}
    \tilde F(\omega=0) = \lambda_\ell
    = \frac{2}{(2\ell-1)!!}\, k_\ell\, R^{2\ell+1}\,,
\end{equation}
where $R$ is the stellar radius and $k_\ell$ the dimensionless tidal
Love number of multipole order $\ell$~\citep{Hinderer:2007mb,Damour:2009vw,Binnington:2009bb}.

At finite
frequency \cref{eq:RW} can still be solved analytically, by the
MST series of hypergeometric
functions~\cite{Mano:1996mf,Sasaki:2003xr}. The construction provides two
pairs of series solutions, $\{X_0^{\nu},
X_0^{-\nu-1}\}$ and $\{X_C^{\nu}, X_C^{-\nu-1}\}$, which differ in where they converge. The first pair converges at
any finite radius but not at spatial infinity, and the second pair
converges everywhere except at $r=2M$, so it covers the whole exterior of
the star. Here $\nu$ is the renormalised angular momentum (not to be
confused with the metric potential of \cref{eq:metric}), which is fixed by the
convergence of the series and reduces to $\ell$ as $M\omega\to0$. Each
solution admits both representations, related by the connection
coefficients $K_\nu$ and $K_{-\nu-1}$,
\begin{equation}
    \label{eq:mst_x0}
    X_0^{\nu} = K_\nu\, X_C^{\nu}\,, \qquad
    X_0^{-\nu-1} = K_{-\nu-1}\, X_C^{-\nu-1}\,.
\end{equation}
For the present application we work with the rescaled solutions
$X_N^{\nu} \equiv N_{\nu}\, X_C^{\nu}$, where the normalization
constants $N_{\nu}$ are chosen so that $X_N^{\nu}$ are real and have
unit asymptotic amplitude. Outside the star, the Regge--Wheeler
function then admits the unique decomposition
\begin{equation}
    \label{eq:X-decomposition}
    X= A_1\, X^{\nu}_N + A_2\, X^{-\nu-1}_N\,,
\end{equation}
with the amplitudes $A_1$ and $A_2$ determined by matching to the
interior solution at the stellar surface. Explicit expressions for the
MST series and the connection coefficients are collected in
Appendix~\ref{app:MST}; see
also~\cite{Mano:1996mf,Sasaki:2003xr,Chakrabarti:2013lua}.

We write $u$ for $X_N^{\nu}$ and $v$ for $X_N^{-\nu-1}$. Evaluated
at the surface, \cref{eq:X-decomposition} and its radial derivative form a
linear system for the two amplitudes, with $X(R)$ and $X'(R)$ supplied by
the Chandrasekhar transformation. Its solution is
\begin{equation}
\begin{aligned}
    A_1&=\frac{\mathcal W[X,v]}{W_*}\,,\qquad
    A_2=\frac{\mathcal W[u,X]}{W_*}\,,\\
    W_*&\equiv\mathcal W[u,v]=\omega\sin(\alpha_\nu-\alpha_{-\nu-1})\,,
\end{aligned}
    \label{eq:A-wronskian}
\end{equation}
where $\mathcal W[f,g]\equiv f\,\partial_{r_*}g-g\,\partial_{r_*}f$ is the
Wronskian of two solutions of the RW equation. The Wronskian is independent
of $r$, so $W_*$ is a constant, which follows from the asymptotic forms of
$u$ and $v$ with the phases $\alpha_\nu$ and $\alpha_{-\nu-1}$ of
\cref{app:normMST}. Both amplitudes are linear in the interior solution, so
their ratio does not depend on how that solution is normalized.

\subsubsection{Matching to the point particle}
\label{sec:eft-matching}

We now restrict to the quadrupolar response, $\ell=2$. The matching
requires the stellar solution and the effective point-particle solution
to describe the same exterior field in the overlap region
$R<r$, $M\ll r\ll1/|\omega|$. Relativistic corrections mix the growing
and decaying radial behaviors, so the coefficient of a decaying term
alone does not define the induced quadrupole. Its identification depends
on the tidal field and source multipole specified by \cref{eq:Stidal}.

Following CDS~\citep{Chakrabarti:2013lua}, the point-particle field is
obtained from the RW equation with the quadrupolar source derived from
this action. The source is regularized with a Riesz kernel and an analytic
continuation in $\ell$, and the regulator poles are removed by minimal
subtraction. Matching this field to the stellar exterior gives
\begin{equation}\label{eq:F-eft-full}
    \tilde F(\omega)=p(\omega)\,\frac{A_2(\omega)}{A_1(\omega)}+q(\omega)\,,
\end{equation}
with
\begin{equation}\label{eq:pq}
\begin{aligned}
    p&=-\frac{107}{42}\,\frac{M}{\omega^{4}}
      \left[1+\epsilon^{2}\!\left(b_{2}
      +\frac{107}{105}\ln\frac{2\omega}{\mu_{0}}\right)\right]\,,\\
    q&=-\frac{112}{321}\,M^{5}
      \left[1+\epsilon^{2}\!\left(b_{1}
      +\frac{107}{105}\ln\frac{\mu_{0}}{c_{\ell}}\right)\right]\,,
\end{aligned}
\end{equation}
and the numerical constants
\begin{equation}
    b_{1}=\frac{13138723}{18874800}-\frac{107}{105}\,\gamma_{E}\,,
    \qquad
    b_{2}=-\frac{33054269}{9437400}\,,
\end{equation}
where $\gamma_E$ is Euler's constant and $\epsilon=2M\omega$.
The coefficients $p$ and $q$ are given for positive real $\omega$ through
relative order $\epsilon^2$. The inverse-length renormalization scale $\mu_0$
enters via a Riesz-kernel representation of the Dirac delta distribution
in the energy-momentum tensor derived from \cref{eq:Stidal}.
The inverse-length ambiguity parameter $c_\ell$ is introduced to compensate for
the fact that the $\ell=2$ source in~\cref{eq:Stidal} was not analytically
continued in $\ell$. Equation~(15) of CDS uses
$\tilde F_{\rm CDS}=2\tilde F$ and
$A_{2,\rm CDS}=A_2/\epsilon^4$. These conversions give
\cref{eq:pq} in our conventions.

The expansion is applied to the exterior matching coefficients.
The ratio $A_2/A_1$ is obtained from the finite-frequency relativistic
stellar solution. It retains
the poles associated with the interior dynamics. Both terms in
\cref{eq:F-eft-full} belong to the matched response. The smooth term $q$
accounts for the source--tide mixing and is independent of the stellar
interior at fixed $M$ and matching scheme. Together they recover
$\tilde F(0)=\lambda_2=2k_2R^5/3$.

The two scales $\mu_0$ and $c_\ell$ enter differently. The parameter
$c_\ell$ arises from the factor $(rc_\ell)^{-(\ell-2)}$ used in the
analytic continuation of the point source. CDS left its interpretation
open and proposed dimensional regularization to improve the matching
\citep{Chakrabarti:2013lua}. At fixed $\mu_0$, changing $c_\ell$ to
$c_\ell'$ shifts the response, at the retained order, by
\begin{equation}
\left.\Delta\tilde F(\omega)\right|_{\mu_0}
=\frac{64}{45}M^7\omega^2\ln\frac{c_\ell'}{c_\ell}\,.
\label{eq:cl-response-shift}
\end{equation}
This analytic term leaves the static deformability, pole positions
and residues unchanged. It has the form of a finite shift of a local
$\dot E_{ab}\dot E^{ab}$ term in the effective action, which enters
as a relative 3PN order in the tidal effects in the Hamiltonian~\cite{Mandal:2023hqa}.
The improved matching in Ref.~\cite{Saketh:2026inprep}
yields a tidal response that depends on the renormalization scale
in dimensional regularization.
This response can be used consistently
with the scale-dependent Hamiltonian in Ref.~\cite{Mandal:2023hqa}.
The scale dependence of the tidal response then cancels with the direct scale
dependence of the 3PN tidal Hamiltonian, resulting in observables that are
independent of the renormalization scale.
Here, we work with the approach of Ref.~\citep{Chakrabarti:2013lua},
fixing $c_\ell = \mu_0$ throughout and omitting relative 3PN tidal
interactions from the Hamiltonian.
As a relative 3PN correction, the effect of this choice should be small,
yet it may still be relevant for the precision required by
next-generation detectors, see Ref.~\citep{Apostolidis:2026qsg}, or
in scattering scenarios~\cite{Schulze:2026ewu}.

The scale $\mu_0$ also enters $p$ and therefore changes the
finite-frequency residues, while leaving the static response and the
zeros of $A_1$ unchanged. Following the oscillator representation of
CDS, we choose one frequency-independent $\mu_0$ for each stellar model
by fitting the sum
$\sum_\alpha\lambda_\alpha\omega_\alpha^2/(\omega_\alpha^2-\omega^2)$
to the full response below the $f$ mode. The same scale is then used for
the response, residues and displacement normalization. The fitted
values are given in \cref{sec:ns-model}.

A pole of the conservative response \cref{eq:F-eft-full}
satisfies $A_1=0$, leaving the standing-wave exterior $v$. A QNM
requires the incoming components of $u$ and $v$ to cancel. With the
time dependence $e^{i\omega t}$, the asymptotic amplitudes and the
outgoing condition are
\begin{align}
    X_N^j&\sim I_j e^{i\omega r_*}+O_j e^{-i\omega r_*},
    \quad j\in\{\nu,-\nu-1\}\,,
    \label{eq:mst-wave-amplitudes}\\
    A_{\mathrm{in}}&=I_\nu A_1+I_{-\nu-1}A_2=0\,.
    \label{eq:mst-qnm-condition}
\end{align}
For the $1.50\,M_\odot$ star of \cref{sec:ns-model}, the response
pole is at $1695.7\,$Hz, whereas
$\omega_{\rm QNM}/(2\pi)=(1690.3+0.68i)\,$Hz. The change of exterior
condition thus produces a real frequency shift as well as damping
(see also Ref.~\cite{Kuan:2021jmk} for an alternative interpretation).
The outgoing root agrees between the phase-amplitude
method~\cite{Andersson:1995wu}, direct RW integration and MST.

\subsection{Mode frequencies and tidal couplings}\label{sec:poles-residues}

We locate each real response pole by bracketing a zero of
$A_1(\omega)$ in a frequency sweep and refining the root. Near an
isolated mode the response takes the single-pole form of one term
of~\cref{eq:F-total},
$\tilde F(\omega) \simeq \lambda_\alpha\omega_\alpha^2/
(\omega_\alpha^2-\omega^2)$, and expanding~\cref{eq:F-eft-full} about
the zero of $A_1$ gives the mode deformability analytically,
\begin{equation}\label{eq:mode-residue}
    \lambda_\alpha
    = \frac{107\,M}{21\,\omega_\alpha^{5}}
    \left[1+\epsilon_\alpha^{2}\!\left(b_{2}
    +\frac{107}{105}\ln\frac{2\omega_\alpha}{\mu_{0}}\right)\right]
    \frac{A_2(\omega_\alpha)}{A_1'(\omega_\alpha)}\,,
\end{equation}
with $\epsilon_\alpha = 2M\omega_\alpha$. Each residue costs two further evaluations of the
MST amplitudes, for the derivative $A_1'$. The mode
parameters are therefore not fitted to samples of $\tilde F(\omega)$. The
one fitted quantity is the scheme scale $\mu_0$ of \cref{sec:eft-matching},
a single number per star that enters every residue through
\cref{eq:mode-residue}. For comparison with Newtonian work we
also quote the dimensionless overlap integral~\citep{Lai:1993di,Reisenegger:1994b}, obtained by
inverting~\cref{eq:lambda-alpha},
\begin{equation}\label{eq:Q-eff}
    \hat I_\alpha \equiv \frac{I_\alpha}{\sqrt{M}\,R}
    = \omega_\alpha
    \sqrt{\frac{15\,\lambda_\alpha}{4\pi M R^{2}}}\,,
\end{equation}
together with $\Lambda_\alpha \equiv \lambda_\alpha/M^5$ and
$f_\alpha = \omega_\alpha/2\pi$.

\section{Stellar models and their tidal response}\label{sec:ns-model}

We apply the matching to a hadronic star and a hybrid star with a
quark core. Both have composition stratification and an elastic crust.
Their spectra show how each restoring force enters the tidal response,
and allow us to test the oscillator approximation used in the binary
calculation.

\subsection{EOS and mass-radius relation}

\begin{figure}
  \centering
  \includegraphics[width=\columnwidth]{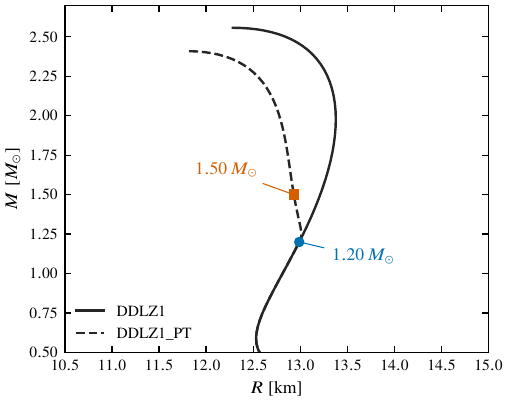}
  \caption{Mass--radius relations for DDLZ1 (solid) and
    DDLZ1$\_$PT (dashed). The markers identify the $1.20\,M_\odot$
    hadronic and $1.50\,M_\odot$ hybrid stars used in the binary.}
  \label{fig:mr}
\end{figure}

\begin{figure}
  \centering
  \includegraphics[width=\columnwidth]{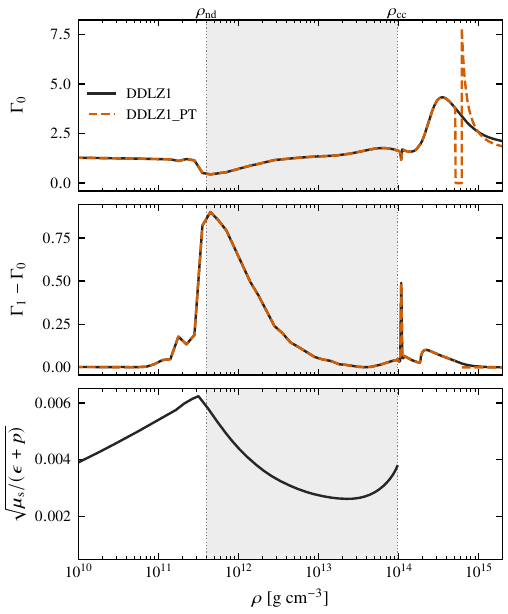}
  \caption{Equilibrium adiabatic index $\Gamma_0$ (top),
    stratification measure $\Gamma_1-\Gamma_0$ (middle), and shear speed
    $v_{\rm s}=\sqrt{\mu_{\rm s}/(\epsilon+p)}$ (bottom), as functions of
    rest-mass density for DDLZ1 (solid) and DDLZ1$\_$PT (dashed).
    The shaded region is the inner crust, between neutron drip
    $\rho_{\rm nd}$ and the crust--core transition $\rho_{\rm cc}$.
    Both EOSs use the same crust.}
    \label{fig:gamma-rho}
\end{figure}

We use the EOS
DDLZ1~\citep{Wei:2020kfb,Xia:2022pja,Xia:2022dvw}, a unified EOS from
density-dependent relativistic mean-field
theory for cold, charge-neutral
matter of neutrons, protons, electrons, and muons
($npe\mu$), and a hybrid variant with a first-order quark--hadron
transition.
In the hybrid EOS the high-density part of DDLZ1 is
replaced by a constant-speed-of-sound (CSS) quark phase, following
\citet{Alford:2013aca}, joined to the hadronic
phase by a Maxwell construction at the transition pressure $p_{\rm tran}$.
There the rest-mass density jumps from $\rho_{\rm tran}$ to
$\rho_{\rm tran}(1+\delta)$ at continuous pressure and chemical potential,
so the energy density jumps from $\epsilon_{\rm tran}$ to
$\epsilon_+ = (1+\delta)(\epsilon_{\rm tran} + p_{\rm tran}) - p_{\rm tran}$,
and above the jump
\begin{equation}
  p(\epsilon) = p_{\rm tran} + c_{\rm s}^{2}\,(\epsilon - \epsilon_+)\,,
  \qquad \epsilon > \epsilon_+ \,.
  \label{eq:css}
\end{equation}
We call this hybrid model DDLZ1$\_$PT and adopt
$(\rho_{\rm tran},\ \delta,\ c_{\rm s}^2)
= (5.13\!\times\!10^{14}\,{\rm g/cm^3},\ 0.2,\ 0.64)$.
\Cref{fig:mr} shows the mass--radius relations of the two EOSs.
Their maximum masses are $2.56\,M_\odot$ and $2.41\,M_\odot$,
respectively, consistent with the observed two-solar-mass pulsars
\citep{Demorest:2010bx,Antoniadis:2013pzd,NANOGrav:2019jur,Fonseca:2021wxt}.

The upper panel of \cref{fig:gamma-rho} shows the equilibrium
adiabatic index $\Gamma_0$ of both
EOSs, computed from the same $\epsilon(p)$ relation used for the
stellar background. It drops sharply at neutron drip, $\rho_{\rm nd} \simeq 4.0\times
10^{11}\,\rm g/cm^3$, rises to about $4$ in the core, and in the hybrid
model vanishes across the density jump. The middle panel shows
$\Gamma_1-\Gamma_0$, the composition gradient that supports the $g$-modes.
It peaks near $\rho_{\rm nd}$ and at $\rho \simeq 2\times 10^{14}\,\rm
g/cm^3$, where muons
appear~\citep{Kantor:2014lja}. The
quark phase is barotropic, $\Gamma_0=\Gamma_1$, and carries no
stratification. The narrow feature just above $\rho_{\rm cc}$
arises from a small density step in the tabulated hadronic EOS.

The crust is elastic because of the
Coulomb lattice of nuclei~\citep{Chamel:2008ca}. For a body-centered cubic
(bcc) lattice the shear modulus
is~\cite{Strohmayer:1991ads}
\begin{equation}
  \mu_{\rm s} = 0.1194 \left(\frac{4\pi}{3}\right)^{1/3} (Ze)^2
        \left(\frac{1-X_{\rm n}}{A}\right)^{4/3} n_{\rm b}^{4/3}\,,
  \label{eq:shear-modulus}
\end{equation}
where $Z$ and $A$
are the mean charge and nucleon numbers of the
nuclei, $X_{\rm n}$ the fraction of free neutrons, and $n_{\rm b}$
 the baryon number density. Elasticity supports polar shear ($s$) modes
\citep{McDermott:1988ApJ,Finn:1990ads} and modifies the $g$ modes by
resisting tangential motion \citep{Gao:2025aqo}. We evaluate the resulting
crustal strain in \cref{sec:strain}. The torsional modes have axial parity and are not
directly excited by the polar tidal field of a nonrotating star~\citep{Schumaker:1983ads,Samuelsson:2006tt}. The lower panel
of \cref{fig:gamma-rho} shows the shear speed
$v_{\rm s} = \sqrt{\mu_{\rm s}/(\epsilon + p)}$,
a few times $10^{-3}$ in the crust.

\subsection{Mode spectrum and tidal response}

We consider the two stars
of the binary studied in \cref{sec:results}, both
described by DDLZ1$\_$PT. The $1.20\,M_\odot$ star, with $R = 12.99\,$km and
$M/R = 0.136$, lies below the onset of the transition at
$1.23\,M_\odot$ and is identical
to a DDLZ1 star. The $1.50\,M_\odot$ star, with $R = 12.93\,$km
and $M/R = 0.171$, lies on the hybrid branch of
\cref{fig:mr} and has a quark core of radius
$R_{\rm PT} \simeq 6.1\,$km.

\begin{table}
  \caption{Quadrupolar ($\ell=2$) oscillation modes of the two stellar
    models and their tidal couplings: mode frequency $f_\alpha$,
    dimensionless mode-specific tidal deformability
    $\Lambda_\alpha = \lambda_\alpha/M^5$, and effective overlap
    $\hat I_\alpha$ of~\cref{eq:Q-eff}, extracted from the poles and
    residues of $\tilde F(\omega)$ with $c_\ell = \mu_0$ and the fitted
    scale $\mu_0R = 0.22$ and $0.38$ of \cref{fig:response}. The last
    two columns give the amplitude damping time of each mode from the
    quadrupole formula, $T_\alpha^{\rm quad}=5/(\lambda_\alpha\omega_\alpha^6)$,
    and from the outgoing quasinormal mode of the same stellar model,
    $T_\alpha^{\rm QNM}=1/\operatorname{Im}\omega_\alpha$; the $g$-mode damping rates are below the numerical resolution. Both stars are
    described by DDLZ1$\_$PT, and the $1.20\,M_\odot$ star, below the onset
    of the transition, has no interface mode $i$. Modes above about $1.3\,{\rm kHz}$ are not crossed
    before the ISCO.}
  \label{tab:modes}
  \begin{ruledtabular}
  \begin{tabular}{l ccccc}
   & \multicolumn{5}{c}{$1.20\,M_\odot$, hadronic} \\
   \cline{2-6}
   Mode & $f_\alpha$ [Hz] & $\Lambda_\alpha$ & $\hat I_\alpha$ & $T_\alpha^{\rm quad}$ [s] & $T_\alpha^{\rm QNM}$ [s] \\
   \hline
   $g_3$ & $64.6$ & $7.19\times10^{-4}$ & $9.59\times10^{-6}$ & $2.2\times10^{14}$ & -- \\
   $g_2$ & $89.8$ & $3.39\times10^{-3}$ & $2.90\times10^{-5}$ & $6.3\times10^{12}$ & -- \\
   $g_1$ & $137.0$ & $1.64\times10^{-2}$ & $9.72\times10^{-5}$ & $1.0\times10^{11}$ & -- \\
   $s_1$ & $533.7$ & $8.55\times10^{-3}$ & $2.73\times10^{-4}$ & $5.7\times10^{7}$ & $5.0\times10^{7}$ \\
   $s_2$ & $906.8$ & $2.21\times10^{-3}$ & $2.36\times10^{-4}$ & $9.2\times10^{6}$ & $7.4\times10^{6}$ \\
   $s_3$ & $1158.1$ & $2.91\times10^{-3}$ & $3.46\times10^{-4}$ & $1.6\times10^{6}$ & $1.2\times10^{6}$ \\
   $s_4$ & $1351.0$ & $1.31\times10^{-2}$ & $8.56\times10^{-4}$ & $1.4\times10^{5}$ & $9.9\times10^{4}$ \\
   $f$ & $1552.7$ & $1610$ & $0.345$ & $0.50$ & $0.35$ \\
   $s_5$ & $1628.6$ & $2.51\times10^{-2}$ & $1.43\times10^{-3}$ & $2.4\times10^{4}$ & $1.9\times10^{4}$ \\
   $s_6$ & $1824.3$ & $2.14\times10^{-3}$ & $4.67\times10^{-4}$ & $1.4\times10^{5}$ & $9.9\times10^{4}$ \\
   \hline
   & \multicolumn{5}{c}{$1.50\,M_\odot$, hybrid} \\
   \cline{2-6}
   Mode & $f_\alpha$ [Hz] & $\Lambda_\alpha$ & $\hat I_\alpha$ & $T_\alpha^{\rm quad}$ [s] & $T_\alpha^{\rm QNM}$ [s] \\
   \hline
   $g_3$ & $59.8$ & $4.09\times10^{-5}$ & $3.32\times10^{-6}$ & $2.0\times10^{15}$ & -- \\
   $g_2$ & $86.2$ & $1.44\times10^{-3}$ & $2.84\times10^{-5}$ & $6.3\times10^{12}$ & -- \\
   $g_1$ & $129.1$ & $3.92\times10^{-3}$ & $7.02\times10^{-5}$ & $2.0\times10^{11}$ & -- \\
   $s_1$ & $645.3$ & $1.37\times10^{-3}$ & $2.07\times10^{-4}$ & $3.7\times10^{7}$ & $3.1\times10^{7}$ \\
   $i$ & $804.0$ & $9.57\times10^{-2}$ & $2.16\times10^{-3}$ & $1.4\times10^{5}$ & $1.1\times10^{5}$ \\
   $s_2$ & $1098.4$ & $3.97\times10^{-4}$ & $1.90\times10^{-4}$ & $5.3\times10^{6}$ & $3.8\times10^{6}$ \\
   $s_3$ & $1424.2$ & $1.13\times10^{-3}$ & $4.16\times10^{-4}$ & $3.9\times10^{5}$ & $2.5\times10^{5}$ \\
   $s_4$ & $1651.8$ & $3.83\times10^{-2}$ & $2.81\times10^{-3}$ & $4.7\times10^{3}$ & $2.2\times10^{3}$ \\
   $f$ & $1695.7$ & $405$ & $0.296$ & $0.38$ & $0.23$ \\
  \end{tabular}
  \end{ruledtabular}
\end{table}

\begin{figure*}
    \centering
    \includegraphics[width=0.9\textwidth]{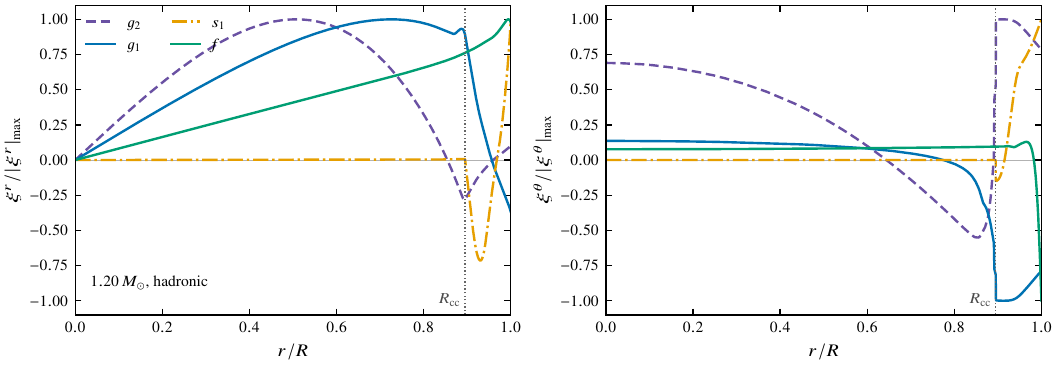}
    \caption{Radial (\emph{left}) and tangential (\emph{right})
      displacement eigenfunctions $\xi^r$ and $\xi^\theta$ of
      representative modes of the $1.20\,M_\odot$ hadronic star, each
      normalized to unit maximum. The vertical dotted line marks the
      crust--core boundary $R_{\rm cc}$. The $f$-mode grows steadily
      towards the surface, the $s_1$ mode is confined to the crust,
      and the $g$-modes peak in the stratified core. The tangential
      displacement of the core $g$-modes slips at the fluid--solid
      boundary $R_{\rm cc}$, where $\xi^\theta$ is discontinuous.}
    \label{fig:eigen1}
\end{figure*}

\begin{figure*}
    \centering
    \includegraphics[width=0.9\textwidth]{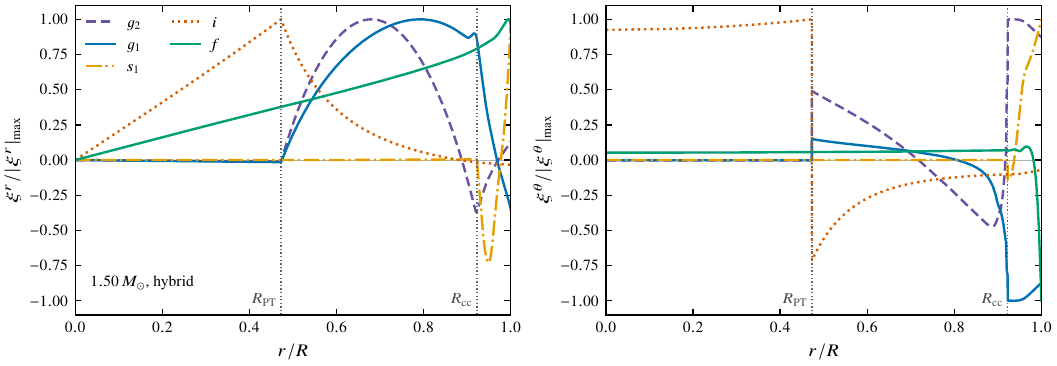}
    \caption{Same as \cref{fig:eigen1}, for the $1.50\,M_\odot$
      hybrid star; the additional vertical dotted line marks the
      quark--hadron interface $R_{\rm PT}$. The quark core is
      barotropic, so the $g$-modes are evanescent below $R_{\rm PT}$
      and their displacement is confined to the hadronic mantle. The
      $i$ mode is sharply peaked at the interface, and its
      $\xi^\theta$ slips across the discontinuity.}
    \label{fig:eigen2}
\end{figure*}

\Cref{tab:modes} collects the resulting mode parameters, and
\cref{fig:eigen1,fig:eigen2} show the radial and tangential
displacements $\xi^r$ and $\xi^\theta$ of representative modes. The
spectra reflect the structure of the two
stars. The $f$ mode dominates the tidal coupling, by
more than three orders of magnitude in $\Lambda_\alpha$~\citep{Hinderer:2016eia,Andersson:2019ahb}.
Its overlap $\hat I_\alpha \simeq 0.3$ is of order unity because $\xi^r$
grows coherently towards the surface. The low-frequency $g$
modes are supported by the composition gradients of
\cref{fig:gamma-rho}. In the hybrid star the quark core is
barotropic, so $g$ modes cannot propagate there~\citep{Jaikumar:2021jbw,Zhao:2022toc}. Their eigenfunctions
decay below the interface, with $\xi^r$ at the percent level of its
maximum, and are concentrated in the stratified hadronic mantle. The couplings of the $g$ modes
depend on the stratification just below the crust,
where the DDLZ1 table contains a density step of about $0.9\%$ near
$n_{\rm b}\simeq0.065\,{\rm fm^{-3}}$, $35$--$55\,{\rm m}$
below $R_{\rm cc}$ in our stars. We compute the perturbations with the
same tabulated $\epsilon(p)$ as the background, so the
buoyancy of this step is included consistently. The shear modes reside in
the crust. Their overlaps grow overall,
though not monotonically, towards the $f$ mode. The interface mode $i$, present only in the phase-transition
model, is peaked at the quark--hadron boundary and carries the largest
coupling below the $f$ mode, $\Lambda_i \simeq 0.1$, the
imprint of the density jump on the tidal response~\citep{Lau:2020bfq,Zhu:2022pja,Counsell:2025hcv,Pereira:2025xsi}.

\begin{figure*}
    \centering
    \includegraphics[width=0.95\textwidth]{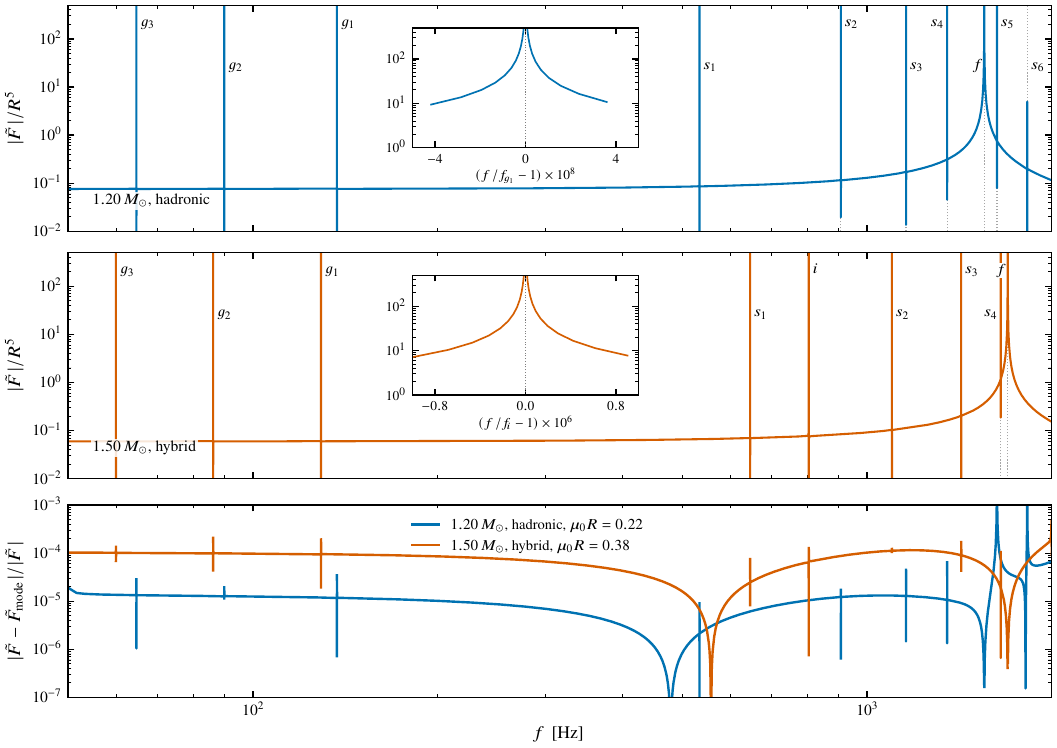}
    \caption{Tidal response of the two stars with $c_\ell = \mu_0$.
      \emph{Top and middle:} $|\tilde F(\omega)|/R^5$ of the
      $1.20\,M_\odot$ hadronic and the $1.50\,M_\odot$ hybrid star across the frequency range containing the modes in \cref{tab:modes}. The dashed vertical lines mark
      the mode frequencies of \cref{tab:modes}, including the shear overtones $s_5$ and $s_6$ of the
      $1.20\,M_\odot$ star above the $f$ mode, which enter only the strain
      reconstruction. The insets resolve the narrow $g_1$ resonance of the
      $1.20\,M_\odot$ star and the $i$-mode resonance of the $1.50\,M_\odot$
      star. \emph{Bottom:} relative
      difference between the full response and the pure mode sum
      $\tilde F_{\rm mode} = \sum_\alpha \lambda_\alpha
      \omega_\alpha^2/(\omega_\alpha^2 - \omega^2)$ at the fitted scale
      $\mu_0R = 0.22$ and $0.38$. Points within $10^{-6}$ of a pole in
      relative frequency ($5\times10^{-4}$ around the $f$ mode) are
      excised, as are narrow windows around the zeros of $\tilde F$ next
      to each resonance, where the relative difference diverges because
      $|\tilde F| \to 0$.}
    \label{fig:response}
\end{figure*}

\Cref{fig:response} shows $|\tilde
F(\omega)|$ of both stars on a grid that clusters points around each
resonance. The response is essentially the adiabatic deformability,
interrupted by very narrow resonances at the
$g$, $s$, and $i$ modes and by
the broad rise towards the $f$ mode. The lower panel
compares the full response with the pure mode sum~\cref{eq:F-total}.
We fit one frequency-independent $\mu_0$ for each star by
minimizing the relative difference below the $f$ mode. This gives
$\mu_0R=0.22$ for the $1.20\,M_\odot$ star and $0.38$ for the
$1.50\,M_\odot$ star. At these scales, the mode sum reproduces the full
response to within $2.1\times10^{-4}$ below the $f$ mode, away from the
neighborhoods of poles and response zeros excluded, as explained in the figure caption.
In this matching scheme, the relativistic response is therefore well
approximated by the Newtonian oscillator structure over the inspiral
band.
The pairs $\{\omega_\alpha,\Lambda_\alpha\}$, or
equivalently $\{\omega_\alpha,\hat I_\alpha\}$, are the input of the mode
Hamiltonian~\cref{eq:H-STF}, and the inspiral of \cref{sec:results}
uses the values of \cref{tab:modes}, obtained at the fitted $\mu_0$ of each
star. 

\section{Dynamical tides in the effective-one-body formalism}
\label{sec:eob-tides}

The mode parameters $\{\omega_\alpha,\lambda_\alpha\}$ of
\cref{sec:ns-model} describe the tidal response of an isolated star.
Coupling the mode to a relativistic orbit brings in three further effects.
First, the tidal field $E_{ij}$ that drives the mode is enhanced relative
to its Newtonian value by post-Newtonian corrections
\citep{Vines:2010ca,Bini:2012gu}. Second, the gravitational
redshift makes the clock of the star run slow relative to the one that
times the orbit, which lowers the mode frequency seen by the orbit. Third,
frame dragging by the orbital angular momentum carries the local frame of
the star forward with the orbit, which raises it. The last two act against
each other, and their net effect on the resonance is given by
\cref{eq:relshift} below.

The orbital model that includes these effects is the EOB description of dynamical tides of
\citet{Steinhoff:2016rfi} and \citet{Hinderer:2016eia}. We construct it in
the same way, with the single $f$ mode replaced by the full set of modes of
\cref{sec:ns-model}, each entering through its pair
$\{\omega_\alpha,\lambda_\alpha\}$. We assume that only body~1 is deformable and neglect the background spins of
both bodies. The second body is included by adding the same contribution with
$1\leftrightarrow2$. The modes enter additively, so we write
the equations for a single mode and suppress its label, writing
$\omega$ and $\lambda$ for its frequency and deformability. The total and
reduced masses are $M=m_1+m_2$ and $\mu=m_1m_2/M$, and we define $\eta=\mu/M$,
$X_A=m_A/M$, and $u=M/r$. From here on $M$ is the total mass, whereas in
\cref{sec:response-action} it denoted the mass of the isolated star. The symbol $\mu$ always denotes the reduced mass, and the
shear modulus of \cref{sec:relativistic-osc} is written $\mu_{\rm s}$.

In the center-of-mass frame the conservative post-Newtonian action reads
\begin{equation}
    \label{Hredaction}
    S
    =
    \int\dd t
    \left[
        \vct{p}\cdot\dot{\vct{r}}
        +P_{ij}\dot Q^{ij}
        -H_{\rm NT}
        -H_{\rm DT}
    \right]\,,
\end{equation}
with $\vct{r}$ and $\vct{p}$ the relative orbital variables,
$\vct{L}=\vct{r}\times\vct{p}$ the orbital angular momentum, and $H_{\rm NT}$
the non-tidal Hamiltonian. Following \citet{Steinhoff:2016rfi}, the
dynamical-tide Hamiltonian is
\begin{equation}
    \label{eq:H_DT_GR}
    H_{\rm DT}
    =
    z_{\rm p}H_{\rm osc}
    +\frac{z_{\rm p}}{2}E_{ij}Q^{ij}
    +\vct{\Omega}_{\rm FD}\cdot\vct{S}_Q\,,
\end{equation}
where the redshift and the gravitomagnetic precession frequency are determined
by the non-tidal Hamiltonian,
\begin{equation}
    \label{eq:z_OmegaFD}
    z_{\rm p}
    =
    \frac{\partial H_{\rm NT}}{\partial m_1}\,,
    \qquad
    \vct{\Omega}_{\rm FD}
    =
    \left.
    \frac{\partial H_{\rm NT}}{\partial\vct{S}_1}
    \right|_{\vct{S}_1=0,\;C_W=0}\,.
\end{equation}
Here $\vct{S}_1$ is a formal variable introduced to extract the coefficient of
the spin--orbit coupling, and $C_W$ denotes the spin-induced multipole Wilson
coefficients. After the derivative is taken, $\vct{S}_1$ is set to zero.

The first term in \cref{eq:H_DT_GR} is the internal oscillator energy. It
retains the form of \cref{eq:H-STF}, with the relativistic $\omega$ and $\lambda$
of \cref{sec:ns-model}. The second term is the coupling to the
post-Newtonian tidal field $E_{ij}=E_{ij}(\vct{r},\vct{p},m_1,m_2)$. The third
term is the tidal spin--orbit coupling, in which
\begin{equation}
    \label{eq:tidal-spin}
    S_Q^{ij}
    =
    4Q^{k[i}P^{j]}{}_k\,,
    \qquad
    S_Q^i
    =
    \frac{1}{2}\epsilon^i{}_{jk}S_Q^{jk}\,,
\end{equation}
is the angular momentum carried by the tidal mode. It couples to the
gravitomagnetic field in the same way as an ordinary spin. Explicit
post-Newtonian expressions for $E_{ij}$ are given in \citet{Vines:2010ca} and
\citet{Steinhoff:2016rfi}.

To identify the frequency shifts, consider a fixed equatorial
circular orbit, so that $z_{\rm p}$, $\Omega_{\rm FD}$ and $\Omega$ are
constant. With $D_t=\partial_t+im\Omega_{\rm FD}$, the mode equation
from \cref{Hredaction} is
\begin{equation}
    \label{eq:Q2m-eom}
    D_t^2 Q_{2m}+z_{\rm p}^2\omega^2Q_{2m}
    =-z_{\rm p}^2\omega^2\lambda E_{2m}\,.
\end{equation}
Its free solutions $Q_{2m}\propto e^{-i\sigma t}$ have frequencies
$\sigma=m\Omega_{\rm FD}\pm z_{\rm p}\omega$. The tidal field varies as
$E_{2m}\propto e^{-im\Omega t}$ and has only $m=0,\pm2$ components.
The prograde $m=2$ resonance therefore satisfies
\citep{Steinhoff:2016rfi}
\begin{equation}
    \label{eq:res-cond}
    \Omega_{\rm res}-\Omega_{\rm FD}=\frac{z_{\rm p}\omega}{2}\,.
\end{equation}
For a circular orbit and vanishing spins the two are unambiguous, and
through 2PN order they reads \citep{Steinhoff:2021dsn}
\begin{align}
    \label{eq:z-OmFD-PN}
    z_{\rm p}
    &=
    1+\frac{x}{2}\left(\eta-3X_2\right)
    +\frac{x^2}{24}\left(5\eta^2-9\eta-6\eta X_2-27X_2\right)\,,
    \nnl
    \frac{\Omega_{\rm FD}}{\Omega}
    &=
    \frac{x}{2}\left(\eta+3X_2\right)
    -\frac{x^2}{24}\left(\eta^2-45\eta+30\eta X_2-27X_2\right)\,,
\end{align}
with $x=(M\Omega)^{2/3}$. Inserted into \cref{eq:res-cond}, they give the
displacement of the resonance from its Newtonian value $\omega/2$,
\begin{equation}
    \label{eq:relshift}
    \frac{2\Omega_{\rm res}}{\omega}
    =
    \frac{z_{\rm p}}{1-\Omega_{\rm FD}/\Omega}
    =
    1+\eta x+\frac{9+4\eta}{6}\eta x^2+\mathcal{O}(x^{5/2})\,.
\end{equation}

The net shift is upward through this order. The companion's mass fraction $X_2$ enters the
1PN terms of $z_{\rm p}$ and $\Omega_{\rm FD}/\Omega$ with opposite signs
and drops out of their sum, so the 1PN displacement is set by the
symmetric mass ratio alone and is the same for both stars and for every
mode. Both coefficients in \cref{eq:relshift} are positive for every
mass ratio, and the displacement vanishes only in the limit $\eta\to0$.
Within this circular 2PN approximation, each resonance lies above its Newtonian position
$\Omega=\omega/2$, by an amount that grows with the mode frequency, since a
higher-frequency mode is crossed at larger $x$.

\subsection{EOB energy map and point-particle sector}
\label{sec:eob-pp}

We now resum the PN dynamics into an EOB Hamiltonian. The real and effective
Hamiltonians are related by the standard energy map~\citep{Buonanno:1998gg,Buonanno:2000ef},
\begin{equation}
    \label{eq:energy-map}
    H_{\rm EOB}
    =M\sqrt{1+2\eta\left(\frac{H_{\rm eff}}{\mu}-1\right)}\,.
\end{equation}
The effective particle obeys the mass-shell constraint
\begin{equation}
    \label{eq:mass-shell}
    \mu^2+\mu_{\rm NG}^2+g_{\rm eff}^{ab}p_ap_b=0\,,
\end{equation}
where the nongeodesic term $\mu_{\rm NG}^2$ collects interactions that cannot be
absorbed into the effective metric. Solving \cref{eq:mass-shell} for
$H_{\rm eff}=-p_0$ gives
\begin{equation}
    \label{eq:Heff-generic}
    H_{\rm eff}
    =\sqrt{A}\sqrt{\mu^2+\mu_{\rm NG}^2
      +\gamma_{\rm eff}^{ij}p_ip_j}+\beta^ip_i\,,
\end{equation}
with $A=-1/g_{\rm eff}^{00}$,
$\beta^i=g_{\rm eff}^{0i}/g_{\rm eff}^{00}$, and
$\gamma_{\rm eff}^{ij}=g_{\rm eff}^{ij}
-g_{\rm eff}^{0i}g_{\rm eff}^{0j}/g_{\rm eff}^{00}$.

For a nonspinning point-particle binary the shift vector $\beta^i$ vanishes and
\cref{eq:Heff-generic} reduces to
\begin{equation}
    \label{eq:Heff-pp}
    H_{\rm eff}^{\rm pp}
    =\sqrt{A_{\rm pp}\left[
       \mu^2+\frac{L^2}{r^2}
       +\frac{A_{\rm pp}}{D_{\rm pp}}\,p_r^2
       +z_3\,u^2\,\frac{\pr^4}{\mu^2}
    \right]}\,,
\end{equation}
where $\pr=(A_{\rm pp}/\sqrt{D_{\rm pp}})\,p_r$ is the tortoise radial momentum
used in the evolution and the quartic term, with $z_3=2\eta(4-3\eta)$, is the
nongeodesic contribution $\mu_{\rm NG}^2$ of Ref.~\cite{Damour:2000we}. The resummed potentials
$A_{\rm pp}$ and $D_{\rm pp}$ are specified in \cref{app:eob-pp}~\citep{Barausse:2009xi,Taracchini:2012ig}.

It remains to resum the tidal sector. The post-Newtonian tidal Hamiltonian
$H_{\rm DT}$ of \cref{eq:H_DT_GR}, with its redshift and frame-dragging
factors, was incorporated into the EOB Hamiltonian by
\citet{Steinhoff:2016rfi}, and we adopt their construction. The next two
subsections outline its basic structure, from which the mode equations and
the driven response are then derived.

\subsection{Corotating frame and the canonical angular momentum}
\label{sec:eob-corot}

We express
the quadrupole in the frame that corotates with the orbit, aligned with the
Newtonian tidal field. Along a circular
orbit the tidal field then has constant components, so the Newtonian coupling
picks out a single fixed component of $Q^{ij}$, and the equilibrium tide is a
fixed point of the mode dynamics rather than an oscillation at twice the
orbital frequency. Because the frame rotates, the transformation to it
introduces Coriolis terms already at Newtonian order and must be carried out
exactly \citep{Steinhoff:2016rfi}. Its net effect is to shift the canonical
orbital angular momentum by the angular momentum carried by the mode, as
shown below. The corotating frame is spanned by
\begin{equation}
    \vct{\Lambda}_1=\vct{n}\,,
    \qquad
    \vct{\Lambda}_2=\vct{\ell}\times\vct{n}\,,
    \qquad
    \vct{\Lambda}_3=\vct{\ell}\equiv\frac{\vct{L}}{L}\,,
\end{equation}
where $\vct{n}=\vct{r}/r$ and $\vct{L}=L\vct{\ell}$ is the orbital angular
momentum. Assuming that the tidal quadrupole remains aligned with the orbital
plane, its corotating components and conjugate momenta are parametrized as
\begin{align}
    \label{eq:Q-corot}
    Q^{IJ}
    &=
    \begin{pmatrix}
        \alpha+\beta & \gamma & 0\\
        \gamma & \alpha-\beta & 0\\
        0 & 0 & -2\alpha
    \end{pmatrix},\\
    \label{eq:P-corot}
    P_{IJ}
    &=
    \frac{1}{2}
    \begin{pmatrix}
        \tfrac{p_\alpha}{3}+p_\beta & p_\gamma & 0\\
        p_\gamma & \tfrac{p_\alpha}{3}-p_\beta & 0\\
        0 & 0 & -\tfrac{2p_\alpha}{3}
    \end{pmatrix}\,.
\end{align}
The corresponding tidal canonical pairs are
$(\alpha,p_\alpha)$, $(\beta,p_\beta)$, and $(\gamma,p_\gamma)$, with
\begin{equation}
    \{\alpha,p_\alpha\}
    =
    \{\beta,p_\beta\}
    =
    \{\gamma,p_\gamma\}
    =1\,,
\end{equation}
and all other brackets vanishing. Physically, $\alpha$ is the axisymmetric
deformation along the orbital axis, $\beta$ measures the tidal bulge aligned
with the binary separation, and $\gamma$ the component rotated by $45^\circ$
in the orbital plane. This is the out-of-phase part of the response, which vanishes
in adiabatic equilibrium and carries the tidal torque.

Because the basis $\vct{\Lambda}_I$ rotates along the orbit, transforming the
tidal components also changes the symplectic potential. In particular
[Eqs.~(5.14)--(5.16) of Ref.~\cite{Steinhoff:2016rfi}],
\begin{equation}
    \label{eq:kin-split}
    P_{ij}\dot Q^{ij}
    =
    P_{IJ}\dot Q^{IJ}
    +\vct{S}_Q\cdot\vct{\Omega},
    \qquad
    \Omega^i
    =
    \frac{1}{2}\epsilon^{ikl}
    \Lambda_I^{\,k}\dot\Lambda_I^{\,l}\,.
\end{equation}
For planar motion the tidal spin of \cref{eq:tidal-spin} is aligned with the
orbital angular momentum, $\vct{S}_Q=S_Q\vct{\ell}$ with
\begin{equation}
    \label{eq:SQ-eob}
    S_Q
    =
    2(\beta p_\gamma-\gamma p_\beta)\,,
\end{equation}
so that $\vct{S}_Q\cdot\vct{\Omega}=S_Q\dot\phi$. The complete symplectic
potential therefore becomes
\begin{equation}
    \label{eq:ptilde-def}
    p_i\dot r^i+P_{ij}\dot Q^{ij}
    =
    p_r\dot r+P_\phi\dot\phi+P_{IJ}\dot Q^{IJ},
    \qquad
    P_\phi\equiv L+S_Q\,.
\end{equation}
Thus, in the corotating chart, $P_\phi$, rather than $L$, is
canonically conjugate to the orbital phase $\phi$. In vector form
the same transformation reads
$\vct{p}_{\rm orb}=\vct{P}-\vct{S}_Q\times\vct{r}/r^2$
[Eq.~(5.18) of Ref.~\cite{Steinhoff:2016rfi}]. The canonical
phase-space variables are
\begin{equation}
    (r,p_r;\,\phi,P_\phi;\,
      \alpha,p_\alpha;\,\beta,p_\beta;\,\gamma,p_\gamma)\,,
\end{equation}
The tidal Hamiltonian of the next subsection involves the quadrupole only
through a few scalar contractions, which in the corotating variables reduce
to
\begin{align}
    \label{eq:corot-contractions}
    Q^{ij}n^i n^j
    &=
    \alpha+\beta,
    &
    Q^{ij}\ell^i\ell^j
    &=
    -2\alpha,
    \nnl
    Q^{ij}Q^{ij}
    &=
    6\alpha^2+2\beta^2+2\gamma^2,
    &
    P^{ij}P^{ij}
    &=
    \frac{p_\alpha^2}{6}
    +\frac{p_\beta^2}{2}
    +\frac{p_\gamma^2}{2}\,.
\end{align}
Together with $S_Q$ of \cref{eq:SQ-eob}, these are the combinations from
which the tidal Hamiltonian is built.

\subsection{Incorporating the tidal interactions}
\label{sec:eob-assign}

With the canonical chart fixed, we incorporate dynamical tides by matching the
PN expansion of the EOB Hamiltonian to the tidal Hamiltonian $H_{\rm DT}$ of \cref{eq:H_DT_GR}.
The placement of each interaction within the effective metric is determined by
its momentum dependence. When $H_{\rm DT}$ is inserted into the mass-shell constraint \cref{eq:mass-shell}
and expanded to first order in the tidal contribution, terms quadratic in $p_0$ deform $A$, terms quadratic
in the spatial momenta $p_i$ modify $\gamma_{\rm eff}^{ij}$, terms linear in $p_i$ contribute
to the shift $\beta^i$, and momentum-independent terms enter the
nongeodesic mass $\mu_{\rm NG}^2$ \citep{Steinhoff:2016rfi}. Accordingly,
the electric tidal coupling $\frac12 E_{ij}Q^{ij}$, with $E_{ab}=C_{acbd}u^cu^d$ and hence quadratic in $p_0$,
contributes through $A$. The 1PN interaction quadratic in the orbital momentum enters the
centrifugal and radial momentum terms within the corresponding bracket,
while the tidal spin--orbit coupling $\vct{\Omega}_{\rm FD}\cdot\vct{S}_Q$,
being linear in the orbital momentum, is incorporated through the shift.
Finally, the oscillator energy is momentum independent and therefore modifies
the effective rest mass according to
$\mu^2\to(\mu+H_{\rm mode})^2\simeq\mu^2+2\mu H_{\rm mode}$,
so that it enters $\mu_{\rm NG}^2$.
For the quasicircular inspiral we use the circular-limit tidal
coefficients of Ref.~\citep{Steinhoff:2016rfi}. The effective Hamiltonian is
\begin{widetext}
\begin{equation}
    \label{eq:Heff-eob}
    H_{\rm eff}
    =\sqrt{A\left[
        \mu^2
        +\frac{(P_\phi-S_Q)^2}{r^2}
        +\frac{A}{A_{\rm pp}^2}\pr^2
        +z_3u^2\frac{\pr^4}{\mu^2}
        +2\mu\left(z_cH_{\rm mode}+C_{\rm int}\right)
    \right]}
    +Z\,S_Q\,,
    \qquad
    A=A_{\rm pp}+\frac{2z_I}{\mu}H_{\rm int}\,,
\end{equation}
\end{widetext}
where
\begin{align}
    H_{\rm int}
    &=\frac12E_{ij}Q^{ij}
    =-\frac{3m_2}{2r^3}(\alpha+\beta)
    =-\frac{3X_2}{2M^2}u^3(\alpha+\beta)\,,
    \label{eq:Hint-eob}
\end{align}
is the Newtonian electric coupling of \cref{eq:H-STF}, with the tidal tensor
\cref{eq:Eij-newton} of the companion
$E_{ij}=-(m_2/r^3)(3n^in^j-\delta^{ij})$ contracted through
\cref{eq:corot-contractions}, and
\begin{equation}
    z_I
    =1-(2X_2-\eta)u
      +\frac{5X_1}{28}(33X_1-7)u^2\,,
    \label{eq:zI-eob}
\end{equation}
is its post-Newtonian correction. The 1PN interaction quadratic in the orbital
momentum is
\begin{align}
    C_{\rm int}
    &=\frac{\mu}{2}\,Q^{ij}\mathcal{C}_{ij}
    =\frac{3X_2}{2M^2}u^3(u+3u^2)
      \left[(\eta-2)\alpha+\eta\beta\right]\,,
    \nnl
    \mathcal{C}_{ij}
    &=\frac{3m_2}{\eta r^4}(1+3u)
    \left(\ell^i\ell^j+\eta n^i n^j\right)\,.
    \label{eq:Cint-eob}
\end{align}
The oscillator energy is
\begin{align}
    H_{\rm mode}
    &=\frac{Q^{ij}Q^{ij}}{4\lambda}+\lambda\omega^2P_{ij}P_{ij}
    \nnl
    &=\frac{6\alpha^2+2\beta^2+2\gamma^2}{4\lambda}
    +\lambda\omega^2
    \left(
       \frac{p_\alpha^2}{6}
       +\frac{p_\beta^2}{2}
       +\frac{p_\gamma^2}{2}
    \right)\,,
    \label{eq:Hmode-eob}
\end{align}
with the redshift factor
\begin{equation}
    \label{eq:zc-circ}
    z_c
    =
    1+\frac{3}{2}X_1u
    +\frac{27}{8}X_1u^2\,,
\end{equation}
and $S_Q$ is the tidal spin of \cref{eq:SQ-eob} with the spin--orbit
coefficient
\begin{equation}
    \label{eq:Zcirc}
    Z
    =
    \left\{
        \left(2+\frac{3}{2}\frac{X_2}{X_1}\right)
        -
        \left[
            \frac{5}{8}\eta
            +\left(\frac{9}{8}+\frac{3}{4}\eta\right)
             \frac{X_2}{X_1}
        \right]u
    \right\}
    \frac{L}{r^3}\,,
\end{equation}
with
$L=P_\phi-S_Q$. Because the centrifugal term keeps $S_Q$ exactly, $Z$
contains only the spin--orbit coupling inherited from
$\vct{\Omega}_{\rm FD}\cdot\vct{S}_Q$ \citep{Damour:2008qf}, and differs
from the $Z$ of Ref.~\cite{Steinhoff:2016rfi}, where the centrifugal
term is linearized in $S_Q$ and the resulting kinematic term
is absorbed into $Z$.  The 2PN terms of
$z_I$ and $z_c$ follow from matching the adiabatic limit to
Ref.~\cite{Bini:2012gu} and from the EOB redshift
$\partial H^{\rm pp}_{\rm EOB}/\partial m_1$, respectively
\citep{Steinhoff:2016rfi}. With several modes, $H_{\rm mode}$, $H_{\rm int}$,
$C_{\rm int}$ and $S_Q$ are each summed over the modes.

With the couplings assigned, the dynamics is fixed. The real Hamiltonian
$H_{\rm EOB}$ of \cref{eq:energy-map}, built from the effective Hamiltonian
\cref{eq:Heff-eob}, generates the evolution of the complete canonical chart
of \cref{sec:eob-corot}. Displaying the mode sector through the
$(\alpha,p_\alpha)$ pair alone, Hamilton's equations read
\begin{equation}
\begin{aligned}
    \dot r&=\frac{A_{\rm pp}}{\sqrt{D_{\rm pp}}}\frac{\partial H_{\rm EOB}}{\partial \pr},
    &\dot p_{r_*}&=-\frac{A_{\rm pp}}{\sqrt{D_{\rm pp}}}\frac{\partial H_{\rm EOB}}{\partial r}+\mathcal F_{r_*},\\
    \dot\phi&=\frac{\partial H_{\rm EOB}}{\partial P_\phi}\equiv\Omega,
    &\dot P_\phi&=\mathcal F_\phi,\\
    \dot\alpha&=\frac{\partial H_{\rm EOB}}{\partial p_\alpha},
    &\dot p_\alpha&=-\frac{\partial H_{\rm EOB}}{\partial\alpha}\,,
\end{aligned}
    \label{eq:hamilton-formal}
\end{equation}
where $(\beta,p_\beta)$ and $(\gamma,p_\gamma)$ obey equations of the same
form, $\Omega$ is the orbital frequency, and the
factor $A_{\rm pp}/\sqrt{D_{\rm pp}}$ comes from the tortoise rescaling of
the radial momentum. The dissipative forces $\mathcal F_{r_*}$ and
$\mathcal F_\phi$ carry the radiation reaction. They have a point-particle
part, for which we adopt the factorized waveform and flux of the
nonspinning SEOBNR model \citep{Damour:2008gu,Pan:2011gk,Taracchini:2012ig},
and a tidal part from the radiation of the dynamical quadrupole, assembled
in \cref{sec:eob-radiation}. The dimensionless form used in the evolution
is given in \cref{sec:eob-eom}.

No dissipative term acts on the mode variables in
\cref{eq:hamilton-formal}. This is an approximation, and we now justify it.
The mode state superposes a driven response, which follows the tide, and a
free oscillation, which is excited when the orbit sweeps through the
resonance and thereafter rings at the mode frequency. The free oscillation
radiates at its own frequency and is damped by this radiation. The quadrupole formula gives the damping time
\begin{equation}\label{eq:T-damping}
  T_\alpha=\frac{5}{\lambda_\alpha\omega_\alpha^6}
  \simeq 0.4\,{\rm s}\;\frac{m_1}{1.5\,M_\odot}
  \left(\frac{\Lambda_\alpha}{400}\right)^{-1}
  \left(\frac{m_1\omega_\alpha}{0.08}\right)^{-6}\,,
\end{equation}
where $\Lambda_\alpha=\lambda_\alpha/m_1^5$ and $m_1$ is the
mass of the deformed star. Here $T_\alpha$ is the amplitude decay time,
consistent with $T_\alpha^{\rm QNM}=1/\operatorname{Im}\omega_\alpha$;
the energy decay time is $T_\alpha/2$.
For the low-frequency modes crossed before the ISCO, both the quadrupole
estimate and the outgoing QNM calculation give damping times of days or
longer (\cref{tab:modes}), compared with seconds to tens of seconds
remaining in the inspiral. The $f$ mode has a much shorter damping time,
but its resonance is not reached. We therefore neglect the
radiation of the free oscillations and let the mode variables evolve
conservatively. Dissipative tides, from the radiation of the $f$
mode or from viscosity~\citep{Cutler:1987ApJ,Cutler:1990ApJ,Kantor:2026dpj}, would enter here as damping terms on the mode
variables, with the lag of the driven response transferring the
dissipated energy from the orbit~\citep{Ripley:2023qxo,HegadeKR:2024slr,HegadeKR:2026iou}. We leave them for future work.

The radiation of the driven response, in contrast, is coherent with
the orbit and is kept. The waveform and flux expressions of
\cref{sec:eob-radiation} take time derivatives at the orbital carrier
frequency, which is valid only for the driven part, so they must be
evaluated with the driven amplitudes alone. This requires separating the
mode state into its driven and free parts. We construct the separation in
\cref{sec:eob-driven-free} from the mode equations, in closed form, and
then assemble the waveform and the radiation reaction in
\cref{sec:eob-radiation}.

\subsection{Mode amplitudes and the driven--free separation}
\label{sec:eob-driven-free}

To derive the mode equations we use that the real EOB Hamiltonian depends on
the canonical mode variables only through the four scalars $H_{\rm mode}$,
$H_{\rm int}$, $C_{\rm int}$, and $S_Q$, and define the four derivatives
\begin{equation}
\begin{aligned}
    g_H&\equiv\frac{\partial H_{\rm EOB}}{\partial H_{\rm mode}},
    &g_I&\equiv\frac{\partial H_{\rm EOB}}{\partial H_{\rm int}},\\
    g_C&\equiv\frac{\partial H_{\rm EOB}}{\partial C_{\rm int}},
    &g_S&\equiv\frac{\partial H_{\rm EOB}}{\partial S_Q}\,.
\end{aligned}
    \label{eq:mode-chain-coefficients}
\end{equation}
The first three are dimensionless, while $g_S$ has dimensions of frequency.
As for any forced oscillator, each canonical pair can be traded for one
complex amplitude whose modulus measures the size of the oscillation and
whose argument is its phase,
\begin{equation}
\begin{aligned}
    b_0&\equiv\frac12\left(\alpha+i\kappa_0p_\alpha\right)\,,
    &\kappa_0&\equiv\frac{\lambda\omega}{3}\,,\\
    b_\pm&\equiv\frac12\left[
       \beta\mp i\gamma+i\kappa(p_\beta\mp ip_\gamma)
    \right]\,,
    &\kappa&\equiv\lambda\omega\,.
\end{aligned}
    \label{eq:b-mode-transformation}
\end{equation}
Substituting \cref{eq:b-mode-transformation} into the mode
sector of \cref{eq:hamilton-formal}, with the scalars
\cref{eq:Hmode-eob,eq:SQ-eob}, we obtain three driven first-order oscillators,
\begin{subequations}\label{eq:b-mode-eom}
\begin{align}
    \dot b_0+i\Delta_0b_0
    &=-\frac{i\kappa_0}{2}F_\alpha\,,
    &\Delta_0&=g_H\omega\,,
    \label{eq:b0-mode-eom}\\
    \dot b_\pm+i\Delta_\pm b_\pm
    &=-\frac{i\kappa}{2}F_\beta\,,
    &\Delta_\pm&=g_H\omega\pm2g_S\,,
    \label{eq:bpm-mode-eom}
\end{align}
\end{subequations}
where $F_\alpha$ and $F_\beta$ are the forces that the orbit exerts on the
mode. They collect the derivatives of the interaction scalars,
$F_q\equiv g_I\,\partial H_{\rm int}/\partial q
+g_C\,\partial C_{\rm int}/\partial q$ for $q=\alpha,\beta$. With
\cref{eq:Hint-eob,eq:Cint-eob} they read
\begin{align}
    F_\alpha
    ={}&-\frac{3X_2}{2M^2}u^3 g_I
    +\frac{3X_2}{2M^2}u^3(u+3u^2)(\eta-2)g_C\,,
    \label{eq:Falpha-eob}\\
    F_\beta
    ={}&-\frac{3X_2}{2M^2}u^3 g_I
    +\frac{3X_2}{2M^2}u^3(u+3u^2)\eta g_C\,.
    \label{eq:Fbeta-eob}
\end{align}
The corresponding EOB driving tidal tensor is defined by
\begin{equation}
    E_{ij}\equiv2\left.
    \frac{\partial H_{\rm EOB}}{\partial Q^{ij}}
    \right|_{H_{\rm mode},\,S_Q}\,,
    \label{eq:Eij-eob-def}
\end{equation}
where holding $H_{\rm mode}$ and $S_Q$ fixed restricts the derivative
to the interactions $H_{\rm int}$ and $C_{\rm int}$. In the corotating
frame of \cref{eq:Q-corot}, its components are
\begin{equation}
    E_{IJ}
    =\begin{pmatrix}
        \tfrac{F_\alpha}{3}+F_\beta & 0 & 0\\
        0 & \tfrac{F_\alpha}{3}-F_\beta & 0\\
        0 & 0 & -\tfrac{2F_\alpha}{3}
    \end{pmatrix}\,,
    \label{eq:Eij-eob}
\end{equation}
so that $\tfrac12E_{ij}Q^{ij}=F_\alpha\alpha+F_\beta\beta$.
There is no independent force term in $\gamma$ for an aligned
quasicircular tidal field. The coefficients $g_H,g_I,g_C,g_S$
are evaluated on the full canonical state, so these equations do not
assume an orbit independent of the modes. On a frozen orbit, the detunings
$\Delta_0$ and $\Delta_\pm$ appear in the denominators of the driven response. In the Newtonian limit $g_H\to1$ and $g_S\to-\Omega$,
they reduce to $\omega$ and $\omega\mp2\Omega$, the mode frequency seen from
the frame corotating with the orbit. The signs of $b_\pm$ label the two
circular phase-space branches of one physical oscillator. During a prograde
inspiral $g_S<0$, so only $\Delta_+$ decreases through zero and undergoes resonance.

The resonance condition $\Delta_+=0$ can be made explicit.
Differentiating \cref{eq:Heff-eob}, with
$\partial H_{\rm EOB}/\partial H_{\rm eff}=M/H_{\rm EOB}$ from
\cref{eq:energy-map}, gives the two coefficients
\begin{align}
    g_H&=\frac{M}{H_{\rm EOB}}\,
        \frac{\mu Az_c}{H_{\rm eff}-ZS_Q}\,,
    \nnl
    g_S&=-\Omega+\Omega_{\rm FD}\,,
    \qquad
    \Omega_{\rm FD}\equiv\frac{M}{H_{\rm EOB}}\,Z\,.
    \label{eq:gH-gS}
\end{align}
The coefficient $g_H$ is the redshift between the proper time of the star and
the time of the orbit, so the mode frequency in the orbital dynamics is
$g_H\omega$. The coefficient $g_S$ is minus the rate at which the tidal
field turns relative to the star. The orbit turns at $\Omega$, which enters
\cref{eq:Heff-eob} through the centrifugal term, and the local inertial frame of
the star is dragged forward at $\Omega_{\rm FD}$, which enters through the
spin--orbit term $ZS_Q$. The tidal field therefore drives the $m=2$ bulge
at the frequency $2(\Omega-\Omega_{\rm FD})$, and the resonance occurs
where this equals $g_H\omega$,
\begin{equation}
    \frac{2\Omega_{\rm res}}{\omega}
    =
    \frac{g_H}{1-\Omega_{\rm FD}/\Omega}\,.
    \label{eq:relshift-eob}
\end{equation}
This is \cref{eq:relshift} with $z_{\rm p}$ and $|\vct{\Omega}_{\rm FD}|$
replaced by their resummed forms $g_H$ and $\Omega_{\rm FD}$. Expanded in
$x$ along the circular orbits of the point-particle Hamiltonian, the two
reproduce \cref{eq:z-OmFD-PN} term by term.

We use the same physical decomposition as the Newtonian composite
solution of \citet{Pnigouras:2025muo}: a driven response that follows
the tide and a free oscillation excited at resonance. Here the detuning
and driving force are supplied by the relativistic EOB dynamics.
For a linear sweep the two contributions can be separated in closed form.
Let $t_n$ denote the resonance crossing, defined by
$\Delta_+(t_n)=0$, and introduce the time measured from the crossing,
$\tau=t-t_n$. Near $t_n$, we linearize the detuning and freeze the
slowly varying tidal force,
\begin{equation}
    \Delta_+(t)\simeq-\dot\Delta_n\tau,
    \qquad
    F_\beta(t)\simeq F_{\beta n}\,,
\end{equation}
where $\dot\Delta_n\equiv-\dot\Delta_+(t_n)>0$ and
$F_{\beta n}\equiv F_\beta(t_n)$. The solution of
\cref{eq:bpm-mode-eom} with no incoming free oscillation is the
Fresnel integral \citep{Lai:1993di,Reisenegger:1994a},
\begin{align}
    b_+(t)
    =&-\frac{i\kappa F_{\beta n}}{2}
      \sqrt{\frac{\pi}{\dot\Delta_n}}\,e^{i\pi z^2/2}
      \biggl\{\left[C(z)+\frac12\right]
      \nnl&
      -i\left[S(z)+\frac12\right]\biggr\},
    \qquad
    z\equiv\sqrt{\frac{\dot\Delta_n}{\pi}}\,\tau\,,
    \label{eq:fresnel-solution}
\end{align}
where $C(z)=\int_0^z\cos(\pi x^2/2)\,\dd x$ and
$S(z)=\int_0^z\sin(\pi x^2/2)\,\dd x$. Before resonance the solution approaches the driven equilibrium
response. At $z=0$ its magnitude is proportional to
$|F_{\beta n}|/\sqrt{\dot\Delta_n}$, so a slower sweep excites a larger
oscillation. After resonance it approaches a free oscillation whose
amplitude is twice the amplitude at the crossing.

We write this solution as
\begin{equation}
\begin{aligned}
    b_+&= b_+^{\rm d}+b_+^{\rm free}\,,\\
    b_+^{\rm d}&=-\frac{\kappa F_\beta(t)}{2}
      \frac{\mathcal J(|z|)}{\Delta_+(t)}\,,\\
    b_+^{\rm free}&=-\frac{\kappa F_{\beta n}}{2}\Theta(\tau)
      \sqrt{\frac{2\pi}{\dot\Delta_n}}\,
      e^{i[\pi/4-\int_{t_n}^t\Delta_+(t')\,\dd t']}\,,
\end{aligned}
    \label{eq:free-b-amplitude}
\end{equation}
where $\Theta$ is the Heaviside step function and
\begin{align}
    \mathcal J(x)
    \equiv{}-i\pi x e^{i\pi x^2/2}
    \biggl\{\left[C(x)-\frac12\right]
      -i\left[S(x)-\frac12\right]\biggr\}
    \label{eq:fresnel-shape-function}
\end{align}
is the resonance shape function. It is the function that
\citet{Pnigouras:2025muo} obtained for the Newtonian problem, and here it
multiplies the relativistic equilibrium response, with the detuning
$\Delta_+$ and the force $F_\beta$ of the EOB Hamiltonian. For a linear sweep the free phase is $\pi z^2/2+\pi/4$.
We retain the integral of the instantaneous detuning when extending the
solution away from resonance. With the force frozen and the detuning
linear, \cref{eq:free-b-amplitude} is an identity, which follows from
\cref{eq:fresnel-solution} because $C$ and $S$ are odd functions. Its
content is in the two limits of $\mathcal J$. At large argument
$\mathcal J(x)=1+i/(\pi x^2)+\mathcal O(x^{-4})$, so far from the
crossing $b_+^{\rm d}$ is the equilibrium response
$-\tfrac{\kappa}{2}F_\beta/\Delta_+$ with its first adiabatic correction,
which is why the frozen force can be replaced by the instantaneous
$F_\beta(t)$. At the resonance
$\mathcal J(x)=\tfrac{\pi x}{2}(1+i)+\mathcal O(x^2)$, so $b_+^{\rm d}$
stays finite where $\Delta_+$ vanishes. The free oscillation is switched
on at the crossing with the constant amplitude of the Fresnel solution and
rings at the mode frequency.
The individual driven and free terms have compensating jumps
at the crossing; their sum is continuous. This split is a convention for
the linear-sweep solution, which we extend to the evolving orbit.

To evaluate $b_+^{\rm d}$ along the orbit we need $z$ in terms of the
instantaneous orbit. Away from a resonance, iterating
\cref{eq:b-mode-eom} about the equilibrium response generates the adiabatic
expansion
\begin{equation}
    b
    =-\frac{\kappa}{2}\,\frac{F}{\Delta}
      \left(1-\frac{iD}{\Delta^2}+\cdots\right)\,,
    \label{eq:adiabatic-expansion}
\end{equation}
for each branch, with the effective sweep rates
\begin{equation}
    D_0\equiv\dot\Delta_0-\frac{\dot F_\alpha}{F_\alpha}\Delta_0\,,
    \qquad
    D_\pm\equiv\dot\Delta_\pm-\frac{\dot F_\beta}{F_\beta}\Delta_\pm\,,
    \label{eq:local-sweep-variables}
\end{equation}
in which the second term accounts for the growth of the drive as the
orbit tightens. Comparing \cref{eq:adiabatic-expansion} with the
large-argument tail of $\mathcal J$ fixes the argument of the shape
function as $x=|\Delta|/\sqrt{\pi|D|}$, with $D<0$ during the inspiral.
At the crossing the second term of \cref{eq:local-sweep-variables}
vanishes, $D_+(t_n)=-\dot\Delta_n$, and $x$ reduces to $|z|$, so this
choice joins \cref{eq:free-b-amplitude} smoothly. The sweep rates follow
from the orbit. The redshifted mode frequency $g_H\omega$ moves slowly,
and the detuning sweeps because the driving frequency the mode sees,
$-2g_S$, grows. With $g_S=-\kappa_S\Omega$, the frame-dragging factor
$\kappa_S$ held fixed, a Keplerian chirp
$\dot\Omega=-\tfrac32\Omega\dot r/r$, and the forces
\cref{eq:Falpha-eob,eq:Fbeta-eob} scaling as $u^3$,
\begin{equation}
    D_\pm\simeq3\left(g_H\omega\pm g_S\right)\frac{\dot r}{r}\,,
    \qquad
    D_0\simeq3g_H\omega\frac{\dot r}{r}\,,
    \label{eq:D-orbit}
\end{equation}
with $\dot r<0$ the radial velocity of \cref{eq:hamilton-formal}.
\Cref{eq:D-orbit} drops the variation of $g_I$ and $g_C$ in $\dot F/F$,
that of $\kappa_S$ and $g_H$ in $\dot\Delta$, and the post-Newtonian
corrections to $\dot\Omega$.

The driven amplitudes evaluated by the model at every instant are
therefore
\begin{equation}
\begin{aligned}
    b_0^{\rm d}
    &=-\frac{\kappa_0F_\alpha}{2}
      \frac{\mathcal J(x_0)}{\Delta_0},
    &x_0&=\frac{|\Delta_0|}{\sqrt{\pi|D_0|}},\\
    b_\pm^{\rm d}
    &=-\frac{\kappa F_\beta}{2}
      \frac{\mathcal J(x_\pm)}{\Delta_\pm},
    &x_\pm&=\frac{|\Delta_\pm|}{\sqrt{\pi|D_\pm|}}\,.
\end{aligned}
    \label{eq:local-driven-amplitudes}
\end{equation}
and we use the matched approximation
\begin{equation}
    b_+\simeq b_+^{\rm d}+b_+^{\rm free},
    \qquad
    b_-\simeq b_-^{\rm d},
    \qquad
    b_0\simeq b_0^{\rm d}\,.
    \label{eq:driven-free-response}
\end{equation}
For the comparisons with numerical mode amplitudes below, we
compute the sweep rates directly from \cref{eq:local-sweep-variables} on
the saved orbit. The waveform prescription uses the local approximation
\cref{eq:D-orbit}. The driven amplitudes of \cref{eq:local-driven-amplitudes} enter the
waveform and the radiation reaction through the real fields
\begin{equation}
    \alpha^{\rm d}=2\operatorname{Re}b_0^{\rm d}\,,
    \,\,
    \beta^{\rm d}=\operatorname{Re}(b_+^{\rm d}+b_-^{\rm d})\,,
    \,\,
    \gamma^{\rm d}=\operatorname{Im}(b_-^{\rm d}-b_+^{\rm d})\,.
    \label{eq:driven-real-fields}
\end{equation}
The free oscillation after the resonance is described by
\cref{eq:free-b-amplitude}. The energy that it carries is evaluated in
\cref{sec:res-energy}.

\subsection{Waveform and radiation reaction}
\label{sec:eob-radiation}

For the tidal contribution to the waveform and radiative torque, we
follow \citet{YuLau2025}, who derived the finite-frequency corrections
to GW radiation through relative 1PN order.
The waveform modes $h_{\ell m}$ at luminosity distance $D_L$ are
assembled in the factorized form, which we write through the
dimensionless amplitudes $\hat h_{\ell m}$,
\begin{equation}
    \label{eq:hlm-master}
    \hat h_{\ell m}
    \equiv\frac{D_L}{M}\,h_{\ell m}
    =h_{\ell m}^{\prime N}
    \left[
        c_{\ell+\epsilon}\,\mathcal H_{\ell m}^{\rm pp}
        +\mathcal H_{\ell m}^{t,(1)}
        +(-1)^m\mathcal H_{\ell m}^{t,(2)}
    \right]\,,
\end{equation}
where $\epsilon=\operatorname{mod}(\ell+m,2)$ labels the parity of
the mode, and the superscripts $(1)$ and $(2)$ label the
contributions from the tidal deformation of the two stars:
$\mathcal H_{\ell m}^{t,(2)}$ equals $\mathcal H_{\ell m}^{t,(1)}$
with the labels $1\leftrightarrow2$ exchanged, and the factor
$(-1)^m$ accounts for the half-orbit rotation that maps one body to
the other. The azimuthal radiation-reaction force, the torque on the
orbit, is built from the full square of the modes and is measured in
the reduced mass,
\begin{equation}
    \label{eq:radiation-reaction-flux}
    \hat{\mathcal F}_\phi
    \equiv\frac{\mathcal F_\phi}{\mu}
    =-\frac{M\Omega}{8\pi\eta}
    \sum_{\ell,m>0}m^2
    \left|\hat h_{\ell m}\right|^2\,.
\end{equation}

The reduced Newtonian basis is
\begin{equation}
    \label{eq:hlm-newtonian}
    h_{\ell m}^{\prime N}
    =\eta\,n_{\ell m}^{(\epsilon)}\,
     v_\phi^{\ell+\epsilon}\,
     Y_{\ell-\epsilon,-m}\!\left(\frac{\pi}{2},\phi\right)\,,
\end{equation}
where $n_{\ell m}^{(\epsilon)}$ are numerical
coefficients~\cite{Damour:2008gu} and
\begin{equation}
    \label{eq:vphi-def}
    v_\phi
    =M\Omega
    \left[M\left.\frac{\partial H_{\rm EOB}}{\partial P_\phi}
    \right|_{p_{r_*}=0}\right]^{-2/3}\,,
\end{equation}
is the non-Keplerian orbital velocity, which reduces to
$v_\omega=(M\Omega)^{1/3}$ for a Newtonian circular orbit.

The point-particle sector is the standard factorized EOB waveform of
the nonspinning SEOBNR model~\cite{Damour:2008gu,Pan:2011gk}. Its
leading mass-ratio dependence is carried by the Newtonian coefficient
\begin{equation}
    \label{eq:cl-def}
    c_{\ell+\epsilon}
    =X_2^{\ell+\epsilon-1}+(-1)^{\ell+\epsilon}X_1^{\ell+\epsilon-1}\,,
\end{equation}
while
\begin{equation}
    \label{eq:Hpp-def}
    \mathcal H_{\ell m}^{\rm pp}
    =\hat S^{(\epsilon)}\,
     T_{\ell m}\,
     e^{i\delta_{\ell m}}
     \left(\rho_{\ell m}\right)^{\ell}
     N_{\ell m}\,,
\end{equation}
collects the relativistic corrections: $\hat S^{(\epsilon)}$ is the
effective source term, $T_{\ell m}$ the tail factor, $\rho_{\ell m}$
and $\delta_{\ell m}$ the residual amplitude and phase corrections,
and $N_{\ell m}$ the non-quasicircular correction applied to the $22$
mode. Their explicit expressions can be found in
Refs.~\cite{Damour:2008gu,Pan:2011gk,Taracchini:2012ig}.

The tidal sector contributes to the waveform in two ways: directly,
as a source of the radiative multipoles, and indirectly, through its
effect on the orbit. For adiabatic tides both contributions were
derived through relative 1PN order in Ref.~\cite{VinesWaveform2011}.
For dynamical tides, we retain the canonical quadrupole as an independent
dynamical variable throughout the derivation.

We first transform the quadrupole to the inertial frame. With the
orbit in the $x$--$y$ plane and $\phi$ the orbital phase,
\cref{eq:Q-corot} becomes
\begin{equation}
\begin{aligned}
    Q^{xx}+Q^{yy}&=2\alpha,\qquad Q^{zz}=-2\alpha\,,\\
    Q^{xx}-Q^{yy}-2iQ^{xy}&=2(\beta-i\gamma)e^{-2i\phi}\,,\\
    Q^{xz}&=Q^{yz}=0\,.
\end{aligned}
    \label{eq:Qij-inertial}
\end{equation}
For every radiative term linear in the quadrupole, we insert the driven
fields of \cref{eq:driven-real-fields}, which carry the superscript
``d''. The coefficients of the inertial
phases $1$, $e^{-2i\phi}$, and $e^{2i\phi}$ are then respectively
$\alpha^{\rm d}$, $\beta^{\rm d}-i\gamma^{\rm d}$, and
$\beta^{\rm d}+i\gamma^{\rm d}$. We evaluate time derivatives of the radiative moments on an
instantaneous circular orbit, holding the driven amplitudes
$\alpha^{\rm d}$, $\beta^{\rm d}$ and $\gamma^{\rm d}$ fixed.
Differentiating the explicit orbital phases then gives
$\dd/\dd t\to-im\Omega$.  The quantities $H_{\rm mode}$ and $S_Q$
below are computed from the full canonical state, including the free
oscillation.

Next, $Q^{ij}$ is propagated to the global source moments of the
binary. The 1PN global moments of each individual body are given in
Eqs.~(B4)--(B5) of Ref.~\cite{Vines:2010ca}; translating them to the
system center of mass and adding the orbital point-particle moments
gives the system mass quadrupole $I^{ij}$ through 1PN order and the
mass octupole $I^{ijk}$ and current quadrupole $J^{ij}$ at Newtonian
order. We write their helicity projections with the common factor
$e^{-im\phi}$ suppressed, and denote by $r_h$ the harmonic separation.
For the deformation of star 1, the $m=2$ mass-quadrupole scalar is
\begin{align}
    \label{eq:Z-fixed-rh}
    \mathcal Z
    ={}&\frac{\mu r_h^2}{2}+(\beta^{\rm d}-i\gamma^{\rm d})
    +\frac{\mu Mr_h}{84}
    \left[42(2\eta-1)+(19-57\eta)\,y\right]
    \nnl&
    +C_{20}\alpha^{\rm d}
    +C_{22}(\beta^{\rm d}-i\gamma^{\rm d})
    +C_{2,-2}(\beta^{\rm d}+i\gamma^{\rm d})
    \nnl&
    +X_2^2r_h^2
    \left(\frac{H_{\rm mode}}{2}
        +\frac{2}{3}\Omega S_Q\right)\,,
\end{align}
where $y=r_h^3\Omega^2/M$ collects the velocity-squared terms of the
1PN moments while $r_h$ and $\Omega$ are still independent variables
($y=1$ only after imposing the Newtonian circular-orbit condition),
the third term is the 1PN point-particle moment, and the 1PN tidal
coefficients are
\begin{align}
    \label{eq:Z-coefficients}
    C_{20}
    &=-\frac{X_2M}{140r_h}
    \left(420X_2^2+18X_2y-658X_2+105\right),
    \nnl
    C_{22}
    &=-\frac{X_2M}{280r_h}
    \left(420X_2^2+376X_2y-1106X_2+945\right),
    \nnl
    C_{2,-2}
    &=-\frac{3X_2M}{8r_h}
    \left(4X_2^2-2X_2+1\right)\,.
\end{align}
The Newtonian octupole and current-quadrupole scalars are
\begin{align}
    \label{eq:lower-multipoles}
    \mathcal O_3
    &=\mu r_h^3(X_2-X_1)
      +6X_2r_h(\beta^{\rm d}-i\gamma^{\rm d}),
    \nnl
    \mathcal O_1
    &=\frac{\mu r_h^3}{5}(X_1-X_2)
    -\frac{2X_2r_h}{5}
      \left[6\alpha^{\rm d}+(\beta^{\rm d}-i\gamma^{\rm d})\right],
    \nnl
    \mathcal J_1
    &=\frac{\mu\Omega r_h^3}{2}(X_2-X_1)
    \nnl&\quad
    +\frac{X_2r_h}{4}
    \left[6\Omega\alpha^{\rm d}
        +4\Omega(\beta^{\rm d}-i\gamma^{\rm d})
        +3S_Q\right]\,,
\end{align}
where $\mathcal O_3$ and $\mathcal O_1$ are the $m=3$
and $m=1$ projections of $I^{ijk}$ and $\mathcal J_1$ the
$m=1$ projection of $J^{ij}$. In each line the first term is the
orbital point-particle moment and the remainder the translated stellar
quadrupole.

These are fixed-$r_h$ expressions. To obtain a fixed-frequency
source, $r_h$ is eliminated in favor of $x=(M\Omega)^{2/3}$
using the conservative radial equation of motion,
Eqs.~(5.9a)--(5.9e) of Ref.~\cite{Vines:2010ca}, specialized to a
circular orbit at fixed canonical quadrupole. Iterating the radial equation around
the Newtonian circular radius $r_0=M/x$ gives
\begin{equation}
    \label{eq:rh-circular}
    r_h=\frac{M}{x}\left[1-\frac{3-\eta}{3}x
        +\delta r^{(1)}+(1\leftrightarrow2)\right]\,,
\end{equation}
where the point-particle term is included once and the tidal
correction from the deformation of star 1 is
\begin{align}
    \label{eq:delta-r-tidal}
    \delta r^{(1)}
    ={}&\frac{3X_2x^2}{2}
    \frac{\alpha^{\rm d}+\beta^{\rm d}}{\mu M^2}
    +\frac{H_{\rm mode}}{3M}
    -\frac{X_2(2+X_2)}{3}x^{3/2}\frac{S_Q}{\mu M}
    \nnl&
    +\frac{X_2x^3}{2\mu M^2}
    \left[(5-8X_2-3X_2^2)\alpha^{\rm d}
    \right.\nnl&\qquad\left.
         +(3-8X_2-3X_2^2)\beta^{\rm d}\right]\,.
\end{align}

Finally, the radiative modes follow from the system moments through
the standard multipole formula~\citep{Thorne:1980ru,Blanchet:2013haa}
\begin{align}
    \label{eq:hlm-from-multipoles}
    h_{\ell m}
    ={}&\frac{1}{D_L}\frac{8\pi}{(2\ell+1)!!}\,
    \mathcal Y^{*\,\ell m}_{\langle L\rangle}
    \left[\sqrt{\frac{(\ell+1)(\ell+2)}{\ell(\ell-1)}}\,
        I^{(\ell)}_{\langle L\rangle}\right.
    \nnl&\hspace{2.2cm}\left.
    +2i\sqrt{\frac{\ell(\ell+2)}{(\ell+1)(\ell-1)}}\,
        J^{(\ell)}_{\langle L\rangle}\right]\,,
\end{align}
with the corotating derivative rule
$(\dd/\dd t)^\ell\to(-im\Omega)^\ell$: the $22$ mode is sourced by
$\mathcal Z$, the $33$ and $31$ modes by $\mathcal O_3$
and $\mathcal O_1$, and the $21$ mode by $\mathcal J_1$.
Substituting \cref{eq:rh-circular} into
\cref{eq:Z-fixed-rh,eq:lower-multipoles} and normalizing each mode
with the reduced Newtonian basis $h_{\ell m}^{\prime N}$ of
\cref{eq:hlm-newtonian}, the orbital point-particle part of each scalar
reproduces the coefficient $c_{\ell+\epsilon}$, the Newtonian limit
of the point-particle bracket in \cref{eq:hlm-master}. The
tidal remainder, expanded to linear order in the tidal
variables and through relative 1PN order for the $22$ mode, yields
the dimensionless multipliers
\begin{align}
    \mathcal H_{22}^{t,(1)}
    ={}&\frac{x^2}{2\mu M^2}\Big[6X_2\alpha^{\rm d}
        +(4+3X_2)(\beta^{\rm d}-i\gamma^{\rm d})
    \nnl&\quad
        +3X_2(\beta^{\rm d}+i\gamma^{\rm d})\Big]
    \nnl&
    +\frac{X_2(2+X_2)}{3}\frac{H_{\rm mode}}{\mu}
    -\frac{2X_2(2-X_2)}{3}x^{3/2}\frac{S_Q}{\mu M}
    \nnl&
    +\frac{X_2x^3}{7\mu M^2}
    \Big[(12-21X_2-34X_2^2)\alpha^{\rm d}
    \nnl&\quad
    -(43+6X_2+17X_2^2)(\beta^{\rm d}-i\gamma^{\rm d})
    \nnl&\quad
    -(1+32X_2+17X_2^2)(\beta^{\rm d}+i\gamma^{\rm d})\Big],
    \label{eq:H22-tidal}\\
    \mathcal H_{21}^{t,(1)}
    ={}&\frac{X_2x^2}{4\mu M^2}
    \Big[(-6+36X_2)\alpha^{\rm d}
         +(-1+18X_2)(\beta^{\rm d}-i\gamma^{\rm d})
    \nnl&\quad
         +(-9+18X_2)(\beta^{\rm d}+i\gamma^{\rm d})\Big]
    +\frac{3X_2}{2}x^{1/2}\frac{S_Q}{\mu M}\,,
    \label{eq:H21-tidal}\\
    \mathcal H_{33}^{t,(1)}
    ={}&\frac{3X_2x^2}{4\mu M^2}
    \Big[(-6+12X_2)\alpha^{\rm d}
         +(5+6X_2)(\beta^{\rm d}-i\gamma^{\rm d})
    \nnl&\quad
         +(-3+6X_2)(\beta^{\rm d}+i\gamma^{\rm d})\Big],
    \label{eq:H33-tidal}\\
    \mathcal H_{31}^{t,(1)}
    ={}&\frac{X_2x^2}{4\mu M^2}
    \Big[(30+36X_2)\alpha^{\rm d}
         +(-1+18X_2)(\beta^{\rm d}-i\gamma^{\rm d})
    \nnl&\quad
         +(-9+18X_2)(\beta^{\rm d}+i\gamma^{\rm d})\Big]\,.
    \label{eq:H31-tidal}
\end{align}
With the time derivatives evaluated as above, the $22$ multiplier is complete through relative 1PN order, the
odd-$m$ multipliers are leading order, and
$\mathcal H_{\ell m}^{t,(A)}=0$ for the $32$ and $44$ modes.
The dimensionless form of these equations that we integrate, and the
initial data, are given in \cref{app:eob-pp}.

\section{Mode excitation and orbital phase}\label{sec:results}

We now integrate the system of \cref{sec:eob-eom} for
the binary of \cref{sec:ns-model}, the $1.20\,M_\odot$ hadronic star and
the $1.50\,M_\odot$ hybrid star, with the modes of
\cref{tab:modes} except the $s_5$ and $s_6$ modes of the lighter star.
These two higher-frequency modes are retained only for strain
reconstruction. Each mode enters through its pair
$\{\omega_\alpha,\lambda_\alpha\}$ alone.
The same evolution supplies the mode amplitudes, their
back-reaction on the GW phase, and the forcing for the crustal-strain
calculation in \cref{sec:strain}.

\subsection{Mode evolution through resonances}\label{sec:res-modes}

\begin{figure*}
    \centering
    \includegraphics[width=\textwidth]{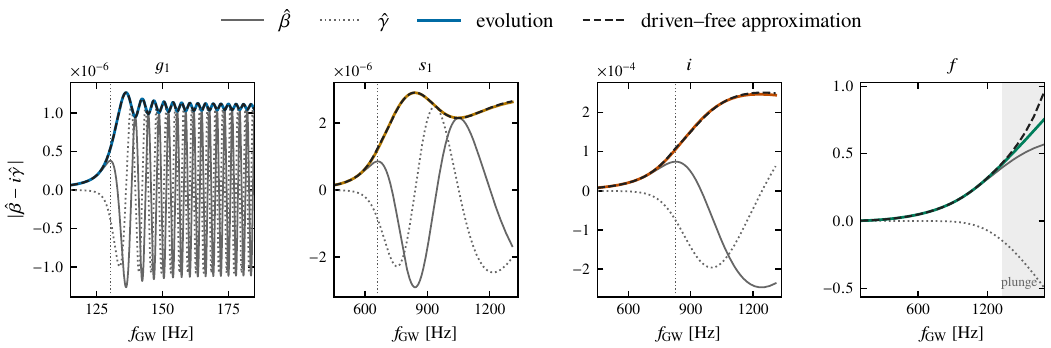}
    \caption{Quadrupole amplitude $|\beta-i\gamma|$ (colored solid curves)
      of the $g_1$, $s_1$, $i$ and $f$ modes of the $1.5\,M_\odot$
      star along the $1.20+1.50\,M_\odot$ inspiral, as a
      function of the gravitational-wave frequency
      $f_{\rm GW}=\Omega/\pi$, in the dimensionless units of
      \cref{sec:eob-eom}. The grey curves are the components $\beta$
      (solid) and $\gamma$ (dotted) of the evolved state, and the dashed
      black curves are the driven--free approximation
      \cref{eq:q-driven-free}. The vertical lines mark the crossings
      $\Delta_+=0$ at $f_{\rm GW}=130$, $661$ and $826\,{\rm Hz}$ for $g_1$,
      $s_1$ and $i$. The $g_1$ panel covers the frequencies around its
      crossing, and the $s_1$ and $i$ panels end at the ISCO,
      $f_{\rm GW}=1336\,{\rm Hz}$. The $f$ panel continues into the plunge
      beyond the ISCO (shaded) and stops before the $f$-mode crossing at
      $1742\,{\rm Hz}$.}
    \label{fig:mode-evolution}
\end{figure*}

Each mode crossed during inspiral passes through three stages,
which \cref{fig:mode-evolution} shows for four modes of the
$1.50\,M_\odot$ star. Long before its crossing the mode follows the orbit
as the equilibrium tide. The bulge then points along the binary
separation, so in the corotating frame of \cref{sec:eob-corot} the
out-of-phase component $\gamma$ vanishes and $\beta$ carries the whole
$|m|=2$ response. Its amplitude grows both with the tidal force and with
the shrinking detuning, as the factor $F_\beta/\Delta_+$ of
\cref{eq:adiabatic-expansion} shows. As the crossing approaches, the
response begins to lag the force and $\gamma$ develops. At $\Delta_+=0$
the tidal field drives the mode at its own frequency, and the mode takes
up energy from the orbit during a time of order $\dot\Delta_n^{-1/2}$.

After the crossing the free oscillation excited at the resonance
persists with constant amplitude. In the corotating frame it rotates at
the rate $|\Delta_+|$, which grows as the orbit tightens, so $\beta$ and
$\gamma$ oscillate with increasing frequency. Their modulus
$|\beta-i\gamma|=|b_++\bar b_-|$ is insensitive to this rotation and
shows only the beat of the free oscillation against the driven part.

We follow the modes up to the innermost stable circular orbit (ISCO). We
define it as the minimum of the binding energy
$H_{\rm EOB}-M$ along the circular orbits with adiabatic tides, which lies
at $r=6.80M$~\citep{Buonanno:2000ef,Damour:2000we}. The inspiral reaches it at the gravitational-wave frequency
$f_{\rm GW}=\Omega/\pi=1336\,{\rm Hz}$. The $g_1$ mode is crossed at
$130\,{\rm Hz}$, far below this frequency, so its resonance is complete and
many beats follow before the ISCO. The $s_1$ and $i$ modes behave
differently. They are crossed at $661$ and $826\,{\rm Hz}$, shortly before
the ISCO, and the inspiral ends within one beat after the crossing. The
$f$ mode has its resonance at $1742\,{\rm Hz}$, above the ISCO frequency,
so it does not reach resonance and only its driven response grows.

The crossings in \cref{fig:mode-evolution} lie slightly above the mode
frequencies, as the resonance condition \cref{eq:relshift-eob}
predicts. The redshift lowers the resonance and the frame dragging raises it, and the
two nearly cancel. The shift is largest at the $s_3$ crossing of the
$1.20\,M_\odot$ star, where $g_H=0.87$ and $1-\Omega_{\rm FD}/\Omega=0.84$.
The redshift alone would lower this resonance by $13\%$ and the frame
dragging alone would raise it by $19\%$, but the crossing lies only $3.5\%$
above the mode frequency $f_\alpha$. A Newtonian resonance condition
therefore remains a fair approximation \citep{Steinhoff:2021dsn}. The 2PN
estimate \cref{eq:relshift} agrees well with the evolution. It reproduces
the crossing frequencies to $0.01\%$ for the $g$ modes and to better than
$1\%$ for the $s$ and $i$ modes.

The dashed curves in \cref{fig:mode-evolution} show the driven--free
approximation of \cref{eq:driven-free-response}, which for the plotted
amplitude reads
\begin{equation}
    \beta-i\gamma
    \simeq b_+^{\rm d}+b_+^{\rm free}+\bar b_-^{\rm d}\,.
    \label{eq:q-driven-free}
\end{equation}
Before the ISCO, the maximum complex-amplitude difference,
normalized by the maximum evolved amplitude in that interval, is
$0.6\%$ for $g_1$ and about $2\%$ for $s_1$ and $i$.
For the $f$ mode the pointwise difference remains below $0.6\%$ up to
$1200\,{\rm Hz}$ and reaches about $3\%$ at the ISCO.
It grows to about $10\%$ at $1500\,{\rm Hz}$ during the plunge, where
the force and detuning vary too rapidly for the local sweep
approximation. We restrict quantitative conclusions to the inspiral.

The amplitudes $b_-$ and $b_0$ have no resonance. The branch $b_-$ is the
bulge that travels against the orbit, and its detuning
$\Delta_-=g_H\omega-2g_S$ only increases during the inspiral. Far below a
crossing $\Delta_-\simeq\Delta_+$, so $b_-$ and $b_+$ contribute equally to
the equilibrium tide. Near and after the crossing $b_+$ dominates, and at
the ISCO $|b_-/b_+|$ lies between $0.05$ and $0.2$ for the four modes of
\cref{fig:mode-evolution}. The amplitude $b_0$ gives the axisymmetric
deformation $\alpha=2\operatorname{Re}b_0$. It is driven by the part of the
tidal field that does not rotate with the orbit, its detuning is
$\Delta_0=g_H\omega$, and in the equilibrium tide $\alpha\simeq\beta/3$.
Before the ISCO both amplitudes follow their driven forms of
\cref{eq:local-driven-amplitudes}, $b_-$ to better than $0.5\%$ and $\alpha$
to better than $2\%$.

\subsection{Effective Love number}
\label{sec:res-keff}

\begin{figure*}
    \centering
    \includegraphics[width=\textwidth]{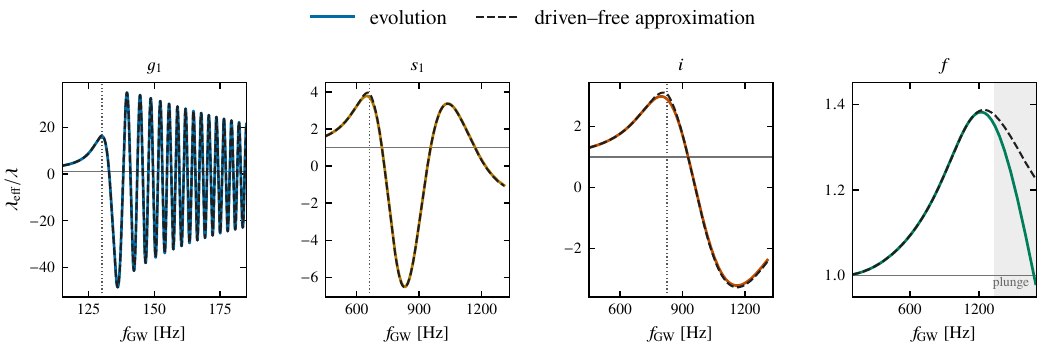}
    \caption{Effective tidal deformability $\lambda_{\rm eff}/\lambda$ of
      \cref{eq:lambda-eff} for the same four modes and the same inspiral as
      in \cref{fig:mode-evolution}. The colored solid curves are obtained
      from the evolved state, and the dashed black curves from the
      driven--free approximation of \cref{eq:driven-free-response}. The
      horizontal lines mark the static value $\lambda_{\rm eff}=\lambda$. The vertical lines, frequency windows and
      shading are as in \cref{fig:mode-evolution}.}
    \label{fig:keff}
\end{figure*}

The adiabatic tide is the static response to the driving field of
\cref{eq:Eij-eob}, for which the restoring force balances the tidal force,
$g_H\,\partial H_{\rm mode}/\partial\alpha+F_\alpha=0$ and
$g_H\,\partial H_{\rm mode}/\partial\beta+F_\beta=0$. With $H_{\rm mode}$ of
\cref{eq:Hmode-eob} this gives
\begin{equation}
    \alpha_{\rm ad}=-\frac{\lambda F_\alpha}{3g_H}\,,
    \qquad
    \beta_{\rm ad}=-\frac{\lambda F_\beta}{g_H}\,,
    \qquad
    \gamma_{\rm ad}=0\,.
    \label{eq:adiabatic-tide}
\end{equation}
Following \citet{Hinderer:2016eia} and \citet{Steinhoff:2016rfi}, we define
the effective tidal deformability of a mode as the projection of its
quadrupole on the tidal field, normalized to the adiabatic tide,
\begin{equation}
    \frac{\lambda_{\rm eff}}{\lambda}
    \equiv\frac{E_{ij}Q^{ij}}{E_{ij}Q^{ij}_{\rm ad}}
    =\frac{F_\alpha\alpha+F_\beta\beta}
          {F_\alpha\alpha_{\rm ad}+F_\beta\beta_{\rm ad}}
    =-\frac{g_H}{\lambda}\,
      \frac{F_\alpha\alpha+F_\beta\beta}{F_\alpha^2/3+F_\beta^2}\,.
    \label{eq:lambda-eff}
\end{equation}
This ratio is also the ratio of the effective Love number to the static
one.

By construction $\lambda_{\rm eff}=\lambda$ for the adiabatic tide. The
effective deformability is therefore the deformability that an adiabatic
tide would need in order to have the same interaction energy with the
orbit, and its deviation from $\lambda$ measures the dynamical tide. It also
allows a dynamical tide to be included in a waveform model with adiabatic
tides, by the replacement $\lambda\to\lambda_{\rm eff}$
\citep{Hinderer:2016eia,Steinhoff:2016rfi,Schmidt:2019wrl,Andersson:2019ahb,Haberland:2025luz}.

\Cref{fig:keff} shows $\lambda_{\rm eff}/\lambda$ for the four modes of
\cref{fig:mode-evolution}. Far below its crossing a mode responds
adiabatically and $\lambda_{\rm eff}=\lambda$. As the crossing approaches,
the small detuning $\Delta_+$ enhances the driven amplitude and
$\lambda_{\rm eff}$ rises. It stays finite at the crossing, whereas the
equilibrium response diverges there as $1/\Delta_+$. The enhancement at the
crossing is of order $\omega/\sqrt{\dot\Delta_n}$, the phase that the mode
accumulates during the resonance. It is a few for $s_1$ and $i$ and about
$16$ for $g_1$, because the inspiral is slowest at low frequency. After the
crossing the free oscillation dominates the quadrupole. It rotates relative
to the tidal field at the rate $|\Delta_+|$, so its projection on the field
changes sign twice in each period, and $\lambda_{\rm eff}$ oscillates
between positive and negative values of the same size.

The $f$ mode is not crossed before the ISCO, and its $\lambda_{\rm eff}$
grows smoothly to about $1.4\lambda$. It decreases again when the orbit
starts to plunge, because the mode can no longer follow the tidal field.
The driven--free approximation (dashed curves) again reproduces $\lambda_{\rm eff}$ of all four modes before the ISCO and
deviates only in the plunge of the $f$-mode panel. The large values of
$\lambda_{\rm eff}/\lambda$ of the other modes do not
imply a large effect on the orbit, because their deformabilities are smaller
than that of the $f$ mode by more than three orders of magnitude
(\cref{tab:modes}). The effective deformability of the whole star is therefore dominated by the $f$ mode. The low-frequency modes affect the orbit
instead through the energy that they take up at the crossing, which we
discuss next.

\subsection{Energy transferred to the modes and phase shift in GWs}\label{sec:res-energy}

In the amplitudes of \cref{eq:b-mode-transformation} the oscillator
energy \cref{eq:Hmode-eob} is
$H_{\rm mode}=(6|b_0|^2+|b_+|^2+|b_-|^2)/\lambda$. Only $b_+$ passes
through a resonance and acquires a free part, so the energy that a
crossing leaves in the free mode, measured in its local oscillator Hamiltonian, is
\begin{equation}
    E_{\rm mode}^{\rm free}
    =\frac{|b_+^{\rm free}|^2}{\lambda}
    =\frac{\pi\lambda\omega^2F_{\beta n}^2}
           {2\dot\Delta_n}\,,
    \label{eq:free-mode-energy}
\end{equation}
with $b_+^{\rm free}$ from \cref{eq:free-b-amplitude}. The mode is
driven at its own frequency for a time of order $\dot\Delta_n^{-1/2}$,
which sets the scaling $F_{\beta n}^2/\dot\Delta_n$.

\begin{figure}
    \centering
    \includegraphics[width=\columnwidth]{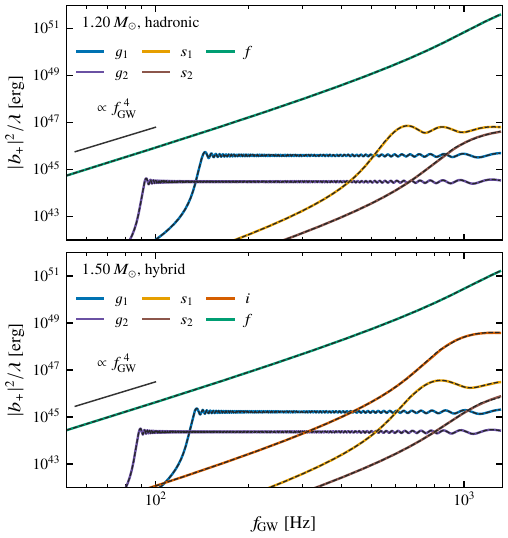}
    \caption{Energy of the resonant branch, $|b_+|^2/\lambda$, for the modes
      of the $1.20\,M_\odot$ hadronic star (top) and the $1.50\,M_\odot$
      hybrid star (bottom) along the $1.20+1.50\,M_\odot$ inspiral,
      with $b_+$ built from the evolved state through
      \cref{eq:b-mode-transformation}. After a crossing the curves settle
      on the plateau $E_{\rm mode}^{\rm free}$ of
      \cref{eq:free-mode-energy}. The short line marks the
      $f_{\rm GW}^4$ growth of the driven energy far below a resonance.
      The dashed curves are the driven--free approximation of
      \cref{eq:driven-free-response}, and the panels end at the ISCO.}
    \label{fig:resonant-energy}
\end{figure}

\Cref{fig:resonant-energy} shows $|b_+|^2/\lambda$ of the evolved state
for the modes of both stars. Below a crossing it is the energy of the
driven tide, $\lambda\omega^2F_\beta^2/(4\Delta_+^2)$, which rises as
$f_{\rm GW}^4$ with the tidal force and steepens as the detuning closes.
Through the crossing the energy grows by the factor
$2\pi\omega^2/\dot\Delta_n$, of order $10^4$ for the $g$ modes and $10^3$
for $s_1$ and $i$, and then stays at $E_{\rm mode}^{\rm free}$, with a
ripple from the beat between the free and the driven parts. The plateaus
lie at $(2$--$4)\times10^{45}\,{\rm erg}$ for the $g_1$ modes and about
ten times lower for $g_2$, at $5\times10^{46}$ and
$2.5\times10^{46}\,{\rm erg}$ for $s_1$ of the $1.20$ and $1.50\,M_\odot$
stars, and at $3\times10^{48}\,{\rm erg}$ for the $i$ mode, which takes
up the most energy because it has both a large deformability and a high
frequency. The $f$ modes are not crossed, and their driven energy grows
to above $10^{51}\,{\rm erg}$ at the ISCO. The dashed curves are the
driven--free approximation of \cref{eq:driven-free-response}.

To obtain the orbital work from the internal oscillator energy
$H_{\rm mode}$, we must include redshift and frame dragging. The prograde
free branch carries angular momentum
$\Delta S_Q=2E_{\rm mode}^{\rm free}/\omega$.
For a narrow resonance on a circular orbit, the work taken from the
orbit is
\begin{equation}
    \Delta E_{\rm orb}
       \simeq\Omega_n\Delta S_Q
       =\frac{2\Omega_n}{\omega}E_{\rm mode}^{\rm free}\,.
    \label{eq:resonance-orbital-energy}
\end{equation}
At the crossing, \cref{eq:relshift-eob} gives
$2\Omega_n/\omega=g_H+2\Omega_{\rm FD}/\omega$, identifying the redshift
and frame-dragging contributions. This factor approaches unity in the
Newtonian limit. The orbit is then
ahead in time by $\Delta t\simeq\Delta E_{\rm orb}/|\dot E_{\rm GW}|$
\citep{Reisenegger:1994a,Lai:1993di,Flanagan:2006sb,Flanagan:2007ix}.
At fixed frequency the GW phase is lower by
\begin{equation}
    \Delta\phi_{\rm GW}
       \simeq-\frac{2\Delta E_{\rm orb}}{|\mathcal F_\phi|}
       =-\frac{4\Omega_n}{\omega}
          \frac{E_{\rm mode}^{\rm free}}{|\mathcal F_\phi|}\,,
    \label{eq:phase-step}
\end{equation}
where the torque is evaluated at resonance and
$|\dot E_{\rm GW}|=\Omega_n|\mathcal F_\phi|$ on a circular orbit.

\begin{figure}
    \centering
    \includegraphics[width=\columnwidth]{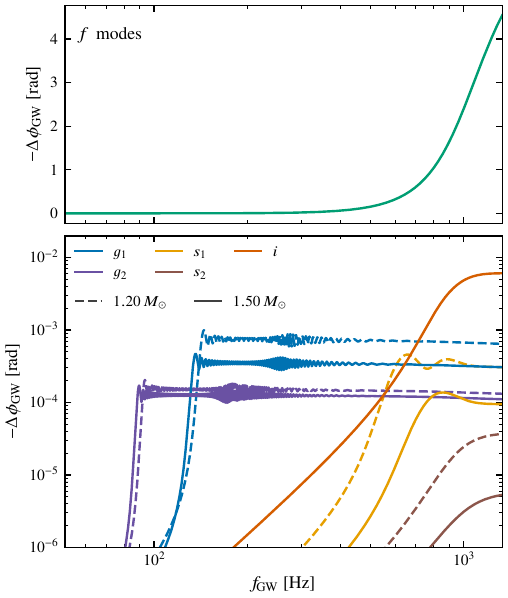}
    \caption{Magnitude of the negative GW phase shift along the
      $1.20+1.50\,M_\odot$ inspiral. We compare evolutions at equal
      frequency, $\Delta\phi_{\rm GW}=2(\phi_A-\phi_B)$, aligned at
      $52\,{\rm Hz}$. Top: the two $f$ modes evolved dynamically,
      relative to an evolution with all modes adiabatic. Bottom: one
      additional low-frequency mode evolved dynamically, relative to
      the evolution with only the $f$ modes dynamical. Dashed and solid
      curves refer to modes of the $1.20$ and $1.50\,M_\odot$ stars.
      Both panels end at the ISCO.}
    \label{fig:phase-shift}
\end{figure}

\Cref{fig:phase-shift} shows the phase shifts measured in the evolution.
The curve of each low-frequency mode grows before its crossing,
where the driven response exceeds the adiabatic one, steps at the
crossing, and is nearly flat afterwards. The ripple on the flat part is
the back-reaction of the free oscillation on the orbit, whose interaction
energy oscillates at the beat frequency $|\Delta_+|$. Its envelope peaks
where the beat frequency equals the orbital frequency, at
$f_{\rm GW}\simeq2f_\alpha$, and the radial oscillation it induces is only
$\delta r/r\sim10^{-8}$. The steps are
$6\times10^{-3}\,{\rm rad}$ for the $i$ mode,
$(1$--$7)\times10^{-4}\,{\rm rad}$ for $g_1$ and $s_1$, and about
$1.5\times10^{-4}\,{\rm rad}$ for $g_2$. \Cref{eq:phase-step} reproduces
the $g$-mode steps to within $6\%$. The measured $s_1$ and $i$
steps are $20$--$26\%$ smaller than the instantaneous-crossing estimate,
since those resonances last a significant fraction of the remaining
inspiral. All low-frequency modes together shift the
phase by less than $10^{-2}\,{\rm rad}$.

The $f$ modes act differently. They are not crossed, and their effect comes
from the growth of $\lambda_{\rm eff}$ in \cref{fig:keff}. At the
ISCO, where $2\Omega$ has reached $0.8$ and $0.86$ of the $f$-mode
frequencies of the two stars, they have shifted the phase by
$-4.6\,{\rm rad}$ relative to the inspiral with adiabatic tides, which
itself differs from the point-particle inspiral by $-11.2\,{\rm rad}$.
The dynamical response of the $f$ modes therefore increases the tidal
phase shift by about $40\%$, as also found by
\citet{Hinderer:2016eia}, and it exceeds the effect of all other modes by
more than two orders of magnitude, because the $f$ modes carry
almost all of the tidal deformability of the stars (\cref{tab:modes}) and
the leading phase contribution at fixed mode frequency scales with $\Lambda_\alpha$.

Both quantities can be estimated in closed form. Expanding
\cref{eq:free-mode-energy} along the point-particle circular orbits about
the Newtonian resonance $\Omega_n=\omega/2$, with $x_n=(M\Omega_n)^{2/3}$,
the resonance shift of \cref{eq:relshift}, the tidal force to 1PN order
\citep{Vines:2010ca} and the chirp to 1.5PN order
\citep{Blanchet:2013haa}, gives
\begin{align}
    E_{\rm mode}^{\rm free}
    &=\frac{15\pi}{128}\,\Lambda_\alpha
     \left(\frac{m_1\omega}{2}\right)^{7/3}
     X_1^{2/3}X_2\,m_1\,\mathcal K_E
    \nnl
    &=\frac{\pi^2}{512}\,\hat I_\alpha^2
     \left(\frac{\omega}{\Omega_0}\right)^{1/3}
     \left(\frac{R}{m_1}\right)^{5/2}
     q\left(\frac{2}{1+q}\right)^{5/3}
     \frac{m_1^2}{R}\,\mathcal K_E\,,
    \label{eq:free-mode-energy-newtonian}
\end{align}
with
\begin{equation}
    \mathcal K_E
    =1+\left(\frac{1751}{336}-\frac{5X_2}{2}+\frac{13\eta}{4}\right)x_n
     -4\pi x_n^{3/2}\,.
    \label{eq:KE}
\end{equation}
where the first form uses the tidal deformability
$\Lambda_\alpha=\lambda/m_1^5$ and the second the tidal overlap
$\hat I_\alpha$ of \cref{eq:Q-eff}, $m_1$ and $R$ are the mass and radius of
the deformed star, $\Omega_0=\sqrt{m_1/R^3}$, and $q=m_2/m_1$. In the
Newtonian limit $\mathcal K_E=1$, and the second form is the resonant
energy transfer of \citet{Lai:1993di}, their Eq.~(6.11), which counts
both $m=\pm2$ components; see also Ref.~\cite{Gao:2025aqo}. With
$M_{1.4}=m_1/1.4\,M_\odot$ and $R_{10}=R/10\,{\rm km}$,
\begin{align}
    E_{\rm mode}^{\rm free}
    &\simeq1.8\times10^{45}\,{\rm erg}\,
    \left(\frac{\Lambda_\alpha}{10^{-2}}\right)
    \left(\frac{f_\alpha}{100\,{\rm Hz}}\right)^{7/3}
    M_{1.4}^{10/3}
    \nnl&\qquad\times
    q\left(\frac{2}{1+q}\right)^{5/3}\mathcal K_E
    \nnl
    &\simeq1.8\times10^{45}\,{\rm erg}\,
    \left(\frac{\hat I_\alpha}{10^{-4}}\right)^2
    \left(\frac{f_\alpha}{100\,{\rm Hz}}\right)^{1/3}
    \frac{R_{10}^2}{M_{1.4}^{2/3}}
    \nnl&\qquad\times
    q\left(\frac{2}{1+q}\right)^{5/3}\mathcal K_E\,,
    \label{eq:free-mode-energy-estimate}
\end{align}
with $x_n\simeq0.027\,M_{1.4}^{2/3}[(1+q)/2]^{2/3}(f_\alpha/100\,{\rm Hz})^{2/3}$
and, for equal masses, $\mathcal K_E\simeq1+4.8x_n-4\pi x_n^{3/2}$.
The same expansion of \cref{eq:phase-step}, with the torque to
1.5PN order \citep{Blanchet:2013haa}, gives
\begin{align}
    \Delta\phi_{\rm GW}
    &=-\frac{75\pi}{2048}\,\Lambda_\alpha\,\frac{X_1^2}{X_2}\,\mathcal K_\phi
    \nnl
    &=-\frac{5\pi^2}{512}\,\hat I_\alpha^2
      \left(\frac{R}{m_1}\right)^{5}
      \left(\frac{\Omega_0}{\omega}\right)^{2}
      \frac{\mathcal K_\phi}{q(1+q)}\,,
    \label{eq:phase-step-newtonian}
\end{align}
where
\begin{equation}
    \mathcal K_\phi
    =1+\left(\frac{1499}{168}-\frac{5X_2}{2}+\frac{29\eta}{6}\right)x_n
     -8\pi x_n^{3/2}\,.
    \label{eq:Kphi}
\end{equation}
or, in the same units,
\begin{align}
    \Delta\phi_{\rm GW}
    &\simeq-5.8\times10^{-4}\,{\rm rad}\,
    \left(\frac{\Lambda_\alpha}{10^{-2}}\right)
    \frac{2}{q(1+q)}\,\mathcal K_\phi
    \nnl
    &\simeq-6.0\times10^{-4}\,{\rm rad}\,
    \left(\frac{\hat I_\alpha}{10^{-4}}\right)^2
    \left(\frac{f_\alpha}{100\,{\rm Hz}}\right)^{-2}
    \nnl&\qquad\times
    \frac{R_{10}^2}{M_{1.4}^{4}}\,
    \frac{2}{q(1+q)}\,\mathcal K_\phi\,,
    \label{eq:phase-step-estimate}
\end{align}
with $\mathcal K_\phi\simeq1+8.9x_n-8\pi x_n^{3/2}$ for equal masses, which lies between
$1.07$ and $1.17$ for mode frequencies between $100\,{\rm Hz}$ and
$800\,{\rm Hz}$. The mode frequency cancels in the deformability
form, so the phase shift of a mode is set mainly by its
deformability. For $\mathcal K_\phi=1$, \cref{eq:phase-step-newtonian} is
twice the orbital phase shift given in Ref.~\cite{Gao:2025aqo}.
The relativistic corrections raise the energy by $7$--$11\%$ for
crossings between $90$ and $700\,{\rm Hz}$, and the estimates reproduce
the plateaus of \cref{fig:resonant-energy} to a few percent up to the
$s_1$ crossings and to about $10\%$ at the $i$ crossing, where the truncated PN chirp becomes less accurate.

\section{Crustal strain}\label{sec:strain}

We now reconstruct the stellar displacement and crustal strain
from the evolved mode quadrupoles. For each star, this requires the
absolute normalization of its interior perturbations.

\subsection{Construction of displacements}\label{sec:displacement-construction}

The orbital evolution supplies the quadrupole of each mode.
We first consider one quadrupolar mode and suppress its mode label.
Let $\xi^a_{2m}(\mathbf x)$ be its displacement profile from
\cref{eq:displacement_vector} at the conservative pole $\omega$,
with the harmonic time factor removed. We choose the numerical normalization
$W(R)=1$ and denote the quadrupole of this profile by $\bar Q$.
For a mode with nonzero tidal coupling, linearity gives
\begin{equation}
\xi^a_{2m}(\mathbf x,t)
=a_m(t)\xi^a_{2m}(\mathbf x),\qquad
a_m(t)=\frac{Q_{2m}(t)}{\bar Q}.
\label{eq:physical-mode-displacement}
\end{equation}
Thus $\xi^a_{2m}/\bar Q$ is the displacement per unit mode
quadrupole. Rescaling the numerical profile rescales $\bar Q$ by the
same factor and leaves the physical displacement unchanged.

To determine $\bar Q$, we match the interior solution to
$X=A_1u+A_2v$, where $u=X_N^\nu$ and $v=X_N^{-\nu-1}$.
For one unit-normalized real STF component, the matching relation
$\tilde F=pA_2/A_1+q=-Q/E$ of \cref{eq:F-eft-full} implies
\begin{equation}
E=e(\omega)A_1,\qquad
Q=-e(\omega)\bigl[p(\omega)A_2+q(\omega)A_1\bigr]\,.
\label{eq:physical-tidal-readout}
\end{equation}
The conversion factor $e(\omega)$ cancels from $-Q/E$; the response
ratio alone therefore does not fix the absolute amplitude. Introduce
exterior basis solutions with $(E,Q)=(1,0)$ and $(0,1)$,
\begin{equation}
X_E=\frac{1}{e}\left(u-\frac{q}{p}v\right),\qquad
X_Q=-\frac{v}{ep}\,.
\label{eq:strain-unit-exterior-fields}
\end{equation}
They give $X=EX_E+QX_Q$. These are exterior basis fields;
the stellar interior fixes their physical combination through
$Q=-\tilde F E$.

Their absolute scale follows from the gravitational action and
the source coupling $-QE/2$ in \cref{eq:Stidal}. In the canonical Zerilli
variable $Z$, varying this coupling with respect to $Z$ produces the
quadrupole source.
Multiplying the equation for the unit-quadrupole field $Z_Q$ by the
homogeneous unit-tide field $Z_E$ and integrating over the source gives
$E[Z_E]/2=1/2$. Green's identity converts the left-hand side to the
boundary term $K_Z\mathcal W[Z_E,Z_Q]$, with
$\mathcal W[f,g]=f\partial_{r_*}g-g\partial_{r_*}f$.
The quadrupolar coefficient $K_Z=3/(8\pi)$ agrees with the standard
Zerilli energy-flux normalization~\citep{Martel:2005ir}.
The Chandrasekhar transformation then gives
\begin{equation}
K_X\mathcal W[X_E,X_Q]=\frac12,\qquad
K_X=\frac{3}{8\pi}\left(1+\frac{\epsilon^2}{16}\right)\,.
\label{eq:strain-source-pairing}
\end{equation}
Here $\epsilon=2m_A\omega$ for the star of mass $m_A$.
Using $W_*=\mathcal W[u,v]$ in
\cref{eq:strain-unit-exterior-fields} yields
\begin{equation}
\mathcal W[X_E,X_Q]=-\frac{W_*}{e^2p},\qquad
e^2=-\frac{2K_XW_*}{p}\,.
\label{eq:tidal-absolute-normalization}
\end{equation}
We choose the branch continuous with
$e\to-\sqrt{3/(10\pi)}\,\omega^3$ in the Newtonian limit.
The response, source coupling and normalization use the same
regularization scheme. The physical readout retains the relative
$\epsilon^2$ matching accuracy of $p$ and $q$; complex spherical
components follow by linear combinations of the real STF components.

At the conservative pole, the numerical eigenfunction has
$A_1=0$. Its matched coefficient $A_2$ therefore gives
\begin{equation}
\bar Q=-e(\omega)p(\omega)A_2\,.
\label{eq:strain-unit-quadrupole}
\end{equation}
This fixes the conversion in
\cref{eq:physical-mode-displacement}. Equivalently, the exterior
coefficient per unit quadrupole is
\begin{equation}
A_{2,Q}(\omega)\equiv\left.\frac{A_2}{Q}\right|_{E=0}
=-\frac{1}{e(\omega)p(\omega)}\,.
\label{eq:strain-absolute-scale}
\end{equation}
Thus $\bar Q=A_2/A_{2,Q}(\omega)$.
The coefficient $A_{2,Q}$ is defined at each frequency, independently of
the stellar mode condition.

The same $a_m$ multiplies the spatial metric profiles
$H_2$ and $K$ and the tractions. Both the displacement and the
metric enter the Lagrangian strain through
$\Delta g_{ab}=\delta g_{ab}+\mathcal L_\xi g_{ab}$.
The radial functions needed for the strain are
\begin{equation}
\bm U(r)=\left(\frac{T_1(r)}{\mu_{\rm s}(r)},\,
\frac{T_2(r)}{\mu_{\rm s}(r)},\,V(r)\right).
\label{eq:strain-response-coefficients}
\end{equation}
Its contribution to the evolving field is $a_m(t)\bm U(r)$.

\subsection{Strain from the reconstructed field}\label{sec:strain-formalism}

Crustal oscillations may break or melt
the crust before merger~\citep{Penner:2011br,Tsang:2011ad,Pan:2020tht,Passamonti:2020fur,Kuan:2021jmk,Gao:2025aqo}, which has the
potential to power precursor
emission~\citep{Troja:2010zm,Tsang:2011ad,Kuan:2021sin,Neill:2021lat,Suvorov:2022ldw,Suvorov:2024cff,Most:2024eig}, and crust
failure is set by the local strain. Following Ref.~\citep{Gao:2025aqo}, we use the scalar strain
\begin{equation}
    \sigma(r,\theta,\varphi,t)
       =\sqrt{\frac32\,\Delta s_{ab}\Delta s^{ab}}\,,
    \label{eq:strain-total-main}
\end{equation}
where $\Delta s_{ab}$ is the Lagrangian strain tensor of
\cref{eq:strain}. We compare it with the reference levels $0.04$
and $0.1$, motivated by crystal calculations and molecular dynamics
\citep{Baiko:2018jax,Horowitz:2009ya} and adopted in
Ref.~\citep{Gao:2025aqo}. These are illustrative strengths in the stated
strain convention; the yield limit also depends on the deformation and
crystal structure. For an $\ell=2$ tide it is built from the $m=0$
component and the conjugate $m=\pm2$ pair,
\begin{equation}
    \Delta s_{ab}=\Delta s_{ab}^{(0)}
       +2\operatorname{Re}\!\left[e^{2i\varphi}\Delta s_{ab}^{(2)}\right]\,,
    \label{eq:strain-real-tensor}
\end{equation}
where $\Delta s_{ab}^{(2)}$ has the factor $e^{2i\varphi}$
removed, and its expression in terms of the tractions
$T_1$, $T_2$ and the tangential displacement $V$ is given in
\cref{app:strain-reconstruction}.

The quadrupole alone does not measure local strain.
A mode with a small tidal overlap can still have a large displacement
gradient in the crust. In particular, the forced $f$ mode and shear
overtones have large, oppositely signed contributions near the surface.
Their cancellation produces a much smoother tidal deformation than any
one of these modes. Resonantly excited free oscillations retain their
own spatial profiles. A strain calculation must therefore combine the
evolved modes with the remaining forced interior response.

In Newtonian gravity, the evolving tidal displacement and strain
tensor can be reconstructed by summing over a complete set of normal
modes, including in stars with an elastic crust
\citep{Passamonti:2020fur,Passamonti:2022yqp}.
A mode-sum description also extends to conservative
first-post-Newtonian theory~\citep{Gittins:2025fhm}.
Completeness is not established for the relativistic modes used here,
and near-zone calculations show that convergence of a mode sum cannot
generally be assumed~\citep{HegadeKR:2026kku}.
We therefore seek an approximate reconstruction that combines the
retained modes with a smooth remainder determined directly from the
stellar perturbation equations.

Restoring the mode label $n$, we write the interior field
of each harmonic in terms of the profiles $\bm U_n(r)$ as
\begin{equation}
\begin{aligned}
\bm U_m(r,t)
&\simeq
\sum_n a_{nm}(t)\bm U_n(r)
+\bm U_{{\rm bg},m}(r,t),\\
a_{nm}(t)
&=
\frac{Q_{n,2m}(t)}{\bar Q_n}\,.
\end{aligned}
\label{eq:strain-reconstruction-main}
\end{equation}
Here $\bm U_n(r)$ is the profile of mode $n$, normalized to
$W_n(R)=1$ at the surface, and $\bar Q_n$ is the quadrupole of that
profile, fixed by the exterior calibration in
\cref{eq:strain-unit-quadrupole}. The displacement and strain profiles
are normalized as described in \cref{sec:displacement-construction}. The
evolved quadrupole of the mode is
\begin{equation}
\begin{aligned}
    Q_{n,22}(t)&=\beta_n(t)-i\gamma_n(t)\,,\\
    Q_{n,20}(t)&=-\sqrt6\,\alpha_n(t)\,.
\end{aligned}
    \label{eq:strain-mode-amplitude-main}
\end{equation}
in the corotating amplitudes of \cref{eq:Q-corot}, so that $a_{nm}$
scales each profile to the evolved quadrupole. The sum runs over the
modes of \cref{tab:modes}, with the residues at the fitted scale
$\mu_0$, and carries their forced and free parts. The smooth term $\bm U_{{\rm bg},m}$ represents the remainder of
the forced interior response, including omitted high-frequency modes. No
omitted resonance is crossed before the ISCO.

We compute the interior response between $50$ and
$1300\,{\rm Hz}$ by dividing each numerical solution by its matched
tidal field $E_{2m}(\omega)=e(\omega)A_1(\omega)$.
We then subtract the stationary contributions of the retained modes
and fit the remainder with an even polynomial in $\omega$,
\begin{equation}
\begin{aligned}
    \frac{\bm U_m(r,\omega)}{E_{2m}(\omega)}
       &\simeq-\sum_n\frac{\lambda_n\omega_n^2}{\omega_n^2-\omega^2}
                     \frac{\bm U_n(r)}{\bar Q_n}\\
       &\quad+\bm B_0(r)+\omega^2\bm B_2(r)+\omega^4\bm B_4(r)\,.
\end{aligned}
    \label{eq:strain-regular-profile-main}
\end{equation}
Including this remainder allows us to reconstruct the forced
deformation without assuming a complete mode expansion. The
approximation lies in representing the remainder by an even polynomial
over the frequency range considered.

In the time domain the background is
\begin{equation}
    \bm U_{{\rm bg},m}={}\bm B_0E_{2m}
           -\bm B_2\mathcal D_m^2E_{2m}+\bm B_4\mathcal D_m^4E_{2m}\,.
\label{eq:strain-background-time}
\end{equation}
where $E_{22}=F_\beta/g_H$ and $E_{20}=-\sqrt{2/3}\,F_\alpha/g_H$ are
the tidal fields that drive the modes and
\begin{equation}
    \mathcal D_m=\frac{1}{g_H}
        \left(\frac{\dd}{\dd t}+im g_S\right)
    \label{eq:strain-time-derivative}
\end{equation}
is the time derivative in the proper time of the star, with the
frame rotation $-img_S$ of \cref{eq:bpm-mode-eom} removed. The $m=0$ tide has no orbital carrier frequency; its amplitude changes on
the inspiral timescale. The coefficient $\bm B_0$ is fixed by the
full static field, which
also contains the static part of the low-order $g$ modes below the
fitting band. We retain modes up to a lowest omitted pole near $2\,{\rm kHz}$,
above the $1336\,{\rm Hz}$ ISCO frequency.  For the $1.20\,M_\odot$ star
this means retaining $s_5$ and $s_6$ in the strain. A truncated mode sum
fails badly. The modes of \cref{tab:modes} alone overestimate the crust
strain by a factor of about $20$, and even $40$ poles up to
$7\,{\rm kHz}$ leave errors of order $50\%$. With the background, the
reconstruction reproduces the direct interior strain to $0.8\%$ (median)
at frequencies not used in the fit, in both stars.

\subsection{Strain during the inspiral}\label{sec:res-strain}

\begin{figure}
    \centering
    \includegraphics[width=\columnwidth]{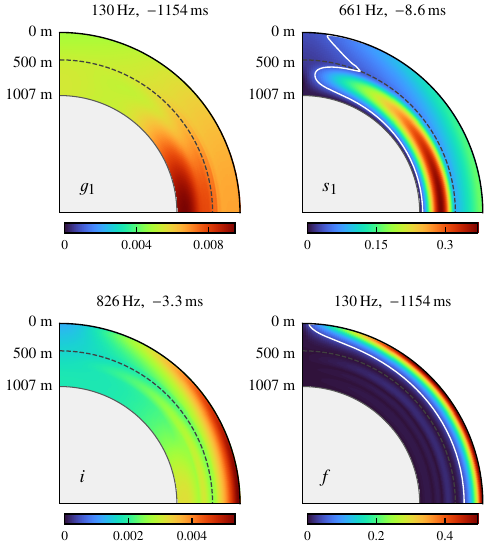}
    \caption{Strain of single modes of the $1.50\,M_\odot$ hybrid
      star during the $1.20+1.50\,M_\odot$ inspiral. Each panel shows
      $\max_\varphi\sigma$ of one term in
      \cref{eq:strain-reconstruction-main}, including $m=0$ and $\pm2$.
      The $g_1$, $s_1$ and $i$ modes are shown at their own resonances,
      where the driving frequency in the local frame of the star equals
      the mode frequency. The $f$ mode is shown at the $g_1$ resonance.
      Headings give $f_{\rm GW}$ and $t-t_{\rm ISCO}$. The polar angle
      runs from the orbital axis at the top to the orbital plane at the
      right. The crust is enlarged to $35\%$ of the displayed radius,
      with physical depths. Each color scale spans the range of its
      panel. The dashed curve marks neutron drip and the white contour
      $\sigma=0.04$.}
    \label{fig:strain-modes}
\end{figure}

\begin{figure*}
    \centering
    \includegraphics[width=\textwidth]{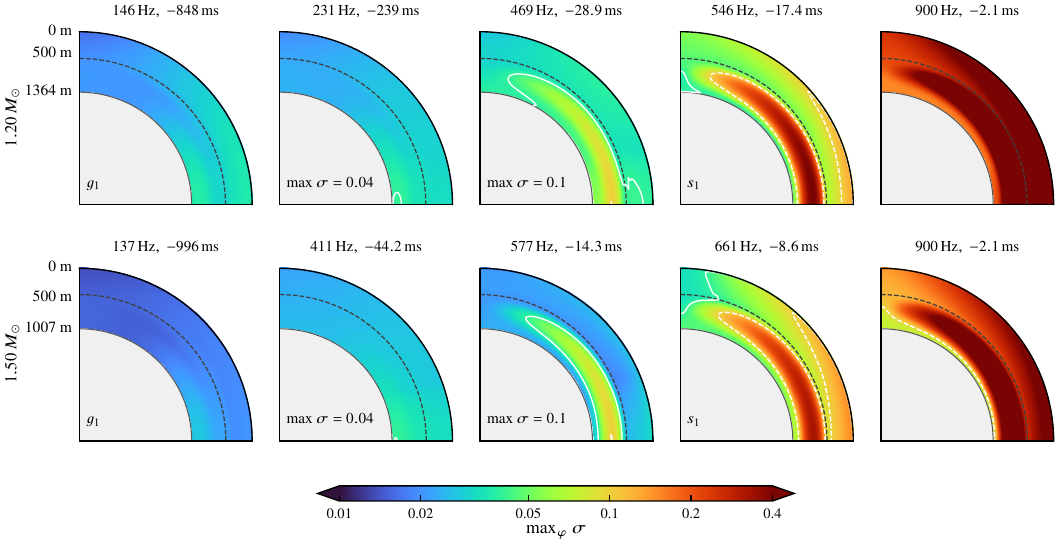}
    \caption{Total strain, the coherent sum of all retained modes and
      the smooth background in \cref{eq:strain-reconstruction-main}, in
      the $1.20\,M_\odot$ (top) and $1.50\,M_\odot$ (bottom) stars of the
      same binary, at five times up to $f_{\rm GW}=900\,{\rm Hz}$. Labels in
      the core mark the times at which the maximum strain first reaches
      $0.04$ and $0.1$ and the $s_1$ resonance. The panels labeled $g_1$
      show the largest strain within $30\,{\rm Hz}$ after the $g_1$
      crossing, when the free $g_1$ oscillation has reached its full
      amplitude and is in phase with $g_2$. Conventions are as in
      \cref{fig:strain-modes}, except that all panels share one logarithmic
      color scale from $0.01$ to $0.4$, saturated outside this range, and
      the solid and dashed white contours
      mark $\sigma=0.04$ and $0.1$. The plotted azimuthal maximum exceeds $0.04$ across the
      crust from about $580\,{\rm Hz}$ in the lighter star and
      $700\,{\rm Hz}$ in the heavier star. The later panels continue the
      linear solution beyond the reference breaking strength.}
    \label{fig:strain}
\end{figure*}

Before we turn to the total field, we look at the modes one by one.
\Cref{fig:strain-modes} shows the strain of single modes of the
$1.50\,M_\odot$ star, which also has the $i$ mode, as
$\max_\varphi\sigma$ at each $(r,\theta)$. A detailed study of the
modal strain fields is given in Ref.~\cite{Gao:2025aqo}, and here we
only recall their main features. Each mode strains the crust where its
displacement varies most rapidly. The $g$ modes strain the base of the
crust, where the stratified fluid below moves horizontally against the
solid, and $g_1$ reaches $\sigma\approx0.01$ at resonance and $0.02$ in
the free oscillation after it. The shear modes are trapped in the crust.
The lowest, $s_1$, shears the inner crust about $670\,{\rm m}$ deep and
reaches $\sigma\approx0.4$ at resonance, and each higher overtone adds
a node and concentrates closer to the surface. The $i$ mode is localized
at the quark--hadron interface, and only a weak tail with
$\sigma\approx0.005$ reaches the crust. The $f$ mode is the tidal bulge of the star, and its strain is
largest in the outer crust, in a thin surface layer. The profile of the
mode is that of an oscillation at the $f$-mode frequency, and in the
outer crust, where the shear speed is a few times $10^{-3}$, a shear wave
at that frequency accumulates one radian of phase within about
$30\,{\rm m}$. The resulting gradient of tangential displacement gives a strain that reaches $\sigma\approx0.5$ already at the
$g_1$ resonance. This does not mean that the crust breaks that early,
because the strain of the crust is set by the total field, in which the
forced part of the $f$ mode cancels against the forced parts of the shear
modes.

\Cref{fig:strain} follows the total field through the inspiral,
first for the $1.50\,M_\odot$ star. Before the first resonance every mode
is forced, and the field is the smooth response of the tide, with
$\sigma$ below $10^{-3}$. After the $g_2$ and $g_1$ crossings, at $87$ and
$130\,{\rm Hz}$, the two modes ring freely at the base of the crust with
a strain of a few $10^{-2}$, beating against each other because they
ring at different frequencies. In this region the $f$ mode alone would
give $\sigma\approx0.5$ in the outer crust, but its layer is cancelled by
the forced shear modes and the total there stays at a few $10^{-3}$. The
strain then grows with the tide until it reaches the lower reference
breaking strength, $\sigma_{\rm b}=0.04$ \citep{Baiko:2018jax}, at about
$400\,{\rm Hz}$. Next comes the $s_1$ resonance. As it approaches, the
forced $s_1$ response is amplified by the closing detuning, and the
strain in the inner crust reaches the higher reference strength, $0.1$
\citep{Horowitz:2009ya}, at about $580\,{\rm Hz}$, before the crossing.
After the crossing at $660\,{\rm Hz}$ the free $s_1$ oscillation
dominates, with $\sigma$ of order unity in its shell, and by $700\,{\rm Hz}$ its azimuthal maximum exceeds $0.04$
across the displayed crust.

The $1.20\,M_\odot$ star goes through the same sequence, but
earlier. Its $g$ modes are excited to larger amplitudes, because their tidal deformabilities are larger
(\cref{tab:modes}) and the companion produces a stronger tide. The total
strain first reaches $0.04$ at $230\,{\rm Hz}$, after the $g_1$ crossing,
about a quarter of a second before the ISCO. Its crust is also thicker, so its
shear modes lie lower, with $s_1$ at $534$ instead of $645\,{\rm Hz}$,
and the strain reaches $0.1$ at $470\,{\rm Hz}$. In both stars the maximum strain agrees to about $2\%$
up to $900\,{\rm Hz}$ with a reconstruction that evolves all $40$
verified poles up to $7\,{\rm kHz}$ and completes them with a constant
remainder. How early the crust reaches the breaking strain thus depends
on the mode couplings, and hence on the EOS and the crust model, and a
survey of these effects is left for future work.

\section{Summary and discussion}\label{sec:summary}

We have connected the relativistic linear dynamical tidal response
of nonrotating neutron stars to mode excitation, GW emission and crustal
deformation in quasicircular binaries. Matching the interior
perturbations to the EFT with analytical MST exterior solutions gives
conservative mode frequencies and deformabilities directly from the
poles and residues. For the two stellar models studied here, one
matching scale per star makes a pure oscillator sum reproduce the
response below the $f$ mode to within $2.1\times10^{-4}$, away from
its poles and zeros. This approximation recovers the Newtonian oscillator
form without assuming completeness of relativistic modes.

We coupled these oscillators to the EOB dynamics of
Refs.~\citep{Hinderer:2016eia,Steinhoff:2016rfi}. Redshift and frame
dragging largely cancel, leaving resonance frequencies within a few
percent of the Newtonian values. An analytical driven--free
decomposition reproduces the evolved amplitudes to a few percent before
the ISCO and supplies the driven quadrupole for the orbital-frequency
waveform. In the $1.20+1.50\,M_\odot$ example, low-frequency resonances
change the GW phase by less than $10^{-2}\,{\rm rad}$. The $f$-mode
resonances are not reached, yet their dynamical response increases the tidal
phase shift by about $40\%$ relative to adiabatic tides at the ISCO.
Our resonance energy and phase estimates reduce to the Newtonian
results of \citet{Lai:1993di}, with PN corrections of order $10\%$.

Local crustal strain requires more information than the induced
quadrupole. The exterior normalization fixes the internal profile
associated with each evolved mode, but a truncated set of modes misses
large cancellations in the forced crustal response. Completing it with
a smooth interior response restores these cancellations. The resulting
strain first reaches the reference strength $0.04$ at $230\,{\rm Hz}$ in
the lighter star, where free $g$-mode oscillations contribute to local
overstrain, and at $410\,{\rm Hz}$ in the hybrid star. The approach to
the $s_1$ resonance raises the strain above $0.1$ in both stars, and its
subsequent excitation produces widespread overstrain. These are
predictions of the linear elastic continuation. They identify where
failure becomes possible, while the actual fracture, melting and
coupling to the magnetosphere require further physics
\citep{Horowitz:2009ya,Baiko:2018jax,Tsang:2011ad,Neill:2021lat,Most:2024eig}.

The present framework can be applied to richer microphysics, including
lepton and hyperon gradients, quark--hadron mixed phases and
superfluidity, which modify buoyancy and tidal couplings
\citep{Kantor:2014lja,Dommes:2015wul,Yu:2017cxe,Jaikumar:2021jbw,Zhao:2022toc}.
In particular, the large GW phase shifts predicted for first-order
phase transitions \citep{Miao:2023jqe,Counsell:2025hcv,Pereira:2025xsi}
warrant reassessment with a consistent relativistic treatment of the
response and orbital dynamics. Dissipation, including viscosity
\citep{Kantor:2026dpj,Storbacka:2026ehf} and GW damping, can be incorporated through a complex response or
dissipative forces in the mode equations given in \cref{eq:hamilton-formal}
\citep{Ripley:2023qxo,HegadeKR:2024slr,HegadeKR:2026iou}.
Our relativistic mode calculations for rotating stars
\citep{Gao:2026lzf} provide a starting point for extending the
tidal response and binary evolution to include stellar spin
\citep{Gupta:2020lnv,Steinhoff:2021dsn}.
The existing EOB Hamiltonian for generic orbits
\citep{Steinhoff:2016rfi} also permits applications to eccentric
inspirals and close encounters \citep{Tsang:2013mca,Vick:2019cun}.
The EOB Hamiltonian can be improved by including the relative 3PN dynamical
tidal corrections from Ref.~\cite{Mandal:2023hqa}, in which the renormalization
scale first appears.
This requires adopting the recently developed improved matching procedure
from Ref.~\cite{Saketh:2026inprep}, which incorporates scale dependence into
the tidal response consistently.

\section{Acknowledgements}

Y.G. thanks Hao-Jui Kuan and Sizheng Ma for helpful
discussions on the scattering-amplitude formalism and for commenting on
the manuscript. He is grateful to Masaru Shibata and Kenta Kiuchi for their
continued encouragement. He also thanks Zhiqiang Miao and Dong Lai for helpful discussions
during his visit to the T. D. Lee Institute, and Arthur Suvorov and
Kostas Kokkotas for useful discussions during his visit to the University
of T\"ubingen.
Numerical computations were performed on the
Sakura cluster at the Max Planck Computing and Data Facility (MPCDF).


\bibliography{refs}

\newpage
\begin{widetext}
\appendix

\section{Interior perturbation equations}\label{app:interior}

We summarize the polar-sector perturbation equations for a nonrotating
two-layer NS consisting of a perfect-fluid core and an elastic solid
crust, following
Refs.~\cite{Finn:1990ads,Detweiler:1985zz,Lindblom:1983ps,Kruger:2014pva,Gao:2025aqo}.
For brevity $n_\ell = (\ell+2)(\ell-1)/2$ and a prime denotes $\dd/\dd r$.
At a density discontinuity we assume slow phase conversion, so
fluid elements remain in their phase over an oscillation period \citep{Tonetto:2020bie}.

\subsection{Equations for perfect fluid}

The master equations for
$(H_1, K, W, V,H_0,H_2)$~\cite{Detweiler:1985zz,Lindblom:1983ps} in the fluid core are,

\begin{align}
H_1' &= \left[\tfrac{1}{2}(\lambda'-\nu') - \tfrac{\ell+1}{r}\right] H_1
        + \frac{e^\lambda}{r}\!\left[H_0 + K - 16\pi(\epsilon+p) V\right]\,,\label{eq:pf-H1}\\
K' &= \frac{1}{r} H_0 + \frac{n_\ell+1}{r} H_1
      + \left(\tfrac{1}{2}\nu' - \tfrac{\ell+1}{r}\right) K
      - \frac{8\pi(\epsilon+p)}{r}\, e^{\lambda/2}\, W\,,\label{eq:pf-K}\\
W' &= -\!\left(\frac{\ell+1}{r} + \frac{p'}{\Gamma_1 p}\right) W
      + r\, e^{\lambda/2}\!\left[\frac{\epsilon+p}{\Gamma_1 p}\!\left(e^{-\nu}\omega^2 V + \tfrac{1}{2} H_0\right)
      - \frac{2(n_\ell+1)}{r^2} V + \tfrac{1}{2} H_0 + K\right]\,,\label{eq:pf-W}\\
V' &= \!\left(-A_{\rm sch} + \nu' - \tfrac{\ell}{r}\right) V
      - \frac{A_{\rm sch}\, e^\nu}{2\omega^2}\!\left(H_0 + \frac{\nu'\, W}{r\, e^{\lambda/2}}\right)
      + r\, H_1 - \frac{e^{\lambda/2}}{r}\, W\,,\label{eq:pf-V} \\
H_0 &= \frac{1}{(n_\ell+1)\, r - \tfrac{r}{2} e^{-\lambda}(r\lambda' + 2)}
    \bigg\{
        r^2 e^{-\lambda}\!\left[\omega^2 r e^{-\nu} - \tfrac{n_\ell+1}{2}\nu'\right] H_1 \nonumber\\
    &\quad
    + \left[n_\ell\, r - \omega^2 r^3 e^{-\nu} - \tfrac{1}{4} r^2 e^{-\lambda}\nu'(r\nu' - 2)\right] K
    + 4\pi r^2 (\epsilon+p)\!\left(e^{-\lambda/2}\nu'\, W + 2\omega^2 r e^{-\nu}\, V\right)
    \bigg\}\,.\label{eq:pf-H0}\\
H_2 &= H_0\,,
\end{align}
where $A_{\rm sch}$ is the Schwarzschild discriminant, defined as
\begin{equation}
    A_{\rm sch} = \frac{p'}{p}\left(\frac{1}{\Gamma_0}-\frac{1}{\Gamma_1}\right)\,.
\end{equation}
The system can equivalently be closed using the auxiliary variable
to replace the differential equation for $V$,
\begin{equation}\label{eq:Y-def}
    Y = -e^{\nu/2}\,\Delta p/r^\ell\,.
\end{equation}
It is a rescaling of the Lagrangian pressure perturbation, related to $V$ by
\begin{equation}\label{eq:V-alg}
    \omega^2(\epsilon+p)\, V = e^{\nu/2}\, Y
    + \frac{p'\, e^{\nu-\lambda/2}}{r}\, W
    - \tfrac{1}{2}(\epsilon+p)\, e^\nu\, H_0\,.
\end{equation}

We use the $Y$ formulation. Then $V$ follows
algebraically from \cref{eq:V-alg}, $W'$ takes the form of the
corresponding equation in the crust below, and
\begin{equation}
\begin{aligned}
    Y' ={}& -\frac{\ell}{r}Y
    +(\epsilon+p)e^{\nu/2}\biggl[
      \frac12\Bigl(\frac1r-\frac{\nu'}{2}\Bigr)H_0
      +\frac12\Bigl(r\omega^2e^{-\nu}+\frac{n_\ell+1}{r}\Bigr)H_1
      +\frac12\Bigl(\frac{3\nu'}{2}-\frac1r\Bigr)K
      -\frac{(n_\ell+1)\nu'}{r^2}V\\
    &-\frac{1}{r}\Bigl(4\pi(\epsilon+p)e^{\lambda/2}
      +\omega^2e^{\lambda/2-\nu}
      +\frac{e^{-\lambda/2}}{2}\bigl[\nu'(\tfrac12\lambda'+\tfrac2r)-\nu''\bigr]\Bigr)W
    \biggr]\,.
\end{aligned}
\label{eq:pf-Y}
\end{equation}
Note that evolving $V$ directly can reduce cancellation errors at very
low frequencies~\cite{Kruger:2014pva}.

\subsection{Equations for elastic solid}

In the solid crust, the system acquires two additional degrees of
freedom, the radial and tangential tractions $T_1$ and $T_2$. We
include the auxiliary variable $Y$ explicitly in the integration, and
the master equations for
$(H_1, K, H_0, W, V, T_1, T_2, H_2, Y)$ read~\cite{Finn:1990ads,Kruger:2014pva,Gao:2025aqo}
\begin{align}
    H_1' &= \left[\tfrac{1}{2}(\lambda'-\nu') - \tfrac{\ell+1}{r}\right] H_1
            + \frac{e^\lambda}{r}\!\left[H_2 + K - 16\pi(\epsilon+p)\, V\right]\,,\\
    K' &= \frac{H_2}{r} + \frac{n_\ell+1}{r}\, H_1
          + \left(\tfrac{\nu'}{2} - \tfrac{\ell+1}{r}\right) K
          - \frac{8\pi(\epsilon+p)\, e^{\lambda/2}}{r}\, W\,,\\
    H_0' &= K' - r\, e^{-\nu}\,\omega^2\, H_1
            - \left(\tfrac{\nu'}{2} + \tfrac{\ell-1}{r}\right) H_0
            - \left(\tfrac{\nu'}{2} + \tfrac{1}{r}\right) H_2 + \tfrac{\ell}{r}\, K
            - \frac{16\pi}{r}\, T_2\,,\\
    W' &= -\frac{\ell+1}{r}\, W
          + r\, e^{\lambda/2}\!\left[\frac{e^{-\nu/2}\, Y}{\Gamma_1 p}
          - \frac{2(n_\ell+1)}{r^2}\, V + \tfrac{1}{2} H_2 + K\right]\,,\\
    V' &= \frac{T_2}{\mu_{\rm s}\, r} + \frac{e^{\lambda/2}}{r}\, W + \frac{2-\ell}{r}\, V\,,\\
    T_2' &= -\tfrac{1}{2} r\, e^\lambda (\epsilon+p)\, H_0
            + r\, e^{\lambda-\nu/2}\!\left(Y + \frac{e^{\nu/2}\, T_1}{2\, r^2}\right)
            + \left[\tfrac{1}{2}(\lambda'-\nu') - \tfrac{\ell+1}{r}\right] T_2 \nonumber\\
         &\quad + \left[\frac{2\, n_\ell\, e^\lambda\, \mu_{\rm s}}{r} - e^{\lambda-\nu}\, r\omega^2 (\epsilon+p)\right] V
            + e^{\lambda/2}\, p'\, W\,,\\
    H_2 &= H_0 + 32\pi\, \mu_{\rm s}\, V\,,\label{eq:h2_solid}\\
   T_1
    &= 2\mu_{\rm s}\!\left[2 e^{-\lambda/2} W - r^2 K + 2(n_\ell+1) V\right]-\frac{4}{3\Gamma_1 p} e^{-\nu/2}\, \mu_{\rm s}\, r^2\, Y \,,\\
Y &= \frac{e^{\nu/2}}{8\pi r^3}
    \bigg\{
        \left[\tfrac{r}{2} e^{-\lambda}(r\nu' - 2) + (n_\ell+1) r\right] H_0
        - r^2 e^{-\lambda}\!\left[\omega^2 r e^{-\nu} - \tfrac{n_\ell+1}{2}\nu'\right] H_1
        \nonumber\\
    &\quad
        - \left[n_\ell r - \omega^2 r^3 e^{-\nu} - \tfrac{1}{4} r^2 e^{-\lambda}\nu'(r\nu' - 2)\right] K
    \bigg\}
    + \frac{e^{\nu/2}}{r^2}\, T_1
    + \frac{2\, e^{\nu/2 - \lambda}}{r^2}\, T_2\,. \label{eq:alg2}
\end{align}

\subsection{Boundary and junction conditions}

We now specify the boundary conditions at the center and surface of
the NS, together with the matching conditions at the fluid--solid
interface. We denote the jump of a quantity $Q$ across an interface
at radius $r$ by
\begin{equation}
    [Q]_r \equiv \lim_{\varepsilon \to 0^+}\!\big[Q(r+\varepsilon) - Q(r-\varepsilon)\big]\,.
\end{equation}

Regularity at $r=0$ requires the perturbation variables to admit a
Taylor expansion about the origin, with all coefficients uniquely
determined by the two free parameters $K(0)$ and $W(0)$ (see details in Ref.~\citep[pp. 27--30, 59--60]{soton384187}).

The continuity of the spacetime metric and of the physical perturbation
variables imposes
\begin{equation}
    [H_0]_r = [H_1]_r = [K]_r = [W]_r = [T_2]_r = 0\,,
    \qquad [r^2e^{-\nu/2}Y-T_1]_r=0\,.
\end{equation}
The last condition enforces continuity of the total normal
traction, $\Delta p+\delta\pi_r{}^r$. At a fluid--fluid interface it
reduces to continuity of $Y$. The horizontal displacement $V$ is not required to be continuous, and
the algebraic relation~\cref{eq:h2_solid} then implies that $H_2$ jumps by
\begin{equation}
    [H_2]_r = 32\pi\,[\mu_{\rm s}\, V]_r\,.
\end{equation}

The stellar surface is a matter--vacuum boundary. For a two-layer NS
with a solid outer surface,
\begin{equation}
    T_2(R) = 0\,, \qquad
    R^2 e^{-\nu(R)/2}\, Y(R) - T_1(R) = 0\,,
\end{equation}
the second condition following from~\cref{eq:alg2} together with the
continuity of $H_0$, $H_1$, and $K$ at $r=R$. For a three-layer NS
with a fluid ocean overlying the crust, the surface lies in the fluid
and the condition reduces to $Y(R) = 0$, equivalent to $\Delta p(R)=0$.

Since the perturbation equations are linear, one overall normalization
remains free. It drops out of the response function, which depends
only on $A_2/A_1$, and the physical amplitude of each mode is fixed by its
exterior quadrupole in \cref{sec:displacement-construction}.

\section{Analytic MST Solutions to the RW Equation}
\label{app:MST}

In this appendix we collect the analytic solutions to the RW
equation due to Mano, Suzuki, and Takasugi~\cite{Mano:1996mf},
which we employ to construct the exterior response function.

\subsection{The Two Pairs of Solutions}

The hypergeometric solutions $\{X_0^\nu, X_0^{-\nu-1}\}$ and confluent-hypergeometric solutions
$\{X_C^\nu, X_C^{-\nu-1}\}$ are given by~\cite{Mano:1996mf,Sasaki:2003xr,Chakrabarti:2013lua}
\begin{align}
X_0^{\nu} &= e^{i(x-1)\epsilon}(-x)^{-i\epsilon}(1-x)^{\nu+i\epsilon+1}
\sum_{n=-\infty}^{\infty}(1-x)^{n}\, a_n^{\nu}\,
\frac{\Gamma(2n+2\nu+1)\,\Gamma(-n-i\epsilon-\nu-2)}{\Gamma(n-i\epsilon+\nu+3)} \nnlq
\times {}_2F_1\!\bigl(-n-i\epsilon-\nu-2,\,-n-i\epsilon-\nu+2;\,-2n-2\nu;\,1/(1-x)\bigr) , \\
X_C^{\nu} &= \left(1-\frac{\epsilon}{z}\right)^{-i\epsilon}
\sum_{n=-\infty}^{\infty} 2^{\nu+n}\, i^{n}\, e^{-iz}\, z^{\nu+n+1}\, a_n^{\nu}\,
\frac{\Gamma(n-i\epsilon+\nu-1)\,\Gamma(n-i\epsilon+\nu+1)}
     {\Gamma(2(n+\nu)+2)\,\Gamma(n+i\epsilon+\nu+3)} \nnlq
\times {}_1F_1\!\bigl(n+i\epsilon+\nu+1;\,2(n+\nu)+2;\,2iz\bigr) \,,
\end{align}
where the variable $x = 1 - r/2M$ is used in the series expansion of
the near-zone solution, while $z = \omega r$ is used for the far-zone solution.
Here $\Gamma$ denotes the Gamma function, and ${}_2F_1$ and ${}_1F_1$ are the Gauss
hypergeometric and confluent hypergeometric functions, respectively. The series
coefficients $a_n^\nu$ and the renormalized angular momentum $\nu$ are determined
as described below.

The two representations describe the same solution and are related by
\cite[Eq.~(4.1)]{Mano:1996mf}
\begin{equation}
X_0^{\nu} = K_\nu\, X_C^{\nu}\,,
\end{equation}
with \cite[Eq.~(4.2)]{Mano:1996mf}
\begin{align}
K_\nu &= -\frac{\pi i^r\, 2^{-\nu-r}\, \epsilon^{-\nu-r-1}\, \csc(\pi(\nu+i\epsilon))}
{\Gamma(r+i\epsilon+\nu-1)\,\Gamma(r+i\epsilon+\nu+1)\,\Gamma(r+i\epsilon+\nu+3)}
\left[\sum_{n=r}^{\infty} a_n^\nu\,
\frac{\Gamma(n+r+2\nu+1)\,\Gamma(n+i\epsilon+\nu-1)}{(n-r)!\,\Gamma(n-i\epsilon+\nu+3)}\right] \nnl
&\quad \times \left[\sum_{n=-\infty}^{r} a_n^\nu\,
\frac{\Gamma(n-i\epsilon+\nu-1)\,\Gamma(n-i\epsilon+\nu+1)}
     {(r-n)!\,\Gamma(n+r+2\nu+2)\,\Gamma(n+i\epsilon+\nu+1)\,\Gamma(n+i\epsilon+\nu+3)}\right]^{-1}\,. \label{KMST}
\end{align}
The integer $r \in \mathbb{Z}$ is in principle arbitrary (and not to be confused with the radial coordinate $r$ entering $x$ above), as can be verified
numerically; for definiteness we take $r=0$.

\subsection{Recurrence Relation and Determination of \texorpdfstring{$\nu$}{nu}}

The coefficients $a_n^\nu$ obey the three-term recurrence
\cite[Eq.~(2.5)]{Mano:1996mf}
\begin{equation}
\alpha_n^\nu\, a_{n+1}^\nu + \beta_n^\nu\, a_n^\nu + \gamma_n^\nu\, a_{n-1}^\nu = 0\,,
\end{equation}
with \cite[Eqs.~(2.6)--(2.8)]{Mano:1996mf}
\begin{align}
\alpha_n^\nu &= -\frac{i\epsilon\,(\nu+n-i\epsilon-1)(\nu+n-i\epsilon+1)(\nu+n+i\epsilon-1)}
                    {(\nu+n+1)\,(2(\nu+n)+3)} \,,\\
\beta_n^\nu &= -\ell(\ell+1) + (\nu+n)(\nu+n+1)
              + \frac{(\epsilon^2+4)\,\epsilon^2}{(\nu+n)(\nu+n+1)} + 2\epsilon^2 \,, \\
\gamma_n^\nu &= \frac{i\epsilon\,(\nu+n-i\epsilon+2)(\nu+n+i\epsilon)(\nu+n+i\epsilon+2)}
                    {(\nu+n)\,(2(\nu+n)-1)} \,.
\end{align}
Following \cite{Mano:1996mf,Chakrabarti:2013lua}, we introduce the continued-fraction
ratios \cite[Eqs.~(2.9), (2.10)]{Mano:1996mf}
\begin{align}
R_n(\nu) &= \frac{a_n^\nu}{a_{n-1}^\nu}
        = -\frac{\gamma_n^\nu}{\beta_n^\nu + \alpha_n^\nu R_{n+1}(\nu)} \,, \label{Rcf} \\
L_n(\nu) &= \frac{a_n^\nu}{a_{n+1}^\nu}
        = -\frac{\alpha_n^\nu}{\beta_n^\nu + \gamma_n^\nu L_{n-1}(\nu)} \,.
\end{align}
Their asymptotic behavior,
\begin{align}
\lim_{n \to +\infty} n\, R_n(\nu) &= -\frac{i\epsilon}{2} \,, \label{Rlim} \\
\lim_{n \to -\infty} n\, L_n(\nu) &= +\frac{i\epsilon}{2} \,, \label{Llim}
\end{align}
identifies the unique minimal solutions in the two limits, provided the continued
fractions converge there. In general, only one of (\ref{Rlim}) or (\ref{Llim}) can
be enforced for a given $\nu$. Requiring both to hold simultaneously fixes the
renormalized angular momentum and is precisely what guarantees convergence of the
analytic series for $X_0^\nu$ and $X_C^\nu$.

In practice, (\ref{Rlim}) is used as a starting value for $R_n(\nu)$ at some
large $n > 0$, from which $R_n(\nu)$ at smaller $n$ is obtained by downward
iteration of (\ref{Rcf}). The ratio $L_n(\nu)$ is treated analogously, starting
from a large negative $n$ and iterating upward. The renormalized angular
momentum $\nu$ is then fixed by the consistency condition
\cite[Eq.~(2.11)]{Mano:1996mf}
\begin{equation}
R_n(\nu)\, L_{n-1}(\nu) = 1 \,,
\end{equation}
which we impose at $n=1$ for definiteness and solve by standard root-finding,
using the small-$\epsilon$ expansion \cite[Eq.~(6.3)]{Mano:1996mf}
\begin{equation}
\nu = \ell + \left[
-\frac{(\ell-2)^2(\ell+2)^2}{2\ell(2\ell-1)(2\ell+1)}
- \frac{4}{\ell(\ell+1)}
+ \frac{(\ell-1)^2(\ell+3)^2}{(2\ell+1)(2\ell+2)(2\ell+3)} - 2
\right]\frac{\epsilon^2}{2\ell+1} + \mathcal{O}(\epsilon^4)\,,
\end{equation}
as an initial guess. Once $\nu$ is fixed, the recurrence determines the
coefficients $a_n^\nu$ up to an overall normalization, which we set by
$a_0^\nu = 1$, following \cite{Mano:1996mf}.

\subsection{Asymptotic Amplitudes and Normalization}
\label{app:normMST}

As in the MST literature, ``in'' and ``out'' in this subsection refer to
the time dependence $e^{-i\omega t}$. Because the RW equation depends only on
$\omega^2$, the same radial functions serve for the $e^{i\omega t}$ of
\cref{sec:relativistic-osc}, in which the roles of $e^{\pm i\omega r_*}$ are
exchanged. The amplitudes of \cref{eq:mst-wave-amplitudes} are therefore
$I_j=N_jA^j_{C\,\text{out}}$ and $O_j=N_jA^j_{C\,\text{in}}$.

The asymptotic behavior of the far-zone solution at spatial infinity is
\begin{equation}
X_C^\nu \;\stackrel{r_* \to \infty}{\sim}\;
A_{C\,\text{in}}^\nu\, e^{-i\omega r_*} + A_{C\,\text{out}}^\nu\, e^{+i\omega r_*}\,,
\end{equation}
with complex amplitudes
\begin{align}
A_{C\,\text{in}}^\nu &= \tfrac{1}{2} i^{-\nu+i\epsilon-1}
\sum_{n=-\infty}^{\infty} -i^{n}\,(2\epsilon)^{-i\epsilon}\,
e^{i\pi(\nu + n/2)}\, a_n^\nu\,
\frac{\Gamma(n-i\epsilon+\nu-1)}{\Gamma(n+i\epsilon+\nu+3)} , \label{N1MST} \\
A_{C\,\text{out}}^\nu &= \tfrac{1}{2} i^{-\nu+i\epsilon-1}
\sum_{n=-\infty}^{\infty} (2\epsilon)^{i\epsilon}\, a_n^\nu\,
\frac{\Gamma(n-i\epsilon+\nu-1)\,\Gamma(n-i\epsilon+\nu+1)}
     {\Gamma(n+i\epsilon+\nu+1)\,\Gamma(n+i\epsilon+\nu+3)}\,. \label{N2MST}
\end{align}
A standing-wave--normalized solution is then obtained as
\begin{equation}\label{NMST}
X_N^\nu \equiv N_\nu\, X_C^\nu ,
\qquad
N_\nu = \tfrac{1}{2}\bigl(A_{C\,\text{in}}^\nu A_{C\,\text{out}}^\nu\bigr)^{-1/2}\,.
\end{equation}
The square root makes $N_\nu$ bi-valued, and the branch is chosen so that
$N_\nu$ is continuous in $\omega$.
With this choice
one finds
\begin{align}
X_N^\nu &\stackrel{r_*\to\infty}{\sim} \tfrac{1}{2}\!\left[
\bigl(A_{C\,\text{out}}^\nu/A_{C\,\text{in}}^\nu\bigr)^{-1/2} e^{-i\omega r_*}
+ \bigl(A_{C\,\text{out}}^\nu/A_{C\,\text{in}}^\nu\bigr)^{+1/2} e^{+i\omega r_*}
\right] \\
&= \tfrac{1}{2}\bigl[ e^{-i(\omega r_* + \alpha_\nu)} + e^{+i(\omega r_* + \alpha_\nu)} \bigr]
 = \cos(\omega r_* + \alpha_\nu) \,,
\end{align}
with the phase
\begin{equation}
\alpha_\nu := \frac{1}{2i}\log\frac{A_{C\,\text{out}}^\nu}{A_{C\,\text{in}}^\nu} \,.
\end{equation}
For real frequencies on the branch used here, the incoming and
outgoing amplitudes have equal moduli, giving a real phase $\alpha_\nu$.
The normalization is continued analytically when locating complex QNMs. Finally, the Wronskian follows directly from the
asymptotic amplitudes,
\begin{align}\label{WronsMST}
W_* &= 2 i \omega\, N_\nu N_{-\nu-1}
\bigl( A_{C\,\text{in}}^\nu A_{C\,\text{out}}^{-\nu-1}
     - A_{C\,\text{out}}^\nu A_{C\,\text{in}}^{-\nu-1} \bigr) \\
    &= \omega\,\sin(\alpha_\nu - \alpha_{-\nu-1}) \,.
\end{align}

\section{\texorpdfstring{EOB potentials, dimensionless equations, and initial data}{EOB potentials, dimensionless equations, and initial data}}\label{app:eob-pp}

Here we collect the two resummed potentials of the point-particle sector,
$A_{\rm pp}$ and $D_{\rm pp}$, the dimensionless equations of motion, and the initial data used to start the
evolution.

\subsection{Point-particle potentials}

For the point-particle sector we adopt the non-spinning limit of the calibrated
effective-one-body model of \citet{Barausse:2009xi} and \citet{Taracchini:2012ig}.
The potentials are functions of the dimensionless inverse separation $u=M/r$ and
the symmetric mass ratio $\eta$. The lapse $A_{\rm pp}$ is taken in the
log-resummed form of \citet{Barausse:2009xi}, which preserves the correct
strong-field behaviour,
\begin{equation}
    \label{eq:App}
    A_{\rm pp}(u) = \left[\frac{1}{(\eta K-1)^2} + \frac{2u}{\eta K-1}\right]
    \left[1 + \eta\Delta_0
    + \eta\ln\!\Big(1 + \sum_{n=1}^{4}\Delta_n\,u^n\Big)\right]\,.
\end{equation}
The calibration function $K(\eta)$ is fixed by comparison to numerical-relativity
waveforms~\cite{Taracchini:2012ig},
\begin{equation}
    \label{eq:Kcal}
    K(\eta) = 1.4467 - 1.7152\,\eta - 3.2463\,\eta^2\,,
\end{equation}
and the logarithmic coefficients $\Delta_n$ are set by matching the PN expansion of
$A_{\rm pp}$~\cite{Barausse:2009xi}. With $m\equiv\eta K-1$ they read

\begin{subequations}\label{eq:Deltas}
\begin{align}
    \Delta_0 &= K(\eta K-2)\,, \\
    \Delta_1 &= -2\,m\,(\Delta_0+K)\,, \\
    \Delta_2 &= \tfrac12\,\Delta_1(\Delta_1-4m)\,, \\
    \Delta_3 &= -\tfrac13\Delta_1^3 + \Delta_1\Delta_2 + m\,\Delta_1^2
                - 2m\,(\Delta_2-m)\,, \\
    \Delta_4 &= \tfrac{1}{96}\Big[24\Delta_1^4 - 96\Delta_1^2\Delta_2
                + 48\Delta_2^2 - 64\,m\,\Delta_1^3 \nnl
             &\quad + 96\Delta_1(\Delta_3+2m\Delta_2)
                - m\big(192\Delta_3 + m(123\pi^2-3008)\big)\Big]\,.
\end{align}
\end{subequations}
The remaining potential enters through its inverse in the log-resummed form
\begin{equation}
    \label{eq:Dpp}
    D_{\rm pp}^{-1}(u) = 1 + \ln\!\big[1 + 6\eta u^2
    + 2(26-3\eta)\eta\,u^3\big]\,.
\end{equation}
In the test-particle limit $\eta\to0$ the calibration terms drop out,
$A_{\rm pp}\to1-2u$ and $D_{\rm pp}\to1$, recovering the Schwarzschild metric, and
the quartic term vanishes with $z_3\to0$.

\subsection{Dimensionless equations of motion}

\label{sec:eob-eom}

For the numerical evolution we cast the dynamics of
\cref{eq:hamilton-formal} in dimensionless variables, keeping the total mass $M$
and the reduced mass $\mu$ as two independent scales. Lengths, times,
and inverse frequencies are measured in $M$, linear momenta in $\mu$,
angular momenta in $\mu M$, and the quadrupole in $\mu M^2$.
Specifically, the orbital sector uses
\begin{equation}
    \hat t=\frac{t}{M},\quad
    \hat r=\frac{r}{M},\quad
    \hat p_{r_*}=\frac{\pr}{\mu},\quad
    \hat P_\phi=\frac{P_\phi}{\mu M}\,,
    \label{eq:dimensionless-orbit}
\end{equation}
so that the orbital frequency is $\hat\Omega=M\Omega$, and the mode
sector uses
\begin{equation}
\begin{aligned}
    (\hat\alpha,\hat\beta,\hat\gamma)
      &=\frac{(\alpha,\beta,\gamma)}{\mu M^2},\\
    (\hat p_\alpha,\hat p_\beta,\hat p_\gamma)
      &=M(p_\alpha,p_\beta,p_\gamma),\\
    \hat\lambda&=\frac{\lambda}{M^5}=X_1^5\Lambda,
    \quad
    \hat\omega=M\omega\,,
\end{aligned}
    \label{eq:dimensionless-mode}
\end{equation}
where $\Lambda=\lambda/m_1^5$ is the isolated-star deformability. The
oscillator energy \cref{eq:Hmode-eob} and the tidal spin \cref{eq:SQ-eob}
then become
\begin{align}
    \hat H_{\rm mode}
    &\equiv\frac{H_{\rm mode}}{\mu}
    =\eta\,\frac{6\hat\alpha^2+2\hat\beta^2+2\hat\gamma^2}{4\hat\lambda}
    \nnl&\quad
    +\frac{\hat\lambda\hat\omega^2}{\eta}
    \left(
       \frac{\hat p_\alpha^2}{6}
       +\frac{\hat p_\beta^2}{2}
       +\frac{\hat p_\gamma^2}{2}
    \right)\,,
    \label{eq:Hmode-hat}\\
    \hat S_Q
    &\equiv\frac{S_Q}{\mu M}
    =2\left(\hat\beta\hat p_\gamma-\hat\gamma\hat p_\beta\right)\,.
    \label{eq:SQ-hat}
\end{align}
Every Hamiltonian is likewise measured in $\mu$,
$\hat H=H/\mu$, and the hatted variables obey canonical equations
with respect to $\hat t$ and $\HEOB$.

In these variables the energy map \cref{eq:energy-map} reads
\begin{equation}
    \HEOB
    =\frac{1}{\eta}\sqrt{1+2\eta\left(\Heff-1\right)}\,,
    \label{eq:energy-map-hat}
\end{equation}
and the effective Hamiltonian \cref{eq:Heff-eob} keeps its form,
\begin{align}
    \Heff
    ={}&\Big[A\Big(
        1+u^2(\hat P_\phi-\hat S_Q)^2
        +\frac{A}{A_{\rm pp}^2}\hat p_{r_*}^2
        +z_3u^2\hat p_{r_*}^4
    \nnl&\quad
        +2z_c\hat H_{\rm mode}
        +2\hat C_{\rm int}
    \Big)\Big]^{1/2}
    +\hat Z\hat S_Q\,,
    \label{eq:Heff-hat}
\end{align}
where $A=A_{\rm pp}+2z_I\hat H_{\rm int}$ is already dimensionless, the
interaction scalars \cref{eq:Hint-eob,eq:Cint-eob} become
\begin{equation}
\begin{aligned}
    \hat H_{\rm int}
    &=-\frac{3X_2}{2}u^3\left(\hat\alpha+\hat\beta\right)\,,\\
    \hat C_{\rm int}
    &=\frac{3X_2}{2}u^3(u+3u^2)
      \left[(\eta-2)\hat\alpha+\eta\hat\beta\right]\,,
\end{aligned}
    \label{eq:int-scalars-hat}
\end{equation}
and $\hat Z=MZ$ is obtained from \cref{eq:Zcirc} through the
replacement $L/r^3\to\eta(\hat P_\phi-\hat S_Q)u^3$.

With the radiation-reaction force $\hat{\mathcal F}_\phi$ of
\cref{eq:radiation-reaction-flux}, the orbital equations are
\begin{align}
    \label{eq:eom-orbit}
    \frac{\dd\hat r}{\dd\hat t}
    &=\frac{A_{\rm pp}}{\sqrt{D_{\rm pp}}}
      \frac{\partial\HEOB}{\partial\hat p_{r_*}}\,, &
    \frac{\dd\phi}{\dd\hat t}
    &=\frac{\partial\HEOB}{\partial\hat P_\phi}
      =\hat\Omega\,,
    \nnl
    \frac{\dd\hat p_{r_*}}{\dd\hat t}
    &=-\frac{A_{\rm pp}}{\sqrt{D_{\rm pp}}}
      \frac{\partial\HEOB}{\partial\hat r}
      +\hat{\mathcal F}_\phi\frac{\hat p_{r_*}}{\hat P_\phi}\,, &
    \frac{\dd\hat P_\phi}{\dd\hat t}
    &=\hat{\mathcal F}_\phi\,,
\end{align}
where the factor $A_{\rm pp}/\sqrt{D_{\rm pp}}$ accounts for the
tortoise rescaling of the radial momentum and the radial force
follows the standard prescription
$\hat{\mathcal F}_{r_*}
    =\hat{\mathcal F}_\phi\hat p_{r_*}/\hat P_\phi$.
With $\hat F_q=M^2F_q$ and $\hat g_S=Mg_S$ ($g_H$ is already
dimensionless), each mode obeys
\begin{equation}
\begin{aligned}
    \frac{\dd\hat\alpha}{\dd\hat t}
    &=g_H\frac{\hat\lambda\hat\omega^2}{3\eta}\hat p_\alpha,
    &
    \frac{\dd\hat p_\alpha}{\dd\hat t}
    &=-\left(\hat F_\alpha
      +g_H\frac{3\eta}{\hat\lambda}\hat\alpha\right),\\
    \frac{\dd\hat\beta}{\dd\hat t}
    &=g_H\frac{\hat\lambda\hat\omega^2}{\eta}\hat p_\beta
      -2\hat g_S\hat\gamma,
    &
    \frac{\dd\hat p_\beta}{\dd\hat t}
    &=-\left(\hat F_\beta
      +g_H\frac{\eta}{\hat\lambda}\hat\beta
      +2\hat g_S\hat p_\gamma\right),\\
    \frac{\dd\hat\gamma}{\dd\hat t}
    &=g_H\frac{\hat\lambda\hat\omega^2}{\eta}\hat p_\gamma
      +2\hat g_S\hat\beta,
    &
    \frac{\dd\hat p_\gamma}{\dd\hat t}
    &=-\left(g_H\frac{\eta}{\hat\lambda}\hat\gamma
      -2\hat g_S\hat p_\beta\right)\,.
\end{aligned}
    \label{eq:eom-mode}
\end{equation}
\Cref{eq:eom-orbit,eq:eom-mode}, together with the waveform and
flux of \cref{sec:eob-radiation} and the initial data of
\cref{app:numerical-initial-data}, form the closed system that we
integrate numerically.

\subsection{Initial data}
\label{app:numerical-initial-data}

The evolution starts at a separation $r_0$ with the modes on the driven
solution of \cref{sec:eob-driven-free}, evaluated with the inspiral rate
$\dot r$ of the orbit. The initial $\gamma_a$, $p_{\alpha_a}$ and
$p_{\beta_a}$ are therefore small but not zero. Given $g_H$, $g_S$, the
forces $F_\alpha$, $F_\beta$ of \cref{eq:Falpha-eob,eq:Fbeta-eob} and
$\dot r$, the amplitudes $b_0^{\rm d}$, $b_\pm^{\rm d}$ of
\cref{eq:local-driven-amplitudes} follow, with the sweep rates of
\cref{eq:D-orbit}, and \cref{eq:b-mode-transformation} inverts them for
each mode $a$ to
\begin{equation}
\begin{aligned}
    \alpha_a&=2\,{\rm Re}\,b_0^{\rm d}\,,&
    p_{\alpha_a}&=\frac{2\,{\rm Im}\,b_0^{\rm d}}{\kappa_0}\,,\\
    \beta_a&={\rm Re}\left(b_+^{\rm d}+b_-^{\rm d}\right)\,,&
    p_{\gamma_a}&=\frac{{\rm Re}\left(b_+^{\rm d}-b_-^{\rm d}\right)}{\kappa}\,,\\
    \gamma_a&={\rm Im}\left(b_-^{\rm d}-b_+^{\rm d}\right)\,,&
    p_{\beta_a}&=\frac{{\rm Im}\left(b_+^{\rm d}+b_-^{\rm d}\right)}{\kappa}\,,
\end{aligned}
    \label{eq:initial-driven-modes}
\end{equation}
where $\kappa=\lambda_a\omega_a$ and $\kappa_0=\kappa/3$. The rate
$\dot r$ is not known before the orbit is built, so the construction
below first uses the equilibrium tide, \cref{eq:initial-driven-modes} at
$\dot r=0$, where the sweep rates vanish, $\mathcal J=1$, and
\begin{equation}
    \alpha_a=-\frac{\lambda_aF_\alpha}{3g_H}\,,\qquad
    \beta_a=-\frac{\lambda_a\omega_a^2\,g_HF_\beta}{\Delta_+\Delta_-}\,,\qquad
    p_{\gamma_a}=\frac{2g_SF_\beta}{\Delta_+\Delta_-}\,,\qquad
    \gamma_a=p_{\alpha_a}=p_{\beta_a}=0\,,
    \label{eq:initial-equilibrium-modes}
\end{equation}
with $\Delta_\pm=g_H\omega_a\pm2g_S$. In the Newtonian limit these are
$\alpha_a=\lambda_am_2/(2r_0^3)$ and
$\beta_a=3\alpha_a/(1-4\Omega^2/\omega_a^2)$. The equilibrium has
$\gamma_a=0$; the imaginary parts of \cref{eq:initial-driven-modes} are
the lag of the tide behind the orbit, which is restored in the last step.

The orbit follows from two conditions. The circular condition
$\partial H_{\rm EOB}/\partial r=0$ at $p_{r_*}=0$ fixes $P_\phi$, and flux
balance fixes the inspiral rate,
\begin{equation}
    \dot r_{\rm pc}
    =\frac{\mathcal F_\phi}
      {\dd P_\phi^{\rm circ}/\dd r}\bigg|_{r_0,p_{r_*}=0}\,,
    \label{eq:post-circular-rdot}
\end{equation}
where $\mathcal F_\phi$ is the torque of \cref{eq:radiation-reaction-flux}
and $P_\phi^{\rm circ}(r)$ the circular angular momentum. The radial
momentum then solves the radial Hamilton equation
\begin{equation}
    \frac{A_{\rm pp}}{\sqrt{D_{\rm pp}}}
    \frac{\partial H_{\rm EOB}}{\partial p_{r_*}}
    =\dot r_{\rm pc}\,.
    \label{eq:numerical-prstar}
\end{equation}

In practice the steps are the following. (i) Solve
$\partial H_{\rm EOB}/\partial r=0$ for $P_\phi$ with the modes at zero, by
a secant iteration from the Newtonian estimate $\mu\sqrt{Mr_0}$. (ii) Evaluate $g_H$,
$g_S$, $F_\alpha$ and $F_\beta$ on that orbit, insert
\cref{eq:initial-equilibrium-modes}, and solve for $P_\phi$ again. Since
$g_H$, $g_S$ and the forces are derivatives of $H_{\rm EOB}$, which
contains the tide, \cref{eq:initial-equilibrium-modes} is a self-consistent
relation. We repeat this step, with the coefficients re-evaluated on the
new state, until the amplitudes and $P_\phi$ change by less than
$10^{-12}$. The tide changes the coefficients by
$\mathcal O(\lambda_au^5)$, so two or three passes suffice. (iii)
Evaluate \cref{eq:post-circular-rdot}, with the derivative by central
differences of steps (i)--(ii) at $r_0(1\pm10^{-4})$. (iv) Replace the
equilibrium tide by \cref{eq:initial-driven-modes} at
$\dot r=\dot r_{\rm pc}$, which sets $\gamma_a$, $p_{\alpha_a}$ and
$p_{\beta_a}$, and solve \cref{eq:numerical-prstar} for $p_{r_*}$ by a
bracketed root search (Brent's method) between zero and twice the value
that the expansion of \cref{eq:numerical-prstar} to first order in
$p_{r_*}$ gives.

\section{Angular structure of the crustal strain}\label{app:strain-reconstruction}

The angular expansion uses
$Y_{20}=\sqrt{5/(16\pi)}(3\cos^2\theta-1)$ and
$Y_{22}=\sqrt{15/(32\pi)}\sin^2\theta\,e^{2i\varphi}$.
The fields $T_{1,m},T_{2,m},V_m$ are the coefficients of these
harmonics and their angular derivatives after the sum in
\cref{eq:strain-reconstruction-main}. In particular,
\begin{equation}
\begin{aligned}
 T_{j,m}(r,t)&=\sum_n a_{nm}(t)T_{j,n}(r)
                         +T_{j,\mathrm{bg},m}(r,t)\,,\qquad j=1,2\,,\\
 V_m(r,t)&=\sum_n a_{nm}(t)V_n(r)+V_{\mathrm{bg},m}(r,t)\,.
\end{aligned}
    \label{eq:strain-harmonic-coefficients}
\end{equation}
The $m=0$ coefficients are real, and the $m=-2$ coefficients are
the complex conjugates of those for $m=2$. Contracting the real
tensor in \cref{eq:strain-real-tensor} according to
\cref{eq:strain-total-main} gives
\begin{equation}
    \sigma^2=\sigma_0^2+\sigma_2^2
                    +I_{22}+I_{02}\,,
    \label{eq:strain-explicit-total}
\end{equation}
where
\begin{widetext}
\begin{align}
 \sigma_0^2&=\frac{45}{256\pi}\left[(3\cos^2\theta-1)^2\frac{T_{1,0}^2}{\mu_{\rm s}^2}
   +48e^{-\lambda}\sin^2\theta\cos^2\theta\,\frac{T_{2,0}^2}{\mu_{\rm s}^2}
   +48\sin^4\theta\,V_0^2\right]\,,
 \label{eq:strain-axisymmetric-explicit}\\
 \sigma_2^2&=\frac{45}{256\pi}\left[3\sin^4\theta\,\frac{|T_{1,2}|^2}{\mu_{\rm s}^2}
   +16e^{-\lambda}\sin^2\theta(1+\cos^2\theta)\,\frac{|T_{2,2}|^2}{\mu_{\rm s}^2}
   +16(1+6\cos^2\theta+\cos^4\theta)\,|V_2|^2\right]\,,
 \label{eq:strain-pair-amplitude-explicit}\\
 I_{22}&=\frac{45\sin^4\theta}{256\pi}\,
   \operatorname{Re}\left\{e^{4i\varphi}\left[3\frac{T_{1,2}^2}{\mu_{\rm s}^2}
    -16e^{-\lambda}\frac{T_{2,2}^2}{\mu_{\rm s}^2}+16V_2^2\right]\right\}\,,
 \label{eq:strain-pair-interference}\\
 I_{02}&=\frac{45\sqrt6\sin^2\theta}{128\pi}\,
   \operatorname{Re}\left\{e^{2i\varphi}\left[(3\cos^2\theta-1)\frac{T_{1,0}T_{1,2}}{\mu_{\rm s}^2}
   -16e^{-\lambda}\cos^2\theta\,\frac{T_{2,0}T_{2,2}}{\mu_{\rm s}^2}
   +16(1+\cos^2\theta)\,V_0V_2\right]\right\}\,.
 \label{eq:strain-axisymmetric-interference}
\end{align}
\end{widetext}
Here $\sigma_2^2$ is the azimuth-independent contribution of the
$m=\pm2$ pair. It equals the square of the strain amplitude defined by
\citet{Gao:2025aqo}, evaluated with the summed complex fields.
The terms $I_{22}$ and $I_{02}$ describe interference within the
$m=\pm2$ pair and with the axisymmetric field. Averaging over $\varphi$
removes these terms, giving
$\langle\sigma^2\rangle_\varphi=\sigma_0^2+\sigma_2^2$.
The local strain retains both interference terms.
\end{widetext}

\end{document}